\pdfoutput=1

\documentclass[numberedappendix,apj]{emulateapj}
\usepackage{apjfonts}
\usepackage[colorlinks=true,urlcolor=blue,linkcolor=blue,citecolor=blue]{hyperref}
\usepackage{amsmath}
\usepackage{natbib}
\usepackage{graphicx}
\usepackage{subfigure}
\usepackage{algorithm}
\usepackage{algpseudocode}

\usepackage{etoolbox}
\makeatletter
\patchcmd{\NAT@citex}
  {\@citea\NAT@hyper@{\NAT@nmfmt{\NAT@nm}\NAT@date}}
  {\@citea\NAT@nmfmt{\NAT@nm}\NAT@hyper@{\NAT@date}}
  {}
  {}
\patchcmd{\NAT@citex}
  {\@citea\NAT@hyper@{%
     \NAT@nmfmt{\NAT@nm}%
     \hyper@natlinkbreak{\NAT@aysep\NAT@spacechar}{\@citeb\@extra@b@citeb}%
     \NAT@date}}
  {\@citea\NAT@nmfmt{\NAT@nm}%
   \NAT@aysep\NAT@spacechar%
   \NAT@hyper@{\NAT@date}}
  {}
  {}
\patchcmd{\NAT@citex}
  {\@citea\NAT@hyper@{%
     \NAT@nmfmt{\NAT@nm}%
     \hyper@natlinkbreak{\NAT@spacechar\NAT@@open\if*#1*\else#1\NAT@spacechar\fi}%
       {\@citeb\@extra@b@citeb}%
     \NAT@date}}
  {\@citea\NAT@nmfmt{\NAT@nm}%
   \NAT@spacechar\NAT@@open\if*#1*\else#1\NAT@spacechar\fi%
   \NAT@hyper@{\NAT@date}}
  {}
  {}
\makeatother

\graphicspath{{.}}

\begin{document}


\shorttitle{Splashback shells and splashback radii of individual CDM halos}
\shortauthors{Mansfield, Kravtsov \& Diemer}

\title{Splashback shells of Cold Dark Matter halos}

\author{Philip Mansfield\altaffilmark{1,2,$\star$}, Andrey V. Kravtsov\altaffilmark{1,2,3} and Benedikt Diemer\altaffilmark{4}}


\altaffiltext{1}{Department of Astronomy \& Astrophysics, The University of Chicago, Chicago, IL 60637 USA}
\altaffiltext{2}{Kavli Institute for Cosmological Physics, The University of Chicago, Chicago, IL 60637 USA}
\altaffiltext{3}{Enrico Fermi Institute, The University of Chicago, Chicago, IL 60637 USA}
\altaffiltext{4}{Institute for Theory and Computation, Harvard-Smithsonian Center for Astrophysics, 60 Garden St., Cambridge, MA 02138, USA}
\altaffiltext{$\star$}{mansfield@uchicago.edu}


\begin{abstract}

The density field in the outskirts of dark matter halos is discontinuous
due to a caustic formed by matter at its first apocenter after infall.
In this paper,
we present an algorithm to identify the ``splashback shell'' formed by these
apocenters in individual simulated halos
using only a single snapshot of the density field.
We implement this algorithm in the code \textsc{Shellfish}
(SHELL Finding In Spheroidal Halos) and demonstrate
that the code identifies splashback shells correctly and measures their
properties with an accuracy of $<5\%$ for halos with more than 50,000 particles
and mass accretion rates of $\Gamma_\textrm{DK14}>0.5$.
Using \textsc{Shellfish}, we present the first estimates for several basic
properties of individual
splashback shells, such as radius, $R_\textrm{sp}$, mass, and overdensity, and
provide fits to the distribution of these quantities as
functions of $\Gamma_\textrm{DK14}$, $\nu_\textrm{200m}$, and $z.$
We confirm previous findings that $R_\textrm{sp}$ decreases with
increasing $\Gamma_\textrm{DK14}$,
but we show that independent of accretion rate, it also decreases with
increasing $\nu_\textrm{200m}$.
We also study the 3D structures of these shells and find that
they generally have non-ellipsoidal oval shapes.
We find that splashback radii
estimated by \textsc{Shellfish} are
$20\%-30\%$ larger than those estimated
in previous studies from stacked density profiles at high accretion rates.
We demonstrate that the latter are biased low
due to the contribution of high-mass subhalos to these profiles and show
that using the median instead of mean density in each radial bin
mitigates the effect of substructure on density profiles and removes the bias.

\end{abstract}

\section{Introduction}
\label{sec:introduction}

In the Cold Dark Matter (CDM) paradigm of structure formation, dark matter
halos form via the collapse of density peaks in the initial random Gaussian
perturbation field. In the commonly used ``tophat model'' the peak density 
contrast profile is approximated as uniform within a given radius \citep[e.g.,][]{tolman_1934}. 
The constant overdensity in such approximations results in a uniform collapse
time for different radial shells and a single well-defined collapse time for
the peak. This, along with the assumption that virial equilibrium is reached
immediately following collapse, allows one to predict the density contrast
within the boundary of the collapsed objects
\citep{gunn_gott_72,heath_1977, lahav_et_al_1991}.

Accordingly, the most commonly used boundary definition for CDM halos
is the radius, $R_\Delta,$ enclosing a given density contrast
$\Delta\equiv \rho(<r)/\rho_{\rm ref}$, where $\rho_{\rm ref}$ is a reference
density, taken to be
either the mean density of the universe, $\rho_{\rm m}$, or the critical
density for closure, $\rho_{\rm crit}$, at the redshift of observation. The
corresponding enclosed mass is given by 
\begin{equation}
    \label{eq:overdensity}
    M(< R_{\Delta})= \Delta\rho_{\rm ref}\ \frac{4}{3}\pi R_\Delta^3.
\end{equation}
The value of $\Delta$ is usually motivated by tophat collapse models.

However, the overdensity profile in real Gaussian peaks is not constant, but
decreases with increasing radius
\citep[see, e.g., Figure 2 in][]{dalal_et_al_2010}. Because 
the overdensity within a given radius controls the timing of the collapse, the
collapse of different radial shells in such peaks is extended in time. Real
halos also undergo mergers during their formation, which further redistribute
mass within them. Real CDM halos thus do not have an edge at the density
contrast predicted by simple uniform peak collapse models
\citep[see, e.g.,][]{kravtsov_borgani_2012,more_et_al_2015},
meaning that $R_\Delta$ radii are a rather arbitrary definition of halo
extent and do not correspond to any particular feature in the density
profile or in the profiles of other physical properties
\citep[e.g.,][]{diemer_et_al_2013b}. This arbitrariness 
may be problematic when this radius is used
to classify objects into groups which are meant to be qualitatively distinct
from one another, such as subhalos and isolated halos. 
Indeed, multiple recent studies have suggested that a significant fraction of
the halo assembly bias effect may be due to the fact that some subhalos
which have orbited larger hosts are misclassified as isolated
halos when $R_\Delta$ is used as a halo boundary for classification 
\citep{wang_et_al_2009,wetzel_et_al_2014,sunayama_et_al_2016,zentner_et_al_2016}.
However, these so-called ``backsplash'' halos would still
necessarily be contained within their hosts' splashback shells, meaning that
switching to a splashback-based definition could help alleviate this issue.

Furthermore, regardless of the choice of $\Delta$ or
$\rho_{\rm ref}$, contrast-based radius and mass definitions encounter several
problems when the mass accretion histories of halos are estimated. First, as
mentioned above, during major mergers there is mass redistribution within halos,
with a non-trivial amount of mass moving to radii outside of $R_\Delta$ for 
typical values of $\Delta$ \citep*{kazantzidis_et_al_2006}. This causes
spherical overdensity masses to be non-additive during mergers in excess to the
degree that would be expected purely from slingshot processes. Second, the
evolution of both $\rho_{\rm m}$ and $\rho_{\rm crit}$ with time causes evolution
in $R_\Delta$ and  $M_\Delta$, even for completely static density profiles.
This ``pseudo-evolution'' of halo radius and mass typically results in the near
doubling of mass of Milky Way-sized halos between $z=1$ and $z=0$, even when
there is no accretion of new mass \citep{diemer_et_al_2013}.

Given the problems with the standard $R_\Delta$ definition, one can ask
whether there is a more physical way to define halo boundary, one which would
separate the matter that has already collapsed (i.e., orbited within
halo at least once) and matter that is still infalling onto halo for the first
time. In collapse models of spherical and ellipsoidal peaks with power law density
profiles, such a boundary exists and is associated with a sudden drop
in the density profile of collapsed halos 
\citep{fillmore_goldreich_1984,bertschinger_1985,adhikari_et_al_2014,shi_2016}. 
The drop is due to the caustic formed by the ``pile up'' of mass elements that
have just reached the apocenter of their first orbits and is thus the maximum
radius of matter that has orbited through halo at least once.

Recently, such drops in the density profile have also been detected in both
simulated and real CDM halos
\citep{diemer_kravtsov_2014,adhikari_et_al_2014,more_et_al_2015,more_et_al_2016,adhikari_et_al_2016}.
The most distant apocenters of orbits in real halos form a surface that we
will call the \emph{splashback shell}. This shell can be viewed as the halo
boundary. Due to the assumption of spherical symmetry, all previous studies
have necessarily been restricted to analyzing the characteristic scale of this
shell, the \emph{splashback radius}, $R_{\rm sp}$.

The primary challenge in using the splashback shell as a physical boundary
definition for halos is that it is technically challenging to detect
and quantify
in individual objects, both in cosmological simulations and in observations. 
The key problem is that splashback shells are generally located at low
densities,
where the presence of individual neighboring halos or filaments can complicate the
interpretation of the density field.  

Consequently, analyses of the splashback radius have so far been carried out
using stacked radial density profiles of either mass or subhalo abundance
\citep{diemer_kravtsov_2014,adhikari_et_al_2014,adhikari_et_al_2016,more_et_al_2015,more_et_al_2016}.
After stacking, $R_{\rm sp}$ for the population is
operationally defined as the radius of the steepest logarithmic slope,
${\rm d} \ln{\rho}/ {\rm d} \ln{r}$
(or ${\rm d} \ln{n_{\rm sub}}/ {\rm d} \ln{r}$). In principle, this procedure
averages out the noise in the individual profiles, allowing for comparisons of
the splashback radius between different halo populations. However, stacking of 
different halo profiles can also ``wash out'' the sharp density gradient
associated with the splashback shells, if such shells exhibit scatter for
individual halos. 

Studies of the splashback radius based on stacked density profiles have shown
that there is a strong relation between $R_{\rm sp}/R_{\rm 200m}$ and 
halo mass accretion rate, $\Gamma_{\rm DK14}$,
\citep{diemer_kravtsov_2014,more_et_al_2015}. Where $\Gamma_{\rm DK14}$ is defined as
\begin{equation}
    \Gamma_{\rm DK14} \equiv \frac{\ln{M_{\rm 200m}(z_{i+1})} - \ln{M_{\rm 200m}(z_i)}}{
    \ln{a(z_{i+1})} - \ln{a(z_i)}},
 \label{eq:Gamma} 
\end{equation}
here $z_i$ come from a set of redshift intervals which are separated by
roughly a dynamical time. Previous studies have used the intervals
$z_i = \{0, 0.5, 1, 2, 4\},$ a convention which we shall continue to use in
this paper (although future studies may benefit strongly from revisiting
this choice in definition). Such a dependence is expected theoretically due
to the contraction of particle orbits in a rapidly deepening potential of
high-$\Gamma_{\rm DK14}$ halos \citep{diemer_kravtsov_2014,adhikari_et_al_2014}. 

Hints of density steepening due to the splashback radius in the mass and
galaxy distribution  around individual clusters have been reported in several 
recent studies \citep{rines_13, tully_15,patej_loeb_16,umetsu_diemer_2017}. 
Interestingly, the first reliable observational estimates of the
splashback radius from
the radial number density profiles of satellite galaxies in clusters are in
tension with the predictions of simulations \citep{more_et_al_2016}.

The operational simplicity of the stacked-profile approach makes it very
useful, particularly when comparing simulations to observations, but it
is not without weaknesses. First, spherical averaging discards all information
about the shapes of the  splashback shells, even though the filamentary nature
of the cosmic web causes accretion to be highly aspherical, which implies
that splashback shells should also be highly aspherical.
Second, the stacking procedure removes information about individual halos,
making it impossible to study the evolution of a single halo's shell over time,
the properties of subhalos contained within shells, or the scatter around mean
relations. Third, the relationship between the splashback radius estimated from
the stacked profiles and the underlying distribution of individual splashback
radii is unknown and can be complicated. In particular, as we show in
section \ref{sec:median_prof}, the contribution of massive subhalos in
a minority of individual density profiles introduces significant bias in the
estimate of the splashback radius derived from stacked profiles.
 
To address these issues and to explore the properties of splashback shells around
individual halos, in this paper we present an algorithm which identifies the
splashback shells around individual halos using single particle snapshots from
cosmological $N$-body simulations, and an implementation of the
algorithm in the code \textsc{Shellfish}
(SHELL Finding In Spheroidal Halos), which we use to
generate halo catalogs with measured splashback shells and perform
analyses of their basic properties, such as radius and shape, and quantify
their relationships to other halo properties, such as mass accretion rate and
peak height. A public version of \textsc{Shellfish}, along with tutorials and
documentation can be found at
\texttt{github.com/phil-mansfield/shellfish} with a Digital Object Identifier
(DOI) given by \citet{SHELLFISH_2017}.

This paper is organized as follows. An overview of our method is shown in
Figure \ref{fig:algo} and our key result, the $\Gamma_{\rm DK14}$ - $R_{\rm sp}$ relation
for individual halos, is shown in Figure \ref{fig:r_centroid}.
In section \ref{sec:methods} we describe our algorithm to identify the
splashback shells from a halo's particle distribution, in section
\ref{sec:tests} we present extensive tests of the correctness and convergence
properties of the shells identified by our implementation of the
algorithm. In section \ref{sec:results} we discuss the shapes of the splashback
shells and present the relation between shell size and mass accretion rate. We
compare the latter relation to that derived from the stacked profiles, and show
that the stacking introduces significant bias in the estimates of the splashback
radius of rapidly growing halos. We summarize our results in section
\ref{sec:summary}. Appendix \ref{sec:intersection} contains a
description of a high performance ray-tracing algorithm that we
developed as a component of \textsc{Shellfish}.

A reader not interested in the details of the algorithm itself, but only in
the properties of identified shells can skip directly to section
\ref{sec:results}.
We caution, however, that proper interpretation of the issues
discussed in section \ref{sec:results} requires at least a basic understanding
of our shell finding algorithm. 

\section{Methods}
\label{sec:methods}

\subsection{Simulations}

The analysis in this paper uses a subset of the suite of simulations first
introduced in \citet{diemer_kravtsov_2014}. These simulations have box
sizes between
$62.5h^{-1}\,\rm Mpc$ and $500h^{-1}\,\rm Mpc$, allowing us to study halos with
a wide range of masses and accretion rates and use the same cosmological parameters
as the Bolshoi simulation suite \citep{klypin_et_al_2011}:
$\Omega_{\rm m} = 1 - \Omega_\Lambda=0.3$, $\Omega_{\rm b} = 0.0469$,
$H_0$ = 70 km s$^{-1}$ Mpc$^{-1}$, $\sigma_8 = 0.82$, and $n_{\rm s} = 0.95$
\citep{komatsu_et_al_2011}.
All simulations followed the evolution of $1024^3$ particles using the Gadget-2
code \citep{springel_2005}, starting at $z=49$, and were run with the timestep
parameter of $\eta=0.025.$ Simulation-specific parameters can be found in Table
\ref{tab:sims}. Halo catalogs were generated using the \textsc{Rockstar} halo
finder \citep{behroozi_et_al_2013a} and main progenitor lines were found through
the merger tree code \textsc{consistent-trees} \citep{behroozi_et_al_2013b}.

\begin{table}
  \centering
  \caption{Simulation parameters}
  \label{tab:sims}
  \begin{tabular}{cccccc}
  \hline
  \hline\\
  Name & $L$  & $m_{\rm p}$ & $\epsilon$ & $M_{\rm 200m, min}$ & $M_{\rm 200m, max}$ \\
  Units & $h^{-1}$ Mpc & $h^{-1}\, M_\odot$ & $h^{-1}$ kpc &   $h^{-1}\, M_\odot$ & $h^{-1}\, M_\odot$\\
  \\
  \hline\\
  L0500 & 500 & $8.7\times 10^9$ & 14.0 & $4 \times 10^{14}$ & - \\
  L0250 & 250 & $1.1\times 10^9$ & 5.8 & $5 \times 10^{13}$ & $2 \times 10^{14}$ \\
  L0125 & 125 & $1.4\times 10^8$ & 2.4 & $7 \times 10^{12}$ & $5 \times 10^{13}$ \\
  L0063 & 62.5 & $1.7\times 10^7$ & 1.0 & $9 \times 10^{11}$ & $7 \times 10^{12}$ \\ 
 \\
 \hline
  \\
  \end{tabular}
  \tablecomments{Parameters of the simulations used for our testing and
  analysis:
   $L$ is the box size, $m_{\rm p}$ is the particle mass,
  and $\epsilon$ is the force softening length. $M_{\rm 200m, min}$ and
  $M_{\rm 200m, max}$ indicate the mass range of halos from each simulation.
  These simulations were first presented in \citet{diemer_kravtsov_2014}.}
\end{table}

\subsection{Algorithm Description}
\label{sec:algo_descr}

\begin{figure*}
   \centering
   \subfigure[]{
        \label{fig:raw_field}
        \includegraphics[width=0.35\textwidth]{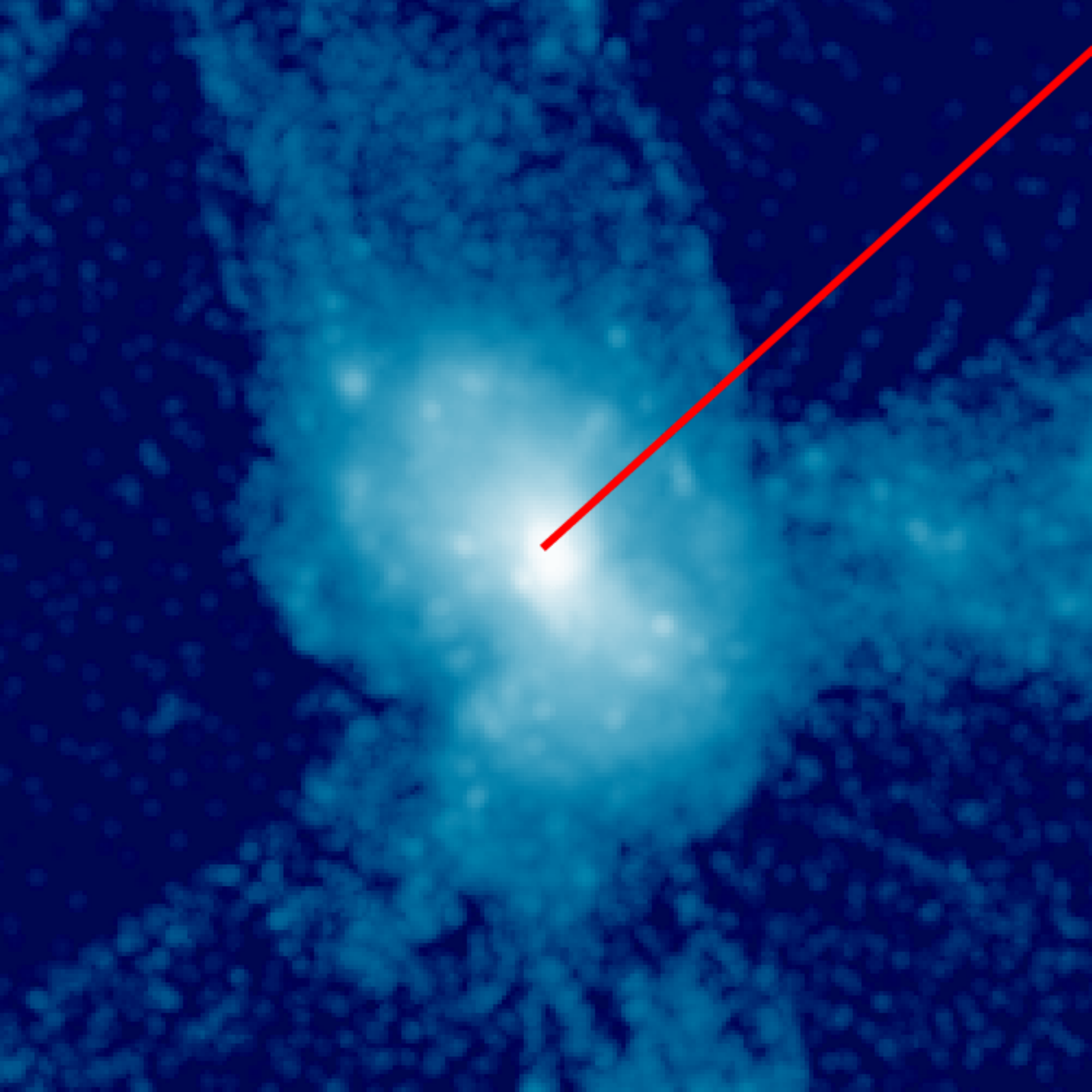}
   }
   \subfigure[]{
        \label{fig:los_prof}
        \includegraphics[width=0.35\textwidth]{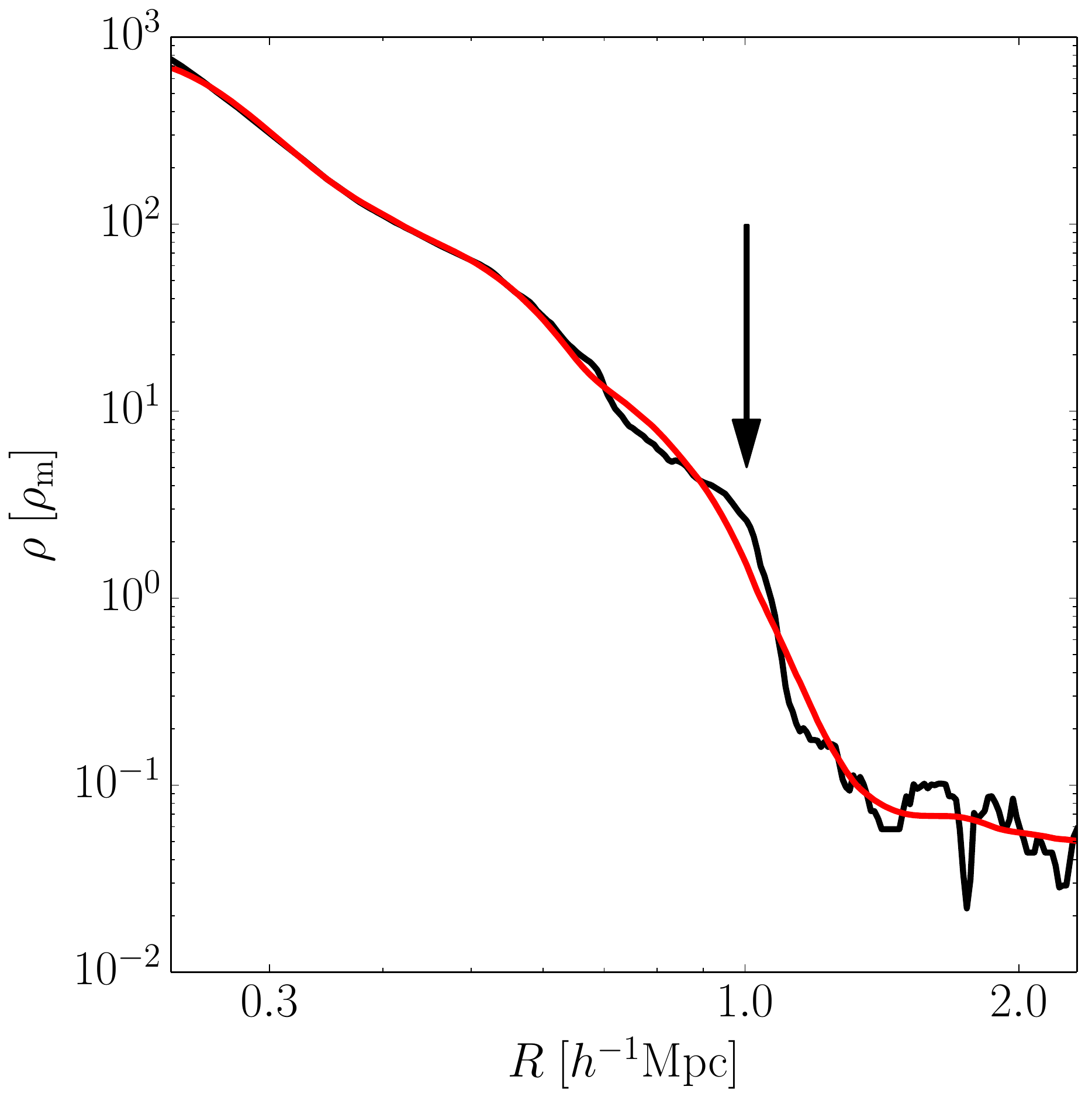}
   } \\
   \subfigure[]{
        \label{fig:filtered_points}
        \includegraphics[width=0.35\textwidth]{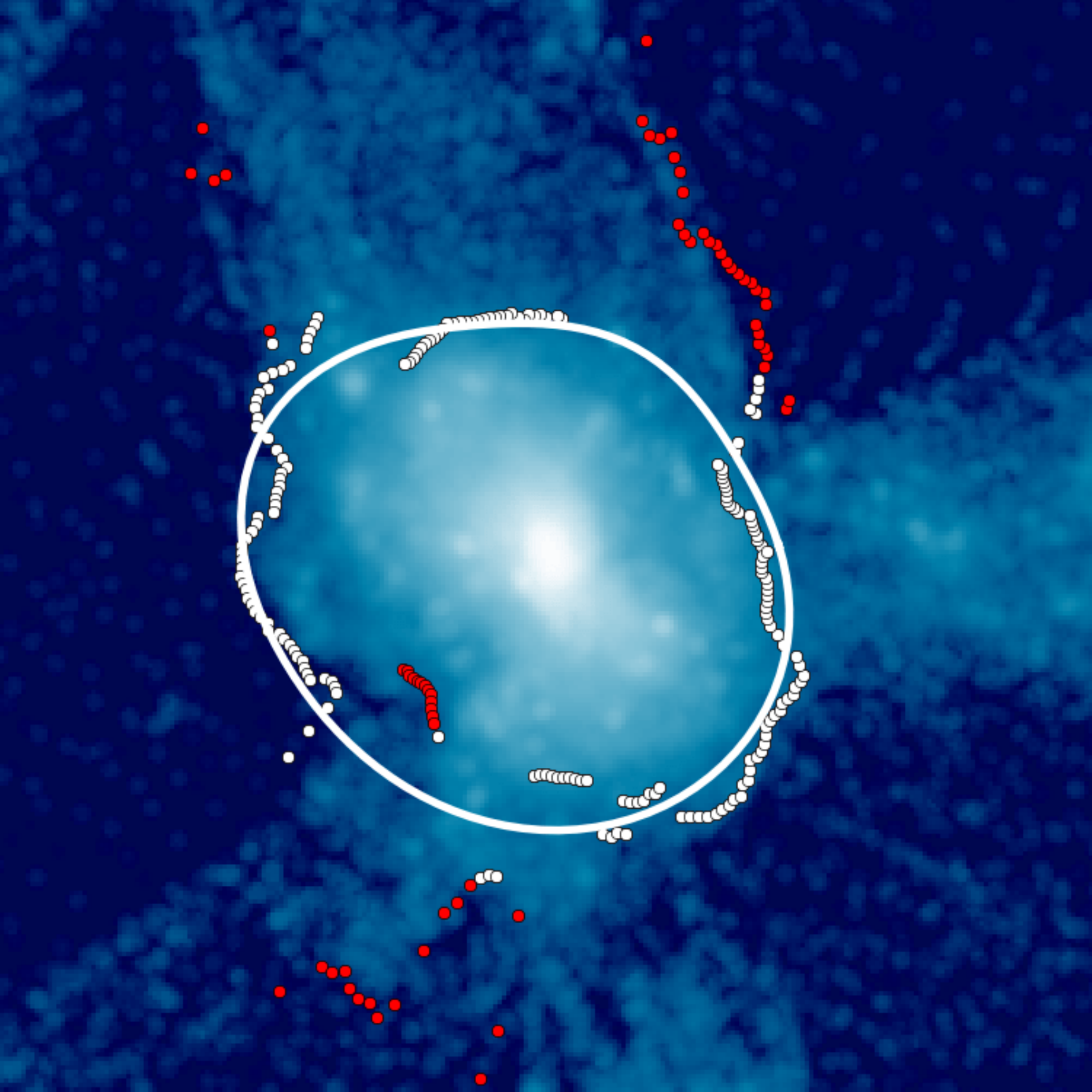}
   }
   \subfigure[]{
        \label{fig:shell_fit}
        \includegraphics[width=0.35\textwidth]{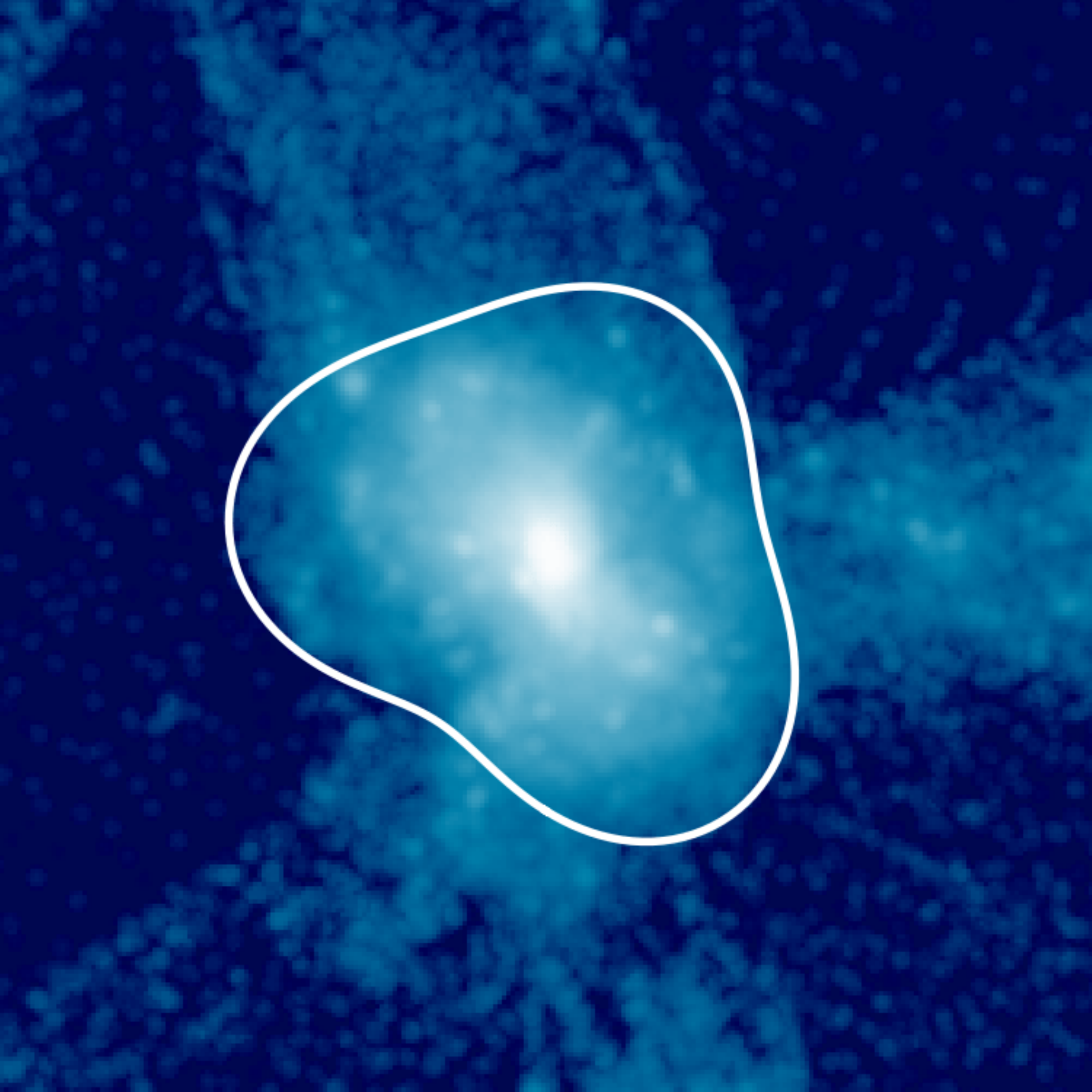}
   } \\
   \caption{An overview of the steps in our shell-finding
   algorithm for a cluster-sized halo (This halo is also shown in
   Figure \ref{fig:pass_4} below).
   Figure \ref{fig:raw_field} shows a random line of sight traced through
   this halo's density field (see \S \ref{sec:density_est} and Appendix
   \ref{sec:intersection}). Figures \ref{fig:los_prof}
   shows a density profile measured along
   along this line of sight before smoothing (black line) and after smoothing
   with a Savitzky-Golay filter (red line). The arrow indicates the point
   of steepest slope in the
   smoothed profile (see \S \ref{sec:smooth}). Figure \ref{fig:filtered_points}
   shows the points of steepest slope for the 256 lines of
   sight in the viewing plane and
   shows the point classification that the algorithm generates for these
   points (see Appendix
   \ref{sec:filter_algo}). The white curve shows the filtering spline created
   during the point selection process. Points which are close enough to this
   curve to pass the filter are shown in white and those which are too far away
   are shown in red. Figure \ref{fig:shell_fit} shows the cross-section of the
   best fit Penna-Dines surface from the overall distribution of splashback 
   points from 100 randomly oriented planes in which such a procedure was
   carried out (see \S \ref{sec:fit}).
   See the text in the corresponding sections for details. All analysis is
   done with
   the parameter values listed in Table \ref{tab:param}, but the underlying
   images are rendered using spherical kernels of radius $0.05R_{\rm 200m}$ to
   make the structures around halos more clear.}
   \label{fig:algo}
\end{figure*}

Our aim is to develop an algorithm which can identify splashback shells around
halos using only their density distribution at a single point in time. In other
words, this will be an algorithm which uses no dynamical information
about the halo's particles and will rely solely on identifying the density
caustic generated by the splashback shell. This restriction would allow such
an algorithm to work on simulations that are only sparsely sampled in time.

Relaxing this restriction allows for alternative measurements of $R_{\rm sp}$
which can leverage the full dynamical information of the simulation.
For example, \citet{diemer_2017} develops an algorithm, \textsc{Sparta},
for finding splashback radii by locating the apocenters of orbiting particles
which requires access to approximately 100 snapshots over the lifetime of the
target halos. An extended comparison between \textsc{Sparta} and
\textsc{Shellfish} can be found in \citet{diemer_et_al_2017}.

Below we describe such an algorithm which does not require any dynamical
information and demonstrate that it identifies correct 
splashback shells, provided that target halos are resolved with a sufficient
number of particles (see section \ref{sec:tests}) and provided that target
halos are not embedded in very dense environments (see section
\ref{sec:median_prof}).

Specifically, our algorithm consists of four steps:
\begin{enumerate}
    \item[1.] The density field is sampled along tens of thousands 1-d lines of
       sight anchored at the center of a halo. The specific design decisions
       governing how the lines of sight are oriented and how densities along
       them are estimated are described in section \ref{sec:density_est} and
       Appendix \ref{sec:intersection}, respectively, and are depicted in
       Figure \ref{fig:raw_field}.
    \item[2.] The locations of the steepest slope in the density profiles
       of each line of sight are estimated using a smoothing filter. This part
       of the algorithm is described in section \ref{sec:smooth} and is
       depicted in Figure \ref{fig:filtered_points}.
    \item[3.] The set of profiles is pruned to remove the profiles where the
       point of steepest slope corresponds to the splashback associated with a
       nearby halo or filament.  The pruning procedure is described in section
       \ref{sec:filter} and Appendix \ref{sec:filter_algo} and is depicted in
       Figure \ref{fig:filtered_points}.
    \item[4.] We fit the 3-d shape of the shell with a smooth, flexible,
       functional form using the locations of the steepest slope in the
       profiles that remain after the pruning step. This is described in
       section \ref{sec:fit} and is depicted in Figure \ref{fig:shell_fit}.
\end{enumerate}
The design choices made in step 1 are the most important for ensuring good
performance of the algorithm and the design choices made in step 3 are the
most important for ensuring that the identified shells are correct.

The free parameters of the algorithm that will be introduced and discussed in
the subsequent sections are summarized in Table \ref{tab:param}. The
logic and procedures of specific parameter choices are discussed in Appendix
\label{sec:setting_params_app}.

\subsubsection{Density estimation along lines of sight}
\label{sec:density_est}

To construct a density profile along a given line of sight we must choose a
way to interpolate particle positions and masses onto that line. For
simplicity, we choose to approximate particles as tophat spheres of radius
$R_{\rm kernel}$ uniform
density.  Other choices, such as tetrahedral, trilinear, or tricubic
tessellations of phase space \citep[e.g.,][]{abel_et_al_12,hahn_angulo_2016},
are also implemented in \textsc{Shellfish} and could in principle be
used in this work. However, we find that these
estimators converge slowly and do not allow splashback shells for halos
with $N_{\rm 200m} \lesssim 10^7$ to be identified reliably and thus do
not use them in practice. A detailed convergence study of phase space density
estimators will be the subject of future work.

The algorithm represents every line of sight as an array of $N_{\rm bins}$
bins logarithmically distributed between the radii $R_{\rm min}$ and
$R_{\rm max}$. The density along a line of sight, $l$, which passes through
a set of constant-density spheres is given by
\begin{equation}
        \label{eq:los_rho}
        \rho_l(r) = \sum_{i=0}^{i<N}\mathbb{I}_{{\rm intr},il}\rho_i
        H(r - r_{\rm in})H(r_{\rm out} - r).
\end{equation}
Here, $i$ indexes over all particles, $\mathbb{I}_{{\rm intr},il}$ is an
indicator function which is $1$ if $l$ intersects with the sphere of particle
$i$ and is $0$ otherwise, $\rho_i$ is the density of sphere $i$, $H$ is the
Heaviside step function, and $r_{\rm in}$ and $r_{\rm out}$ are the
distances to entrance and exit intersection points of $l$ for a given sphere,
respectively.

Evaluating Equation~\ref{eq:los_rho} is easy if a
conventional estimator (such as cloud-in-cell or SPH) is used to write
densities to an intermediate grid before they are translated onto the
lines of sight, since the grid cell that corresponds to a point
at radius $r$ of given ray can be calculated in $O(1)$ operations. However,
using an intermediate grid has a number of disadvantages. First, 
maintaining the high-resolution grid required to accurately measure
the contours of the splashback shell consumes
a large amount of memory. This restricts the number of halos which
can be maintained in memory at once; when generating large catalogs
of shells, this can force particle catalogs to be read many times,
leading to a significant performance cost. Second, 
writing the density estimate to a grid is expensive as it
involves either an exact rasterization scheme (see, for example,
\citealt{powell_abel_14}) of the objects, or Monte Carlo sampling of
each solid with sufficiently many points to eliminate shot noise in each
cell. Both approaches also require that density estimates are
calculated for grid cells which are not intersected by any line of sight.
Third, introducing an intermediate grid reduces the fidelity of the
line of sight density estimates due to pixelation. This is most
apparent as small radii.

We find that in practice these three disadvantages, particularly the second,
are significant and make the use of grids for density estimation undesirable.
For this reason we evaluate Equation~\ref{eq:los_rho} by \emph{directly}
computing the intersection radii between every line of sight and every sphere
with no intermediary grid.
Attempting this evaluation naively would be computationally intensive, so we
use a specialty ray-tracing algorithm, described in the Appendix
\ref{sec:intersection}, which takes advantage of the fact that the vast
majority of the terms in Equation~\ref{eq:los_rho} are zero. This algorithm
speeds up density assignment by several orders of magnitude compared to both
the brute-force geometric approach and the grid-based approach, while still
maintaining a comparatively light memory footprint.

The nature of the ray-tracing algorithm requires that the lines of
sight are confined in $N_{\rm planes}$ planes and are uniformly
spaced in polar angle within these planes. Each plane then contains
$N_{\rm los}$ lines of sight within it.
This means that the line shown in in Figure \ref{fig:raw_field} could not
be evaluated alone and would need to be evaluated simultaneously along with
several hundred other other profiles within the viewing plane. This
turns out to be a convenient configuration for later steps in the shell
finding algorithm.

\subsubsection{Measuring the Point of the steepest slope for line of sight profiles}
\label{sec:smooth}

After the density estimation step, we smooth the density profiles of each
line of sight using a fourth order Savitzky-Golay filter 
\citep{savitzky_golay_1964} with a window length of
$N_{\rm SG}$ bins in $\log r$ - $\log \rho$ space.
A filter is necessary because a high precision determination of
$r_{\rm steep}$ requires that $N_{\rm bin}$ be large, but using a large number of bins allows for noise in
low-density regions. For bins in which $\rho(r) = 0$,
the density is set equal to a small background density value,
$\rho_{\rm bg}$. Once the density profile of a line of sight is smoothed, we
find the radius of the steepest logarithmic slope, $r_{\rm steep}$.

We choose to use a Savitzky-Golay filter because it is effective at removing
small scale noise and because it generally doesn't move the location of the
point of steepest slope, even for large window sizes.

We find that the best results are obtained
for $N_{\rm SG} \approx N_{\rm bin}/4$ to $N_{\rm bin}/2$, as this allows the
filter to remove even moderately large features, such as subhalos. The exact
value chosen is given in Table \ref{tab:param}. For most
lines of sight, the density drop associated with crossing the splashback shell
is the most prominent feature in the profile, and thus such an aggressive
filter window does not remove it. The smoothing process will flatten 
the slope at $r_{\rm steep}$, but the actual value of the slope is not used
by our algorithm.

This process is illustrated in Figure \ref{fig:los_prof}, which shows the line
of sight highlighted in Figure \ref{fig:raw_field}. The black curve shows the
raw profile after the density estimation step, the red curve shows the
profile after applying a Savitzky-Golay filter with a window size of
$N_{\rm SG} = N_{\rm bin}/2$. The vertical arrow shows
$r_{\rm steep}$ for the smoothed profile. This figure
demonstrates several key points. First, the discontinuity due the
splashback shell is very strong. Second, the unsmoothed profile contains
several points with slopes steeper than the splashback discontinuity due to
particle noise. Lastly, the location of $r_{\rm steep}$ has not moved
significantly between the smoothed and unsmoothed profiles.

As mentioned in section \ref{sec:density_est} (see also Appendix
\ref{sec:intersection}), the density estimation step of our algorithm
requires that lines of sight are confined to a set of planes.
The locations of $r_{\rm steep}$ for 256 such lines of sight
are shown in Figure \ref{fig:filtered_points}. This illustrates that, generally,
the values of $r_{\rm steep}$ found by this step are in good agreement with the
visual appearance of density discontinuities. However, some of the density 
discontinuities are clearly not associated with the halo itself but are due to 
nearby filaments or nearby halos. Although this happens in the minority of
lines of sight, these can bias the shape of the inferred splashback shell
significantly.  Therefore, the algorithm makes an additional step in which
lines of sight for which the steepest slope points are likely associated with
other halos and filaments are pruned from the set. 

\subsubsection{Filtering out problematic steepest slope points}
\label{sec:filter}

We remove lines of sight with points of steepest slope that are likely to be
associated with other halos and filaments candidate points through an
additional filtering step. Filaments have their own elongated splashback
shells which are created by the apocenters of 
matter accreted onto filaments from surrounding void regions. The density 
jumps associated with these surfaces are comparable to those found around halos.
Therefore, it is difficult to differentiate between steepest slope points
caused by central halos splashbacks and points caused by filament splashbacks
using only the information contained in a single line of sight profile.
We experimented with a number of different heuristic approaches of this type
and found that
they generally require extensive fine-tuning and are, at best, modestly
effective at removing filament points.

To classify the splashback points, we consider all of the splashback points
within a given plane simultaneously and filter out points which deviate too
sharply from the locations of their neighbors. We do this by heuristically
constructing a \emph{filtering loop}, a curve which smoothly passes close
to most of the plane's candidate points but which is too stiff to accommodate
sharp changes in radius. We then remove points which are too far away from the
filtering loop.
 
Our filtering algorithm employs a spline curve to approximate the shape of the
splashback in a given slice and is described in detail in Appendix 
\ref{sec:filter_algo}. The algorithm introduces two new free parameters,
$\eta$, which controls the strictness of the filter and the
``stiffness'' of the loop, and $N_{\rm rec}$, which affects the angular
resolution of the filtering loop. Larger values of $\eta$ will remove outliers
more aggressively, but would also likely prune a larger number of points
associated with halo. Qualitatively, points which come from features that
deviate by more than $R_{\rm max}/\eta$ from neighboring regions on angular
scales of $2\pi/2^{N_{\rm rec}}$ will be removed from the set of lines of sight.

\subsubsection{Fitting the shape of the splashback shell}
\label{sec:fit}

After the filtering step, we fit the remaining points using a family of
spheroidal functions introduced by
\citet[][hereafter ``Penna-Dines functions'']{penna_dines_2007}.
A Penna-Dines function of order $P$ is defined by
$2P^2$ coefficients, $c_{ijk}$, where $i$ and $j$ range from 0 to $P-1$ and
$k$ ranges from 0 to 1. The shape of a shell with a particular set of
coefficients is given by the function
\begin{equation}
    r(\phi, \theta) = \sum_{i, j = 0}^{P-1}\sum_{k=0}^1 c_{ijk}
                      \sin^{i+j}\theta\ \cos^k\theta\
                      \sin^j\phi\ \cos^i\phi,
\end{equation}
where $\theta$ is the
 polar angle and $\phi$ is the azimuthal angle.
Penna-Dines functions are similar to spherical harmonics in that
adding higher order terms allows for the representation of increasingly
aspherical shells. We choose to fit these functions because their low order
forms are qualitatively similar to the shapes found in splashback shells (this
class of functions is specifically designed to represent lobed shapes) and
because an optimal fit can be found through the relatively simple
and efficient pseudoinverse matrix operation. 

Namely, for a set of $N$ points with coordinates given by
$r_n = \sqrt{x_n^2 + y_n^2 + z_n^2}$, the best fit coefficients can be
computed by the operation
\begin{equation}
    c_{ijk} = r_n^{2P - 1} M^T(MM^T)^{-1}.
\end{equation}
Here, $r_n^{2P -1}$ is a height $N$ vector containing the radii of every point
and $M$ is a $N\ \times\ 2P^2$ matrix with elements
\begin{equation}
   M_{i + jP + kP^2, n} = r_n^{2P - 1 - i - j - k}x_n^iy_n^jz_n^k.
\end{equation}

\subsection{Definitions of basic splashback shell properties}
\label{sec:def}

While a full set of Penna-Dines coefficients is necessary for computing
subhalo/particle membership and for visualizing shells, it is also useful to
encapsulate key properties of the splashback shells in a few representative
parameters. To this end, we use a set of properties
which parameterize the shape of the splashback shells: $R_{\rm sp}$, the
volume-equivalent splashback radius; $\rho_{\rm sp}$,
the net density of shell; $a_{\rm sp}$, $b_{\rm sp}$, and
$c_{\rm sp}$, the inertia tensor equivalent major axes of the shell;
$E_{\rm sp}$, the shell ellipticity; and $A_{\rm sp}$, the shell
asphericity:
\begin{align}
    \label{eq:r_def}
    R_{\rm sp} &\equiv \left(\frac{3V_{\rm sp}}{4\pi}\right)^{1/3} \\
    \label{eq:rho_def}
    \rho_{\rm sp} &= M_{\rm sp}/V_{\rm sp}\\\
    \label{eq:axes_def}
    a_{\rm sp},~b_{\rm sp},~c_{\rm sp} &\equiv {\rm Axes}(I_x,~I_y,~I_z) \\
    \label{eq:E_def}
    E_{\rm sp} &\equiv \frac{a_{\rm sp}}{c_{\rm sp}} - 1 \\
    \label{eq:A_def}
    A_{\rm sp} &\equiv 1 - \frac{S_{\rm sp}}{(36\pi V_{\rm sp}^2)^{1/3}}
\end{align}
Here, $V_{\rm sp}$ is the volume enclosed by the shell, $M_{\rm sp}$ is the mass
of all the particles contained within the shell, $S_{\rm sp}$ is the surface area
of the shell, and Axes($I_x$, $I_y$, $I_z$) is a function which computes the axes
of a uniform density ellipsoidal shell which has the moments
of inertia $I_x$, $I_y$, and $I_z$. The construction of this function is
described in Appendix \ref{sec:ellipsoid}. In Equation~\ref{eq:E_def}, we take the
standard convention that $a_{\rm sp}$ is the major axis and $c_{\rm sp}$ is
the minor axis.

$E_{\rm sp}$ is defined such that it is zero for a sphere and increases for
increasingly elliptical shells. $A_{\rm sp}$ is defined such that it is zero for
a sphere and increases for increasingly aspherical shells. Our numerical
experiments with randomly-shaped shells indicate that it is probable that
prolate ellipsoids are the surfaces which minimize $A_{\rm sp}$ for a given
value of $E_{\rm sp}$. 

\begin{figure*}
   \centering
   \subfigure[]{
        \label{fig:pass_1}
   \includegraphics[width=0.7\columnwidth]{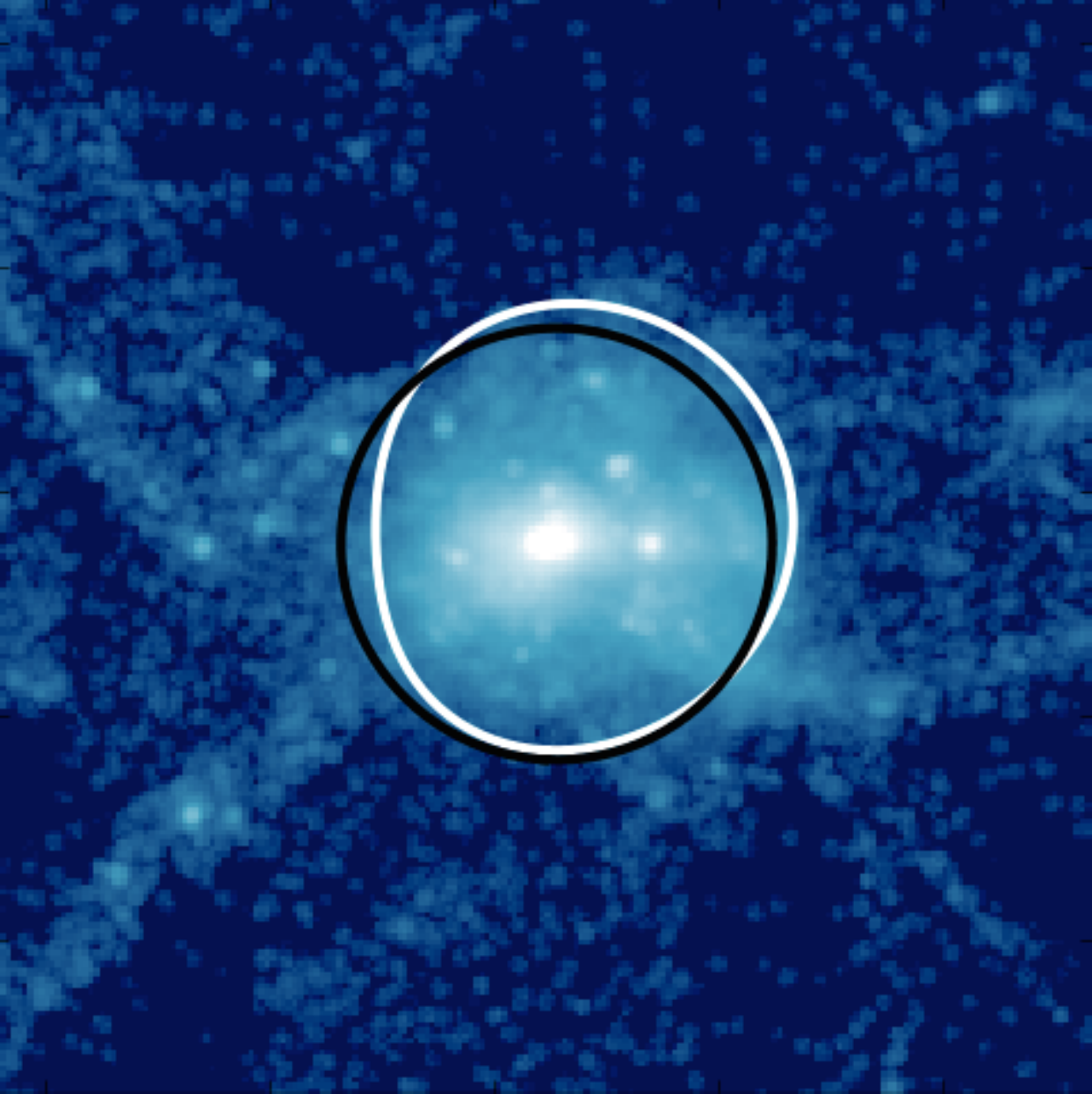}}
   \hspace{0.4cm}
   \subfigure[]{
        \label{fig:pass_2}
   \includegraphics[width=0.7\columnwidth]{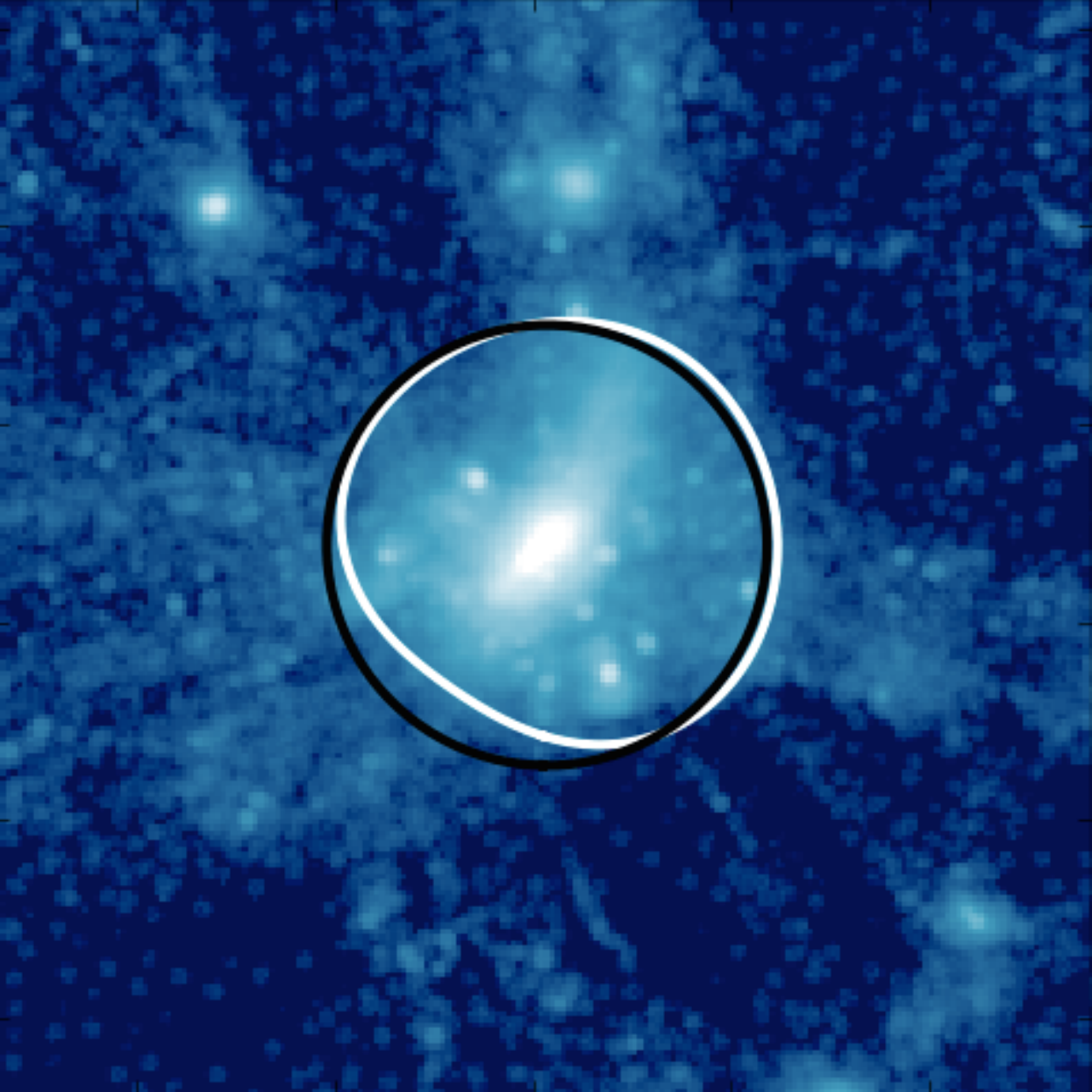}}\\
   \vspace{0.5cm}
   \subfigure[]{
        \label{fig:pass_3}
   \includegraphics[width=0.7\columnwidth]{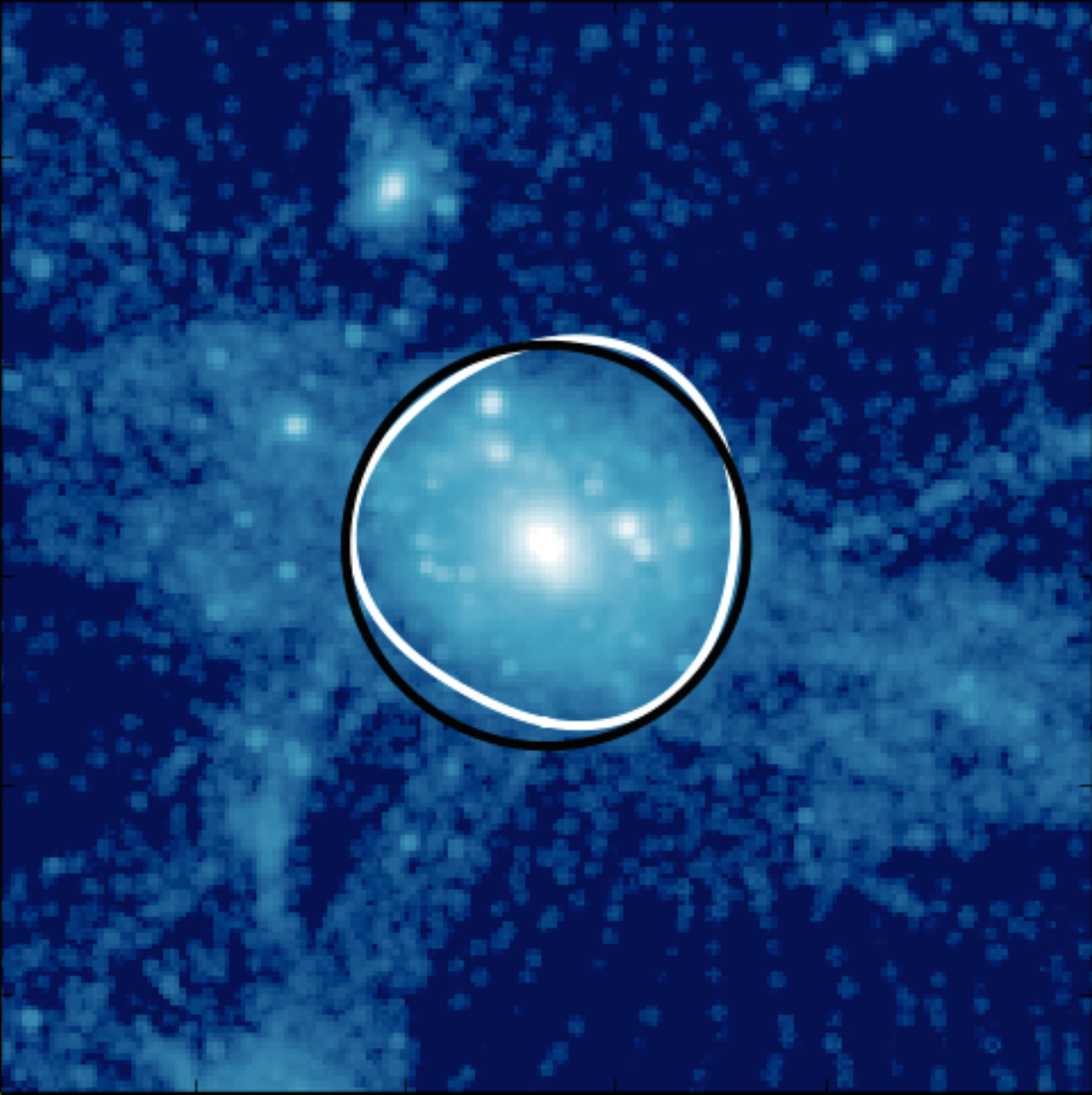}}
   \hspace{0.4cm}
   \subfigure[]{
        \label{fig:pass_4}
   \includegraphics[width=0.7\columnwidth]{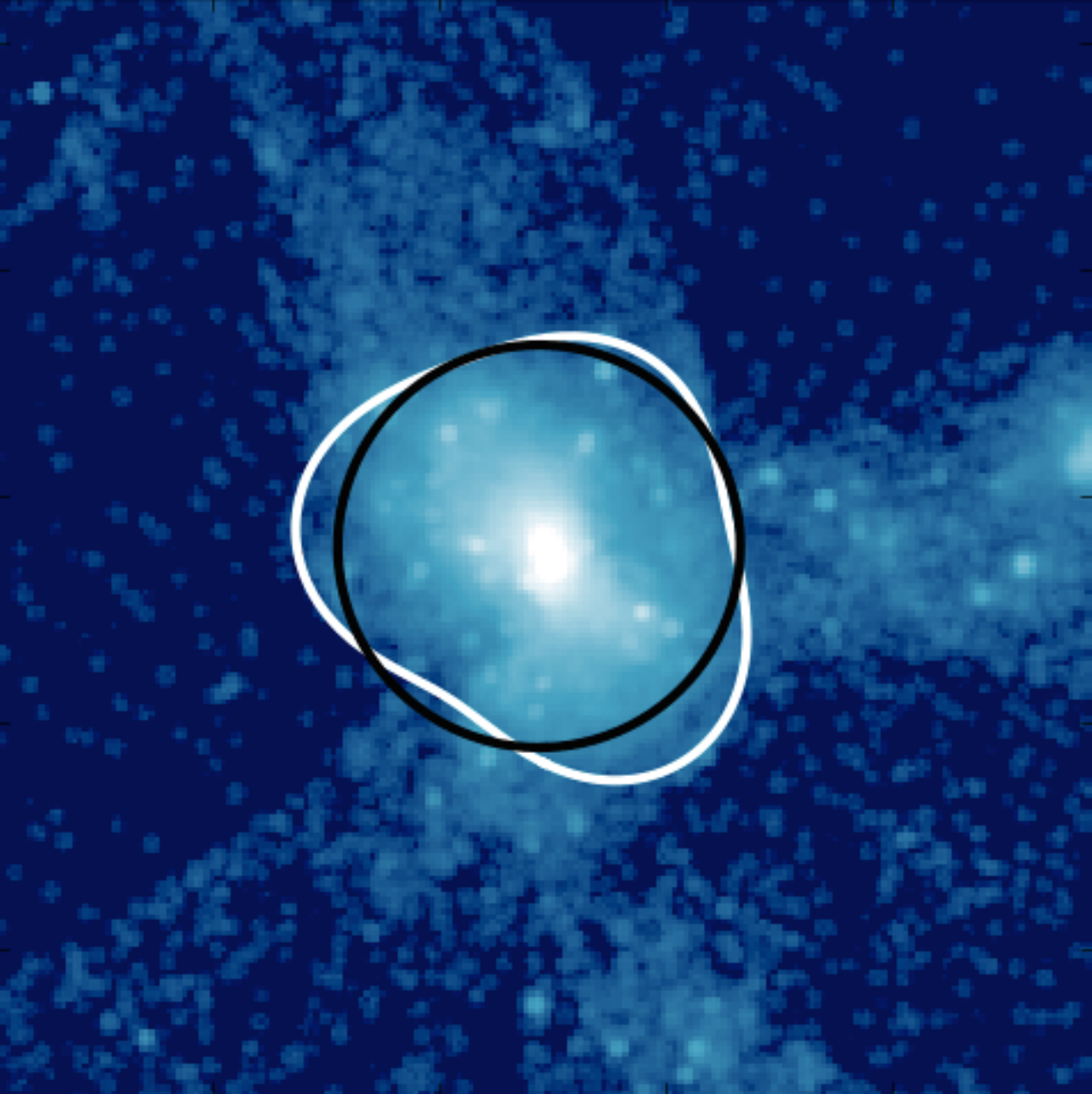}}\\
   \vspace{0.5cm}
   \subfigure[]{
        \label{fig:pass_5}
   \includegraphics[width=0.7\columnwidth]{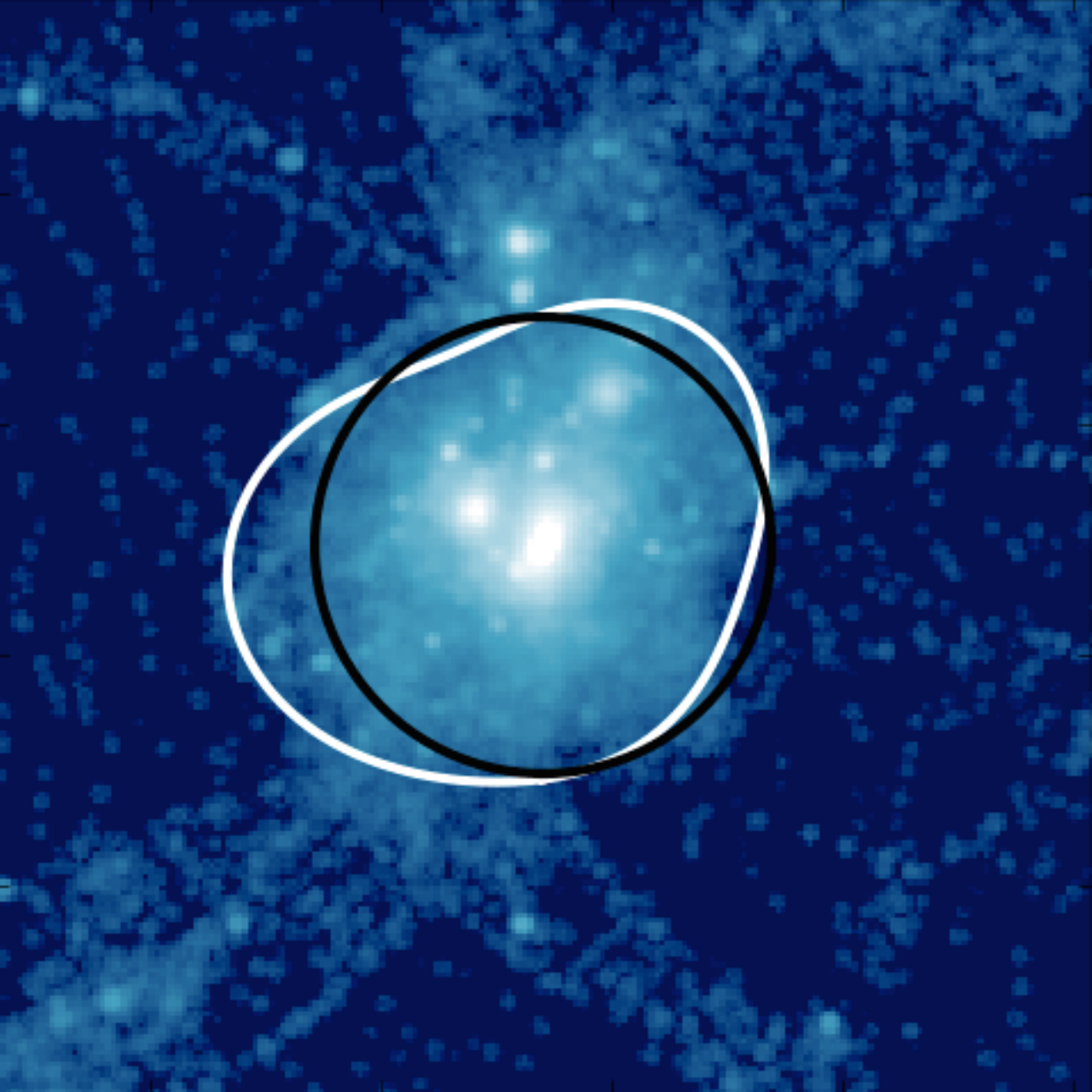}}
   \hspace{0.4cm}
   \subfigure[]{
        \label{fig:pass_6}
   \includegraphics[width=0.7\columnwidth]{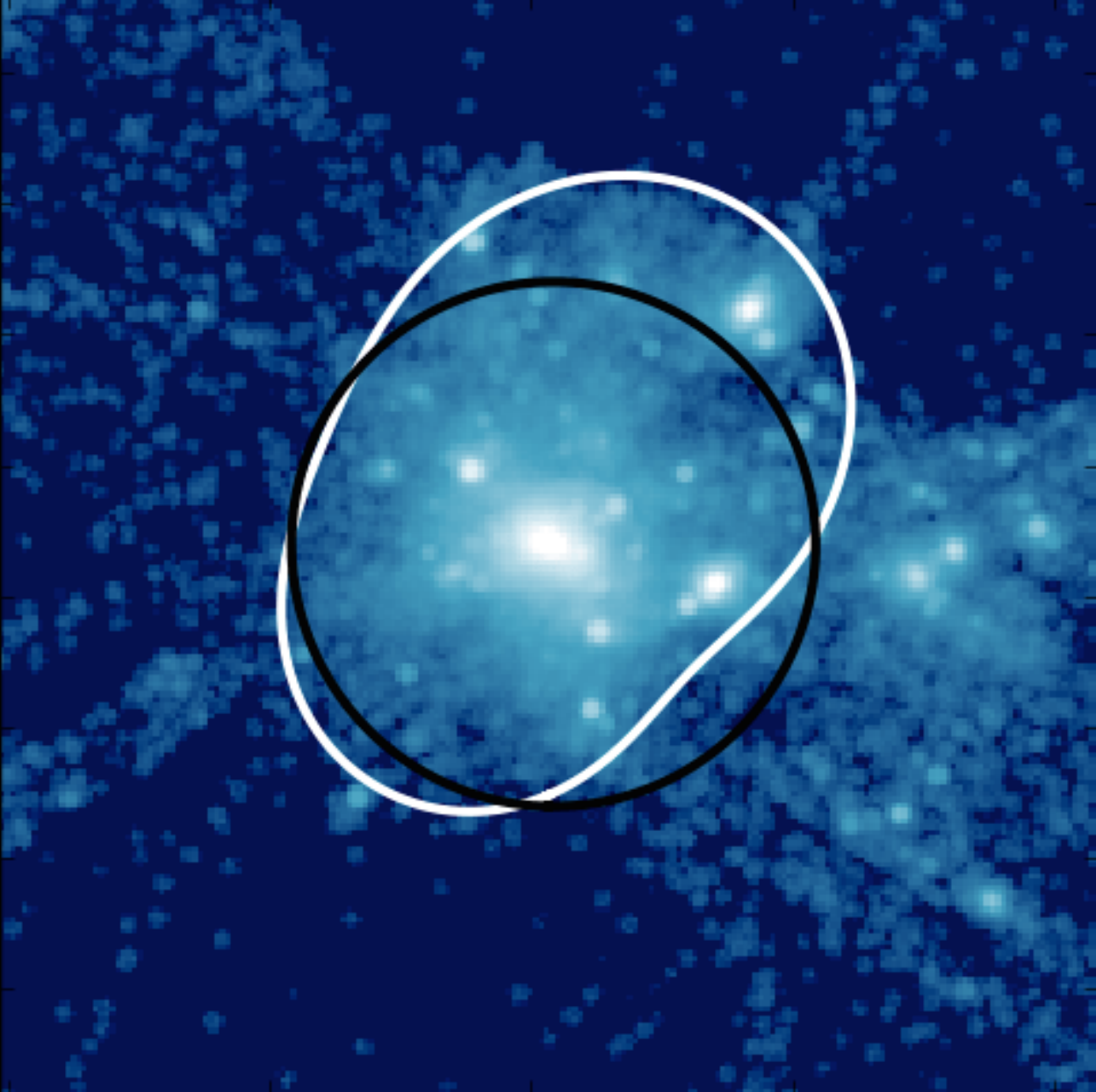}}
   \caption{Density slices of six halos are shown within boxes of size
   $5 R_{\rm 200m}$ along with cross-sections of each halo's splashback
      shell identified by our algorithm (white lines) and cross-sections
      of spheres with the same volume as the splashback shell (black circles).
      The six halos were picked randomly by sampling halos uniformly from
      within in the the $\log{M_{\rm 200m}}~-~\Gamma_{\rm DK14}$ plane in our L0063
      simulation box. Note that Figure \ref{fig:pass_4} shows the halo used
      to illustrate our algorithm in Figure \ref{fig:algo}.}
      \label{fig:pass_ex}
\end{figure*}

\subsection{Summary of the Algorithm Parameters}
\label{sec:setting_params}

\begin{table}
  \centering
  \caption{Parameters values used by \textsc{Shellfish}}
  \label{tab:param}
  \begin{tabular}{cccc}
  \hline
  \hline
  \\
  Parameter & Definition & Value & Optimization Method\\
  \\
  \hline
  \\
  $R_{\rm min}$ & \S \ref{sec:density_est} & $0.3\ R_{\rm 200m}$ & A \\
  $R_{\rm max}$ & \S \ref{sec:density_est} & $3\ R_{\rm 200m}$ & A \\
  $R_{\rm kernel}$ & \S \ref{sec:density_est} & $0.2\ R_{\rm 200m}$ & \S
      \ref{sec:kernel_test} \\
  $\rho_{\rm bg}$ & \S \ref{sec:density_est} & 0.5 $\rho_{\rm m}$ & B \\
  $N_{\rm planes}$ & \S \ref{sec:density_est} & 100 & \S \ref{sec:ring_test} \\
  $N_{\rm los}$ & \S \ref{sec:density_est} & 256 & A \\
  $N_{\rm bins}$ & \S \ref{sec:density_est} & 256 & A \\
  $N_{\rm SG}$ & \S \ref{sec:smooth} & 121 & B \& \S \ref{sec:smooth} \\
  $\eta$ & \S \ref{sec:filter_algo} & 10 & C \\
  $N_{\rm rec}$ & \S \ref{sec:filter_algo} &  3 & C \\
  $P$ & \S \ref{sec:fit} & 3 & C \\
  \\
  \hline
  \\
  \end{tabular}
  \tablecomments{The first
  column gives the parameter name, the second column gives the section where
  we define this parameter, the third column is the adopted fiducial value of
  each parameter within \textsc{Shellfish}, and the fourth column
  indicates the method used to identify the fiducial value. Methods A, B, and
   C are described in Appendix \ref{sec:setting_params_app}.}
\end{table}

The splashback shell finding algorithm described above has 11 free
parameters. The parameters and their adopted fiducial values in
in \textsc{Shellfish} are summarized in Table
\ref{tab:param}. Fortunately, there are three
empirical properties of this parameter family, which allow for a 
fairly straightforward way of choosing their values. First, the shapes of
the final splashback shells depend only weakly on most of these parameters.
Second, the optimal set of parameters does not appear to change for different
halo masses or different halo accretion rates. Third, the optimal value of a
particular parameter generally does not change as other parameters are changed
or can be easily rescaled to reflect such changes.

A discussion on the procedure we use for choosing specific parameter values
can be found in the Appendix \ref{sec:setting_params_app}.

\section{Tests}
\label{sec:tests}

\begin{figure}
   \centering
   \includegraphics[width=\columnwidth]{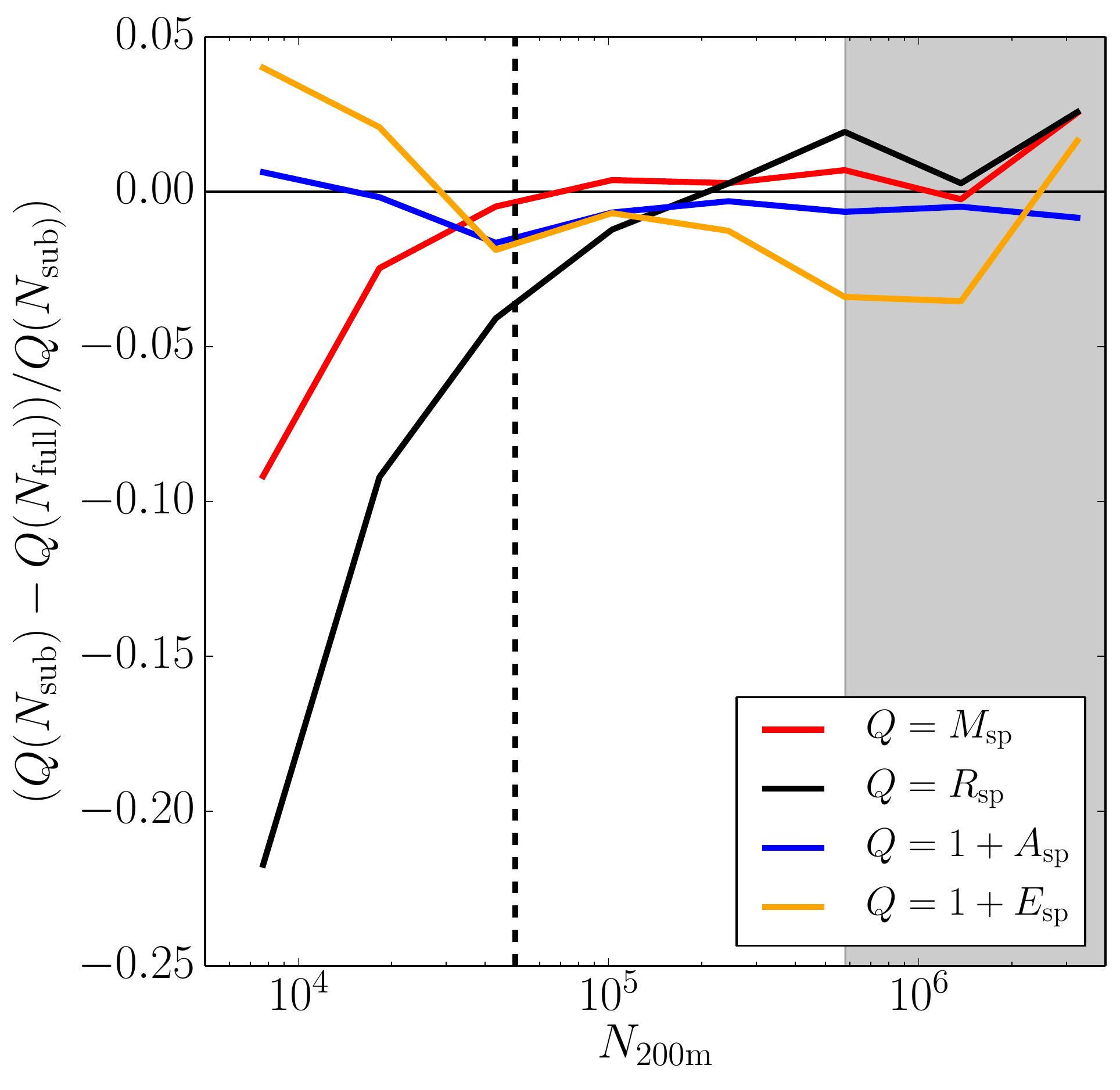}
   \caption{Convergence tests for the properties of splashback shells defined
       in Equation~\ref{eq:r_def} - Equation~\ref{eq:A_def} : enclosed mass, $M_{\rm sp}$,
       radius of the sphere of equivalent radius, $R_{\rm sp}$, ellipticity,
       $E_{\rm sp}$, and asphericity, $A_{\rm sp}$ as a function of the
       number of dark matter particles within $R_{\rm 200m}$, $N_{\rm 200m}$.
       The vertical dashed line corresponds to $N_{\rm 200m} = 50,000$,
       the lower limit used for the analysis in this paper, and the
       shaded vertical region indicates bins which contain two or fewer halos
       and are therefore dominated by individual halo error. Within the
       converged particle count range there is typically a scatter of
       $\approx 2\%$ about the median relation, which has not been plotted here
       for visual clarity. See section
       \ref{sec:tests} for details and discussion on this figure.}
   \label{fig:n200_rsp_convergence}
\end{figure}

In this section we present several tests of the algorithm described in the 
previous section. The parameters of the algorithm have been set to the default
values listed in Table \ref{tab:param}.

The first basic test is a qualitative visual assessment of the 
correctness of the splashback shells identified by \textsc{Shellfish}.

We find that, in general, the identified shells trace the sharp discontinuities
in the density field around halos. We illustrate this for six randomly-selected
example halos in Figure \ref{fig:pass_ex}, where the white curves show the
cross-sections of the identified shells and the black circles show
cross-sections of spheres with radii $R_{\rm sp}$ for those halos. Here
$R_{\rm sp}$ corresponds to the volume-equivalent definition given in
Equation~\ref{eq:r_def}. While we found that this type of simple visual
inspection proved to be very effective in identifying ineffective filtering
algorithms and parameter sets, it is necessarily a qualitative test and cannot
provide a quantitative error estimate.

In our second test, we compare the values of $R_{\rm sp}$ measured by
\textsc{Shellfish} to halos which have an unambiguous steepening in their
profiles relative to the asymptotic high-$R$ NFW slope due to the splashback
shell. \textsc{Shellfish} is unambiguously
incorrect for any halos where it measures $R_{\rm sp}$ outside of this
steepening region. The difficulty with this test is that is that it is hard to
programmatically detect the extent of this steepening region in a robust way.
Additionally, large substructure and dense filaments can create steepening
regions in the outskirts of host halos which appears similar to the steepening
caused by the splashback shell, but occurs in the wrong locations. For these
reasons, we resort to manual inspection of halos to perform this test.

\begin{figure}
   \centering
   \includegraphics[width=0.8\columnwidth]{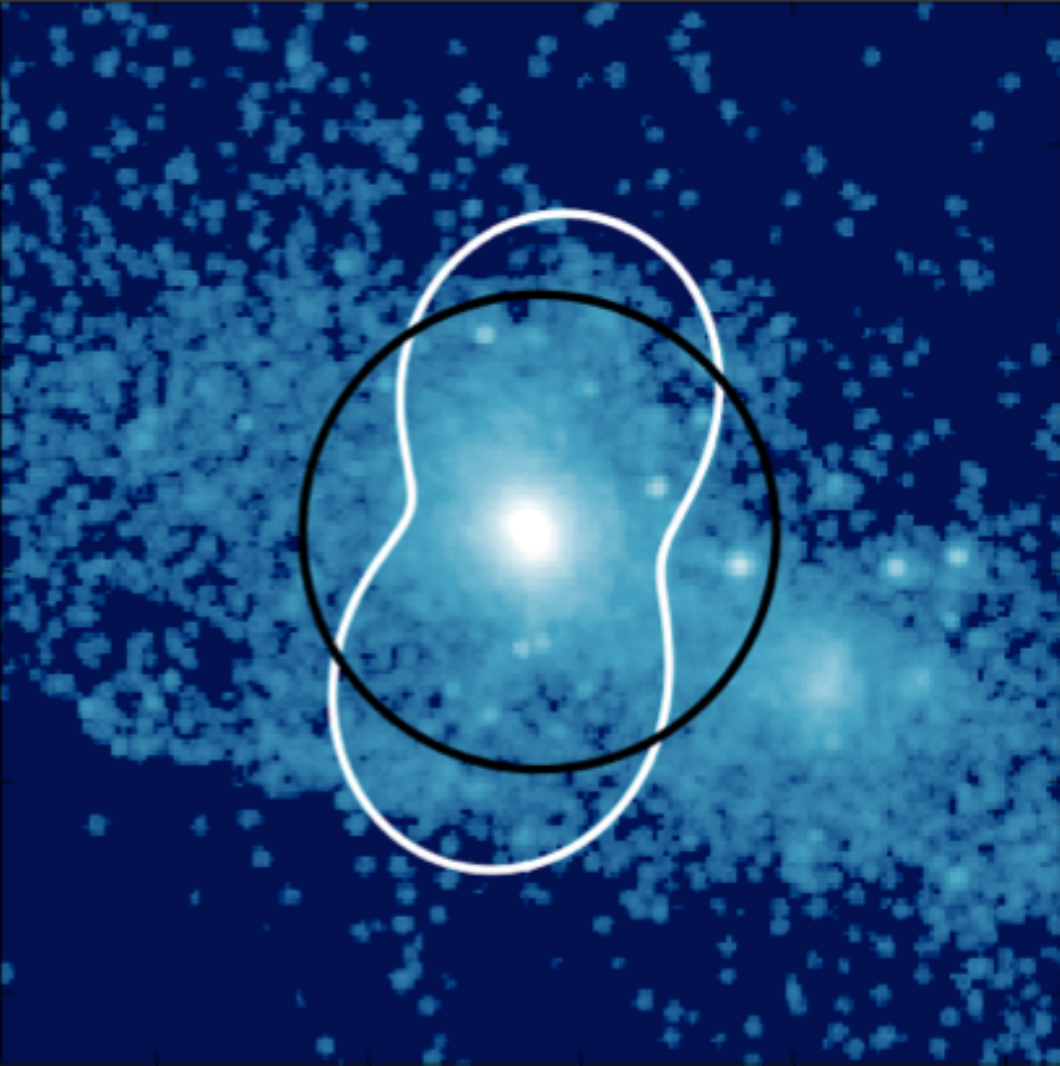}
   \caption{A density slice around one of the halos which fails the second test
   described
in section \ref{sec:tests} (i.e. a ``catastrophic failure''). The image dimensions and the
meanings of the white and black curves are identical to those in Figure
\ref{fig:pass_ex}.
We found that these halos can be very common for improperly calibrated
filtering algorithms, but when the parameters shown in Table \ref{tab:param}
are used, these halos
make up only $\approx 0.5\%$ of our total halo population.}
   \label{fig:catastrophic}
\end{figure}

We inspected the outer profiles of roughly 5,000 $z=0$ halos with 
$N_{\rm 200m} > 50,000$ and identified 906 which had a clear steepening of
the density profile in their outskirts and did not have a significant subhalo
presence in that region. We then identified the starting and ending radii that
bracketed the steepening region of each of these halos,
$R_{\rm start}$ and $R_{\rm end}$, by eye. We then
compared these radial ranges to $R_{\rm sp}$ calculated through
Equation \ref{eq:r_def}. We found that only four halos had $R_{\rm sp}$
measurements outside of the ranges measured from the profiles, corresponding
to a {\it minimum} failure rate of $\approx 0.5\%$. $R_{\rm start}$ and
$R_{\rm end}$ can span a wide range of radii (see, e.g., Figure
\ref{fig:sph_comp} and Figure \ref{fig:kernel_test}), so
this test is not effective at catching $\approx 20\%$ errors. This test is
chiefly sensitive to catastrophic failures, which we found could be as common as
$25\%$ for poorly constructed filtering algorithms or improperly set parameters.
Figure \ref{fig:catastrophic} shows an example of a typical catastrophic failure.
In this case,
there is no strong feature in the surrounding density field which forces
\textsc{Shellfish} to generate an unphysical shell.
Achieving a low failure rate on this test is a necessary, but not sufficient,
condition for any accurate splashback-measuring code.

As a third test, we also carried out a convergence study of the shell properties
defined in Equations~\ref{eq:r_def} - \ref{eq:A_def} with respect to the number
of dark matter particles within a halo, $N_{\rm 200m}$. These were performed by
generating a representative sample of halos and fitting two Penna-Dines shells
to each of them. The first shell is calculated using only one eighth of the
halo's particles and the second is calculated using all the halo's particles.
We use the notation that the number of particles in subsampled halos is
$N_{\rm 200m}/8 = N_{\rm sub}$, and that the number of particles in fully
sampled halos is $N_{\rm 200m} = N_{\rm full}$. The results of this
test are shown in Figure \ref{fig:n200_rsp_convergence}.

Figure \ref{fig:n200_rsp_convergence} shows that for $N_{\rm 200m} > 50,000$,
the systematic error due to particle count in $M_{\rm sp}$ is at the
per cent to sub per cent level, and that the error in $R_{\rm sp}$,
$1 + E_{\rm sp}$, and $1+A_{\rm sp}$ in the same range is at the few per cent
level. The shaded region in Figure \ref{fig:n200_rsp_convergence} indicates
bins in which our simulation suite produced two or fewer halos.
Figure \ref{fig:n200_rsp_convergence} indicates
that to identify splashback shells reliably, halos need to be resolved with
at least $5\times 10^4$ particles. It is not clear to what extent there is
a second order trend in radius after the first order convergence at
$N_{\rm 200m}$. It would not be unreasonable to see a trend of this type: as
$N_{\rm 200m}$ increases, \textsc{Shellfish} may be able to resolve and fit
smaller scale features in halos which could result in small changes in volume.
For this reason, we cannot yet rule out that there is a systematic
$\lesssim 5\%$ trend with mass for $R_{\rm sp}$.

\subsection{Comparison to Particle Trajectories}
\label{sec:trajectories}

As a fourth test of the algorithm, we
inspect the trajectories of individual particles near the splashback shell.
Particles near the correctly identified splashback shells can be expected 
to be either infalling for the first time or to be at the apocenter of their
first orbit. Trajectories of the infalling particles should be roughly
perpendicular to the shell locally and should not show any deflection when
crossing the shell. The trajectories of the particles that have orbited
through the halo should show a sharp turnaround at the shell location. The
relative fractions of particles of these two types will depend
on the mass accretion rate of each specific halo, but the apocenters of
particles of the second type should coincide with the identified splashback
shell. Given that our algorithm does not use any information about particles
trajectory, this test is a useful independent check on whether our algorithm
identifies shells corresponding to the actual outermost apocenters of particle
orbits.

To perform this test on a target halo, we first use \textsc{Shellfish}
to identify a splashback shell around the halo at some redshift
$z_1 > 0$. We then find all particles within some small distance $\delta$ of
this shell and track their trajectories through a redshift range
$z_0 < z_1<z_2$. 

\begin{figure*}
    \centering
    \subfigure[]{
    \label{fig:trj_A}
    \includegraphics[width=0.8\columnwidth]{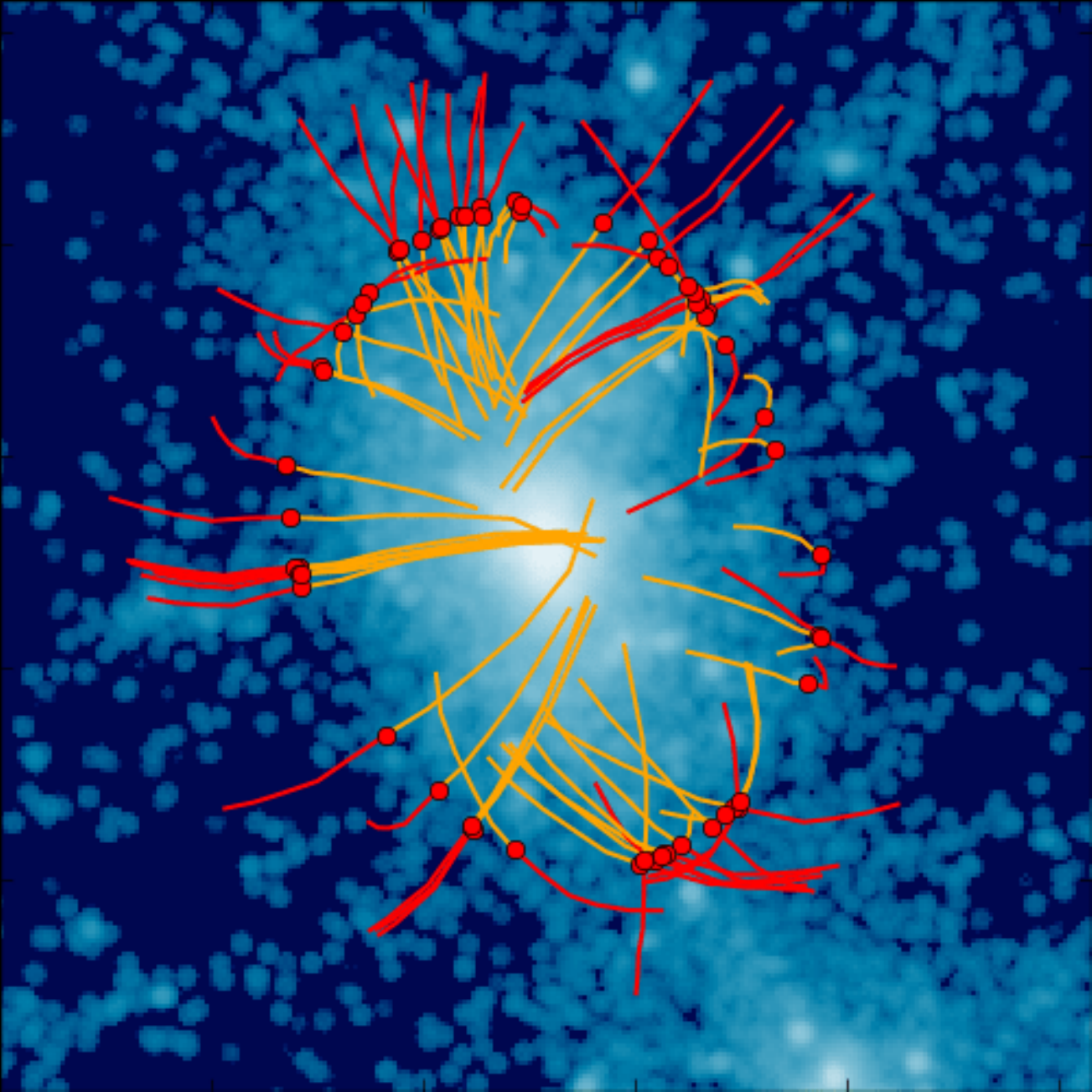}
    }
    \hspace{0.4cm}
    \subfigure[]{
    \label{fig:trj_B}
    \includegraphics[width=0.8\columnwidth]{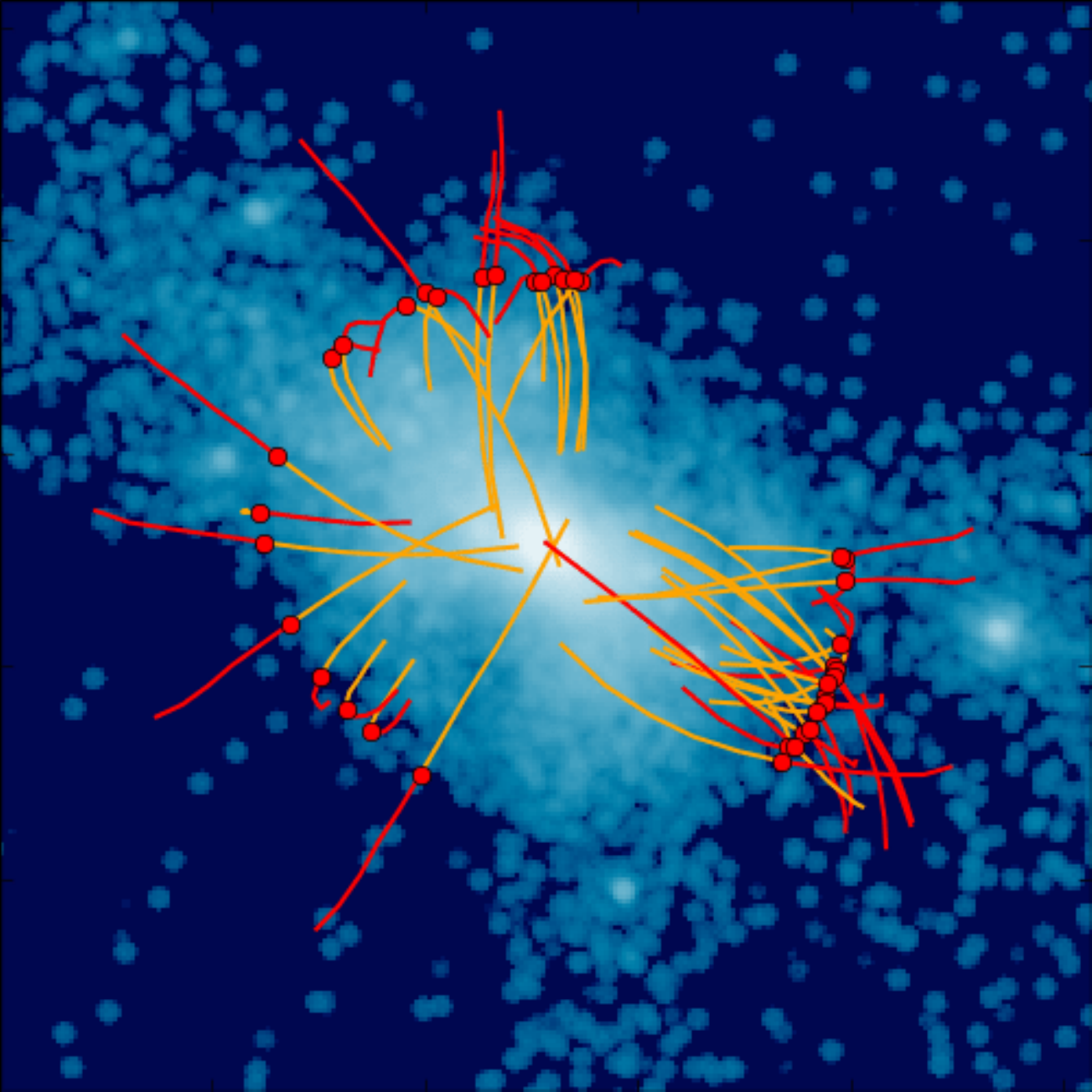}
    }\\
    \subfigure[]{
    \label{fig:trj_C}
    \includegraphics[width=0.8\columnwidth]{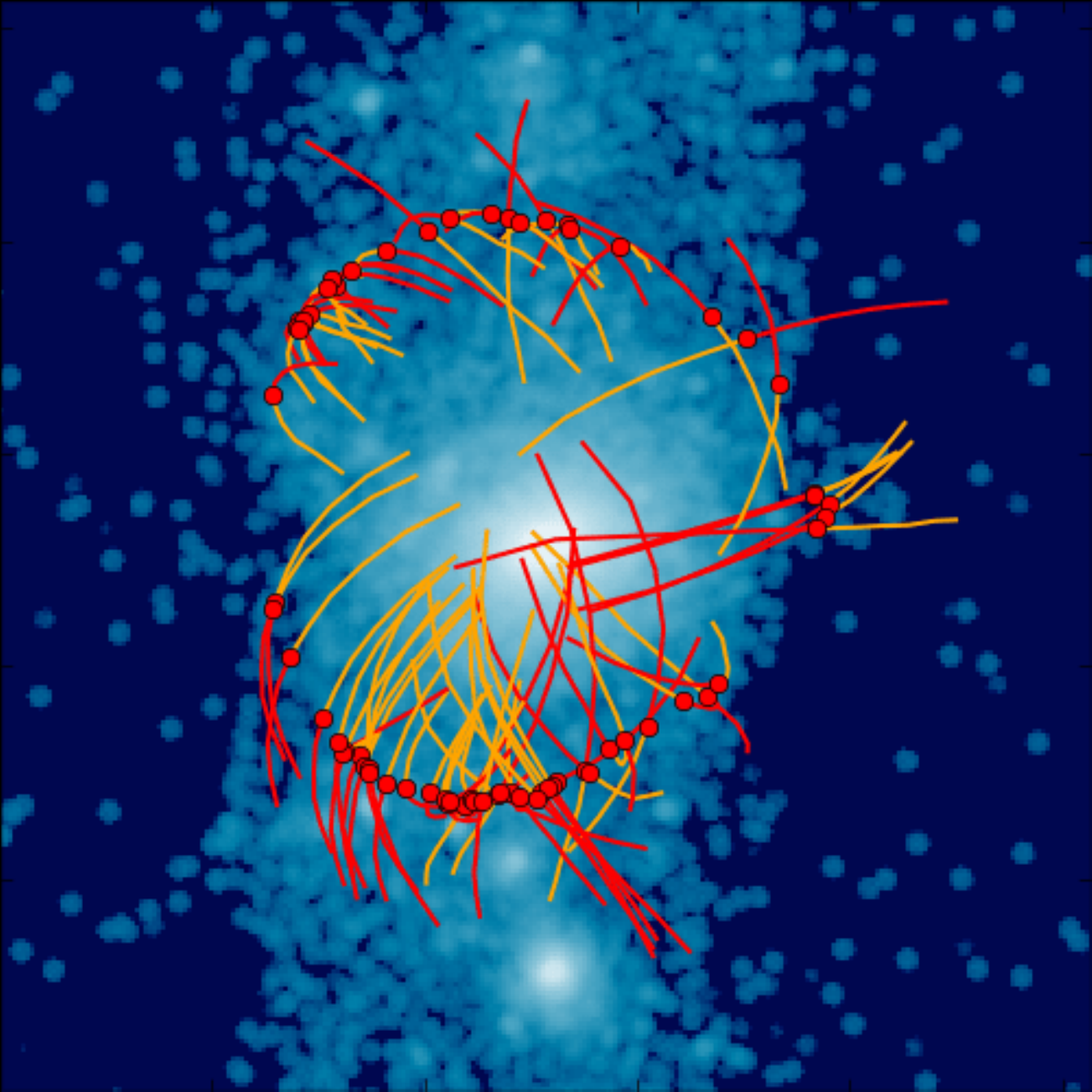}
    }
    \hspace{0.4cm}
    \subfigure[]{
    \label{fig:trj_D}
    \includegraphics[width=0.8\columnwidth]{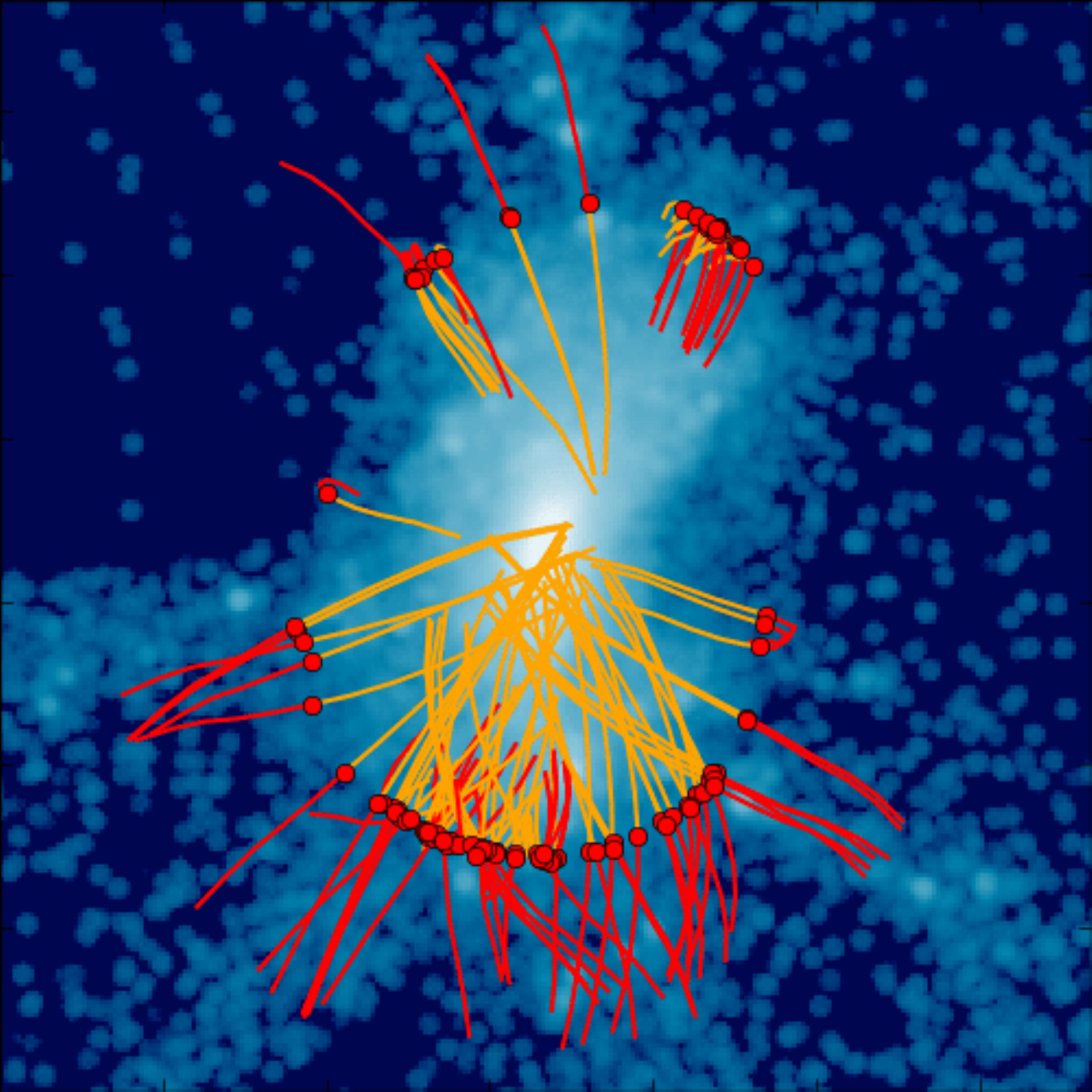}
    }
    \caption{Trajectories for particles during the redshift interval
    $z\in [0.32,0]$ near the splashback shell of four clusters from the 
    L0250 simulation with $M_{\rm 200m}\approx 10^{14}h^{-1}\ M_\odot$
    identified at $z_1=0.13$. Each figure shows a slice through the density
    field in a region centered on the halo with a width of $5R_{\rm 200m}$
    and a depth of $R_{\rm 200m}/5$. Every particle in this slice located
    within $R_{\rm 200m}/50$ of the splashback shell identified by
    \textsc{Shellfish} at $z_1=0.13$ is shown as a red point. The
    trajectory of each particle during the redshift interval $[0.31,0.13]$ is 
    shown by red line, while the trajectory during the redshift interval
    $[0.13,0]$ is shown by yellow lines. See section 
    \ref{sec:trajectories} for details.}
    \label{fig:trajectories}
\end{figure*}

The results of such a test are
shown for four representative clusters with $M_{\rm 200m} \approx 10^{14}h^{-1}\ M_\odot$ from the L0250 simulation in Figure \ref{fig:trajectories},
where we used  $\delta = R_{\rm 200m}/50$, $z_0 = 0.32$, $z_1=0.13$, and
$z_2 = 0$. The location of the particles at $z=z_1$ is shown by red points.
The trajectories of particles from $z_0$ to $z_1$ are shown as 
red curves and the trajectories from $z_1$ to $z_2$ are shown as yellow
curves. Infalling particles have red curves pointing outside of the halo
and yellow curves pointing inside the halo. Particles moving outwards have
reversed colors: yellow curves pointing to the outside and red
curves pointing to the inside. Particles at their apocenters will have both
curves pointing to the inside.

Figure \ref{fig:trajectories} shows that for the cluster-sized halos shown,
most particles around the splashback shell are infalling, as can be expected
for rapidly accreting halos. At the same time, there is a
fraction of particles that exhibit a sharp turnaround near the identified
splashback shell: i.e., the apocenters of their orbit coincide with the
splashback shell identified from the density field. 

Figure \ref{fig:trj_C} does show several trajectories in the southern portion
of the halo which travel outside the identified shell. It is not clear
whether this is because \textsc{Shellfish} was unable to identify
the correct splashback shell due to the high-density filament or whether
those particles were perturbed from their orbits in later time steps by the
nearby subhalo. Such trajectories, however, are a small fraction of the total.

We have carried out such visual inspection of trajectories for a
large number of halos and found results qualitatively similar to those shown
in Figure \ref{fig:trajectories}. This indicates that our algorithm
is reliably picking out splashback shells that coincide with the most distant
apocenters of particle orbits. This analysis has been confirmed by comparison
with an alternative splashback-measuring code, \textsc{Sparta}, which showed
that the radii measured by \textsc{Shellfish} correspond to high-percentile
moments of a halo's apocenter distribution \citep{diemer_et_al_2017}.

\section{Results}
\label{sec:results}

\subsection{Sample Selection}
\label{sec:sample_selection}

To analyze the properties of splashback shells identified using our
algorithm we construct a  sample of halos drawn from the
halo catalogs of all the simulations listed in Table \ref{tab:sims}.
Based on the convergence test results reported in section \ref{sec:tests}
(see Figure \ref{fig:n200_rsp_convergence}), we select halos with
$N_{\rm 200m} > 50,000,$ so that shell properties are converged to the level
$\lesssim 5\%$. We also restrict the maximum mass of halos drawn from
the smaller box simulations so that the  $\Gamma_{\rm DK14}$ distribution of the largest
halos in those simulations is similar to that of halos of the same mass in the
larger boxes. This limit is imposed because small box size may limit the mass
accretion time of the largest halos, as evolution becomes nonlinear on scales
comparable to the box size. The mass ranges sampled by each box are given in
Table \ref{tab:sims}. 

With these mass limits in place, we construct the halo sample for analysis by
subsampling all host halos within the mass range of each box in such a way as
to obtain a uniform distribution of halos in both $\log M_{\rm 200m}$
and $\Gamma_{\rm DK14}$. This procedure is repeated for $z=0$, $z=0.5$, $z=1$, and $z=2$,
resulting in a total sample sizes of 1095, 1198, 846, and 467 halos,
respectively.

\subsection{Comparison With Stacked Radial Density Profiles}
\label{sec:stacked_comparison}

\begin{figure}
   \centering
   \includegraphics[width=\columnwidth]{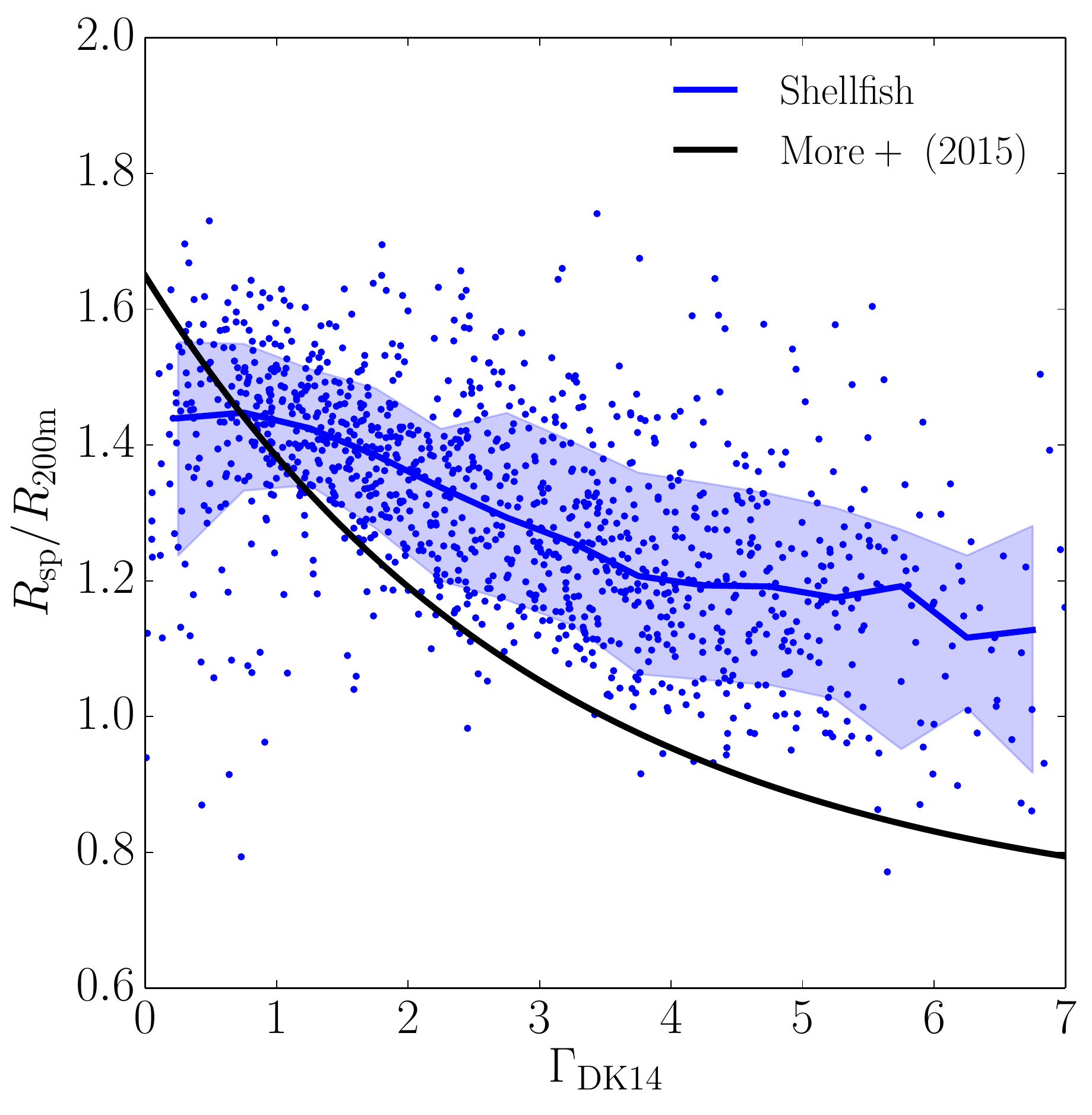}
   \caption{Comparison between the distribution of
   $R_{\rm sp}/R_{\rm 200m}$ values measured by \textsc{Shellfish}
   to the prediction of stacked density profile analysis at $z=0.5$.
   The black curve shows the best fit to location of steepest slope in the
   stacked density profiles as a function of accretion rate, $\Gamma_{\rm DK14}$. We use
   the parameterization for this fit reported in \citet{more_et_al_2015}.
   The blue points show \textsc{Shellfish} $R_{\rm sp}/R_{\rm 200m}$
   measurements for individual halos, the blue curve shows the median
   measurement, and the blue contours show the 68\% envelope. The
   \textsc{Shellfish} curve differs from stacked profiles in both amplitude
   and shape, becoming $\approx$30\% larger for halos with $\Gamma_{\rm DK14} > 4$.
   A qualitatively similar difference can be seen at all redshifts. We
   argue that this difference is due to stacked profiles splashback
   measurements being artificially biased inwards by massive subhalos in
   section \ref{sec:stacked_comparison}.}
   \label{fig:points_only_comp}
\end{figure}

Figure \ref{fig:points_only_comp} presents a comparison between the distribution
of $R_{\rm sp}/R_{\rm 200m}$ values measured by \textsc{Shellfish} and the
predictions of stacked profile analysis as a function of accretion rate. In
particular, we choose to compare against the $\Gamma_{\rm DK14}$ vs.
$R_{\rm sp}/R_{\rm 200m}$ fit reported in \citet{more_et_al_2015}.
We have chosen $z=0.5$ for illustration in this figure, because the $z=0.5$
halo sample contains a good  mix of well-converged, high particle-count halos
which become more abundant as redshift decreases, and halos with large
accretion rates, which become more abundant as redshift increases.

The figure shows that at $\Gamma_{\rm DK14}\lesssim 1.5$ our algorithm 
estimates splashback radii similar to those from stacked profiles, while for 
$\Gamma_{\rm DK14}\gtrsim 1.5,$ \textsc{Shellfish} estimates progressively larger
$R_{\rm sp}$ values compared to the values from the stacked profiles. The
discrepancy in $R_{\rm sp}/R_{\rm 200m}$ is $\approx 30\%$ for
$\Gamma_{\rm DK14} \approx 4$. This discrepancy exists at all redshifts.

Given that the tests presented in section \ref{sec:tests} indicate that our
code identifies splashback shells reliably and estimates their properties to
better than $5\%$ accuracy at the resolution level shown in
Figure \ref{fig:points_only_comp}, it is highly unlikely that the discrepancy is
due to any issue of our algorithm. In particular, a systematic overestimation of
$R_{\rm sp}$ by 30\% would be immediately apparent in the visual comparison of
the identified splashback shells and the underlying density field. Instead, we
find a good agreement in such comparisons. Additionally, we were able to
independently reproduce the results of \citet{more_et_al_2015} using the halo
sample described in section \ref{sec:sample_selection}. Thus, the discrepancy
shown in in Figure \ref{fig:points_only_comp} is the real difference between
the two methods. 

To better understand the origin of this difference, we visually inspected the
radial density profiles of all the halos in our sample and classified them into
one of three qualitative classes. First, we flagged every halo as either
containing a visually distinct steepening region in its outskirts or as
containing no such region. Halos of the latter type we classify as
``featureless''-type profiles. The red curve in Figure \ref{fig:prof_comp} is
an example of such a halo.

The remaining halos contain distinct regions in the density profiles where the
logarithmic slope steepens considerably over a limited range of radii. For
these halos we visually identify the starting radii, $R_{\rm start}$, and
ending radii, $R_{\rm end}$ of their respective steepening regions. We find
that almost all such halos separate neatly into one of two classes: 1) halos
which have relatively sharp and narrow steepening regions that closely
correspond to the radial range of the splashback shell found by
\textsc{Shellfish} for that halo; and 2) halos which have a relatively
shallow and wide steepening region with an $R_{\rm start}$ value significantly
smaller than the minimum radius of the shell found by \textsc{Shellfish}.
We refer to halos of the first type as ``short''-type profiles and
halos of the second type as ``long''-type profiles, respectively. The blue and
yellow curves in Figure \ref{fig:prof_comp} are examples of these two types of
profiles, respectively. The number of halos is roughly similar in the three
classes of  ``featureless'', ``short'', and ``long'' profile types, but the
exact fractions of halos in each class changes with accretion rate and with
mass. 

We find that when we derive splashback radii from the stacked density profiles
using only halos of the short and featureless types, the difference from the
median $R_{\rm sp}$ measured by \textsc{Shellfish} decreases to
$\lesssim 5\%$ at high $\Gamma_{\rm DK14}.$ This is not surprising, given that we noted
that the steepening range in the short-type profiles is consistent with the
radial range of the splashback shells derived by \textsc{Shellfish},
but demonstrates that the difference in $R_{\rm sp}$ is due almost entirely to
the effect of the halos with the long-type profiles on the stacked density
profile. 

Our analysis shows that the steepening region in the density profiles of
long-type halos is not caused by the splashback shell, but by the presence of
massive subhalos. Specifically, visual inspection of the density fields of
long-type halos generally reveals that no portion of the splashback shell can
be found as far inwards as $R_{\rm start}$ for these halos. Instead, we almost
always find that a massive subhalo is present at  $R\approx R_{\rm start}$ for
these halos. Thus, the steepening region is associated with the presence of
subhalo, not the splashback. Given that subhalos in different halos with the
same accretion rate will be located at different $R$, the combined effect of
the massive subhalos on the stacked profile is to ``wash out'' the signature of
the splashback shell and to bias the start of the steepening region to smaller
radii. 

Thus, halos with no massive subhalos in the outskirts have the short-type
profiles, while those that do have such subhalos have long-type profiles. Halos
that either have large neighboring halos outside their splashback shells or
which exist in dense filaments have the steepening due to splashback shell
erased completely and thus have featureless-type profiles. The expectation is
then that if contribution of massive subhalos is removed from the density
profiles the $R_{\rm sp}$ derived from the stacked density profiles
should be consistent with the values estimated by \textsc{Shellfish}. 
We demonstrate that this is the case in the next subsection.

\subsection{Angular Median Density Profiles of Halos}
\label{sec:median_prof}

\begin{figure*}
    \centering
    \subfigure[]{
    \label{fig:sph_comp}
    \includegraphics[width=\columnwidth]{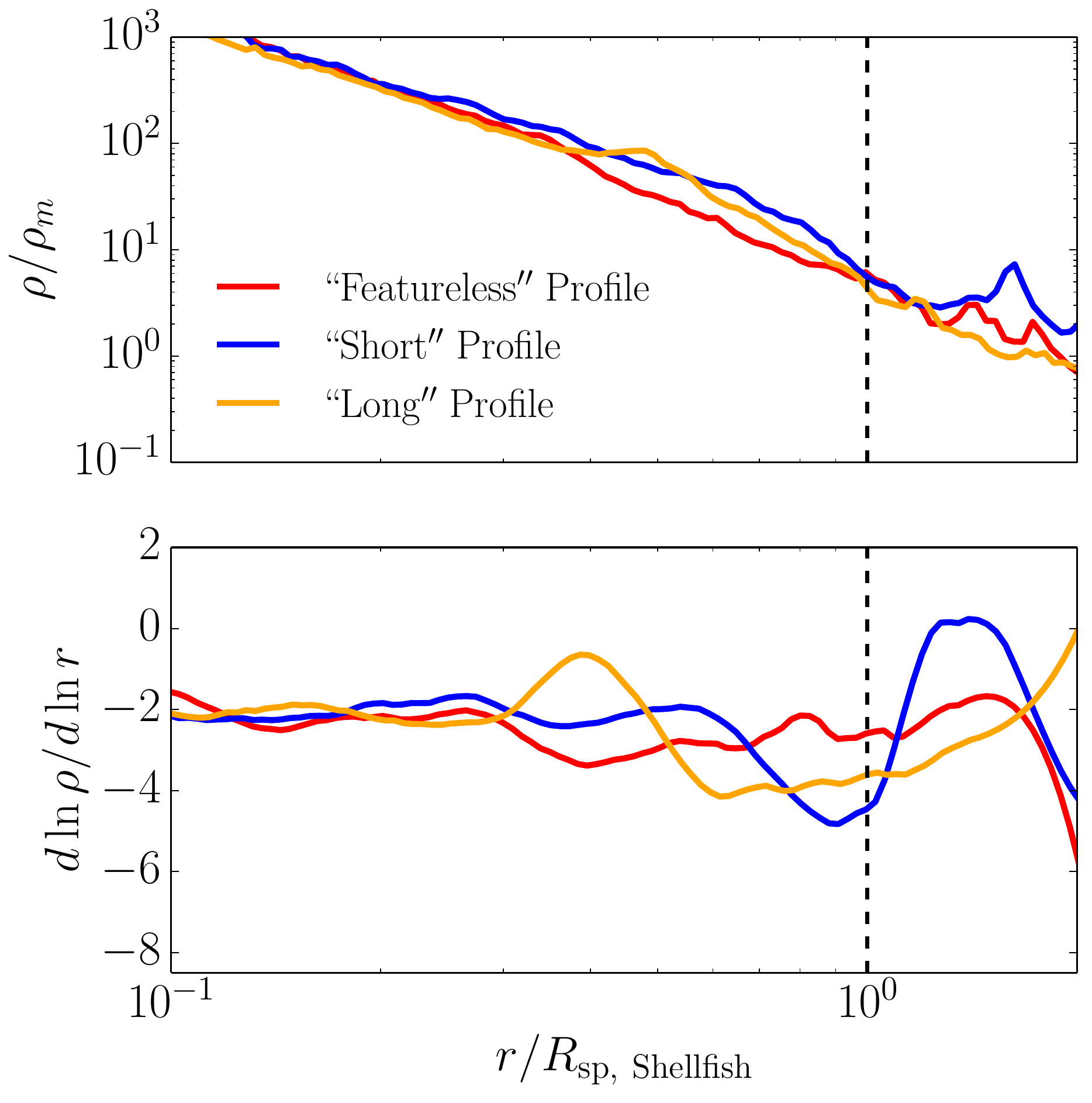}
    }
    \hspace{0.3cm}
    \subfigure[]{
    \label{fig:med_comp}
    \includegraphics[width=\columnwidth]{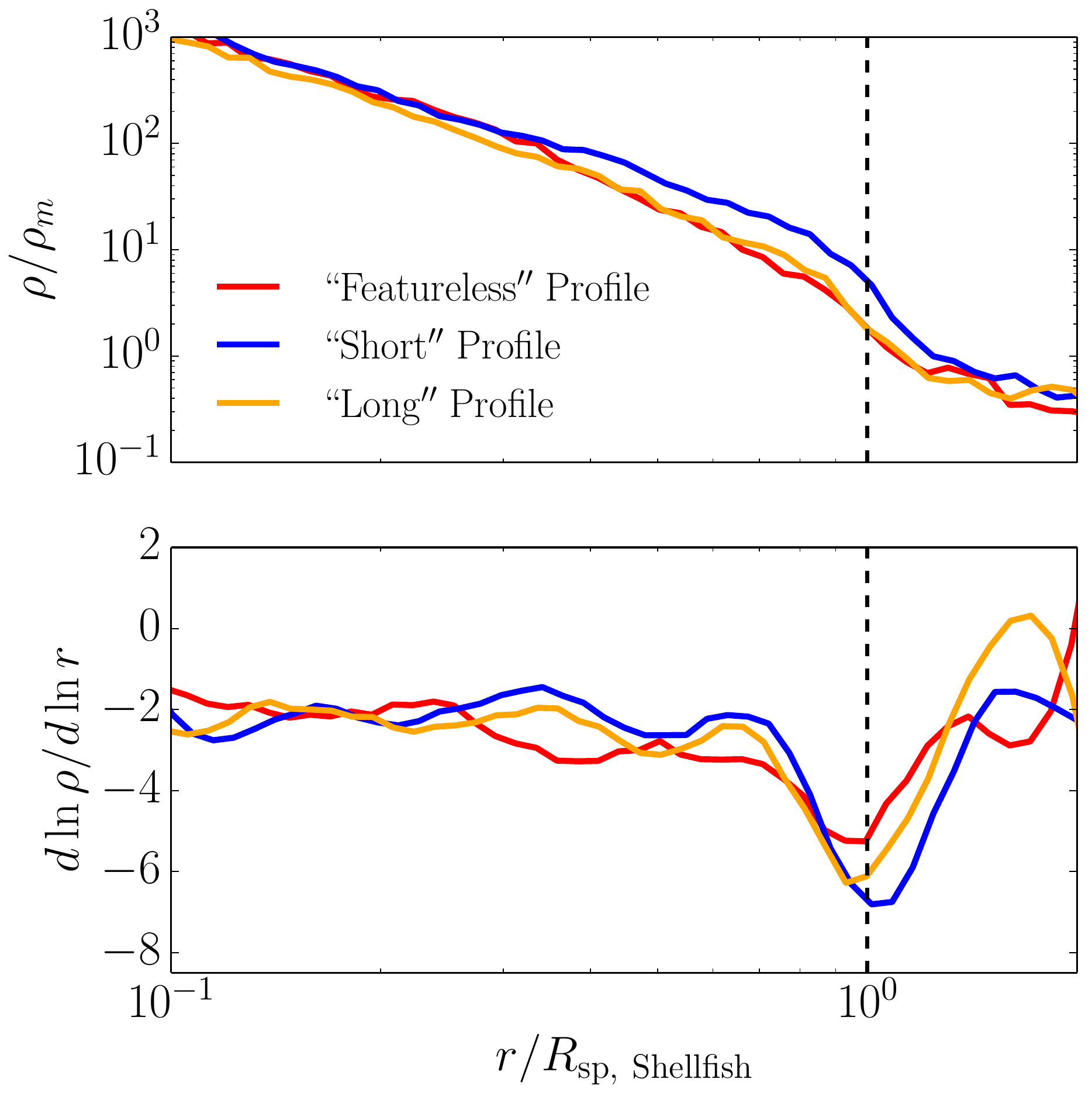}
    }
    \caption{Comparison between spherically averaged radial density profiles
    (Figure \ref{fig:sph_comp}) and the angular median density profiles
    described
    in \ref{sec:median_prof} (Figure \ref{fig:med_comp}). The top panels show
    density and the bottom panels show logarithmic slope after the density
    profiles have been smoothed with a fourth-order Savitzky-Golay filter with
    smoothing windows a third of a decade wide. Both density and slope profiles
    have had their radii normalized by $R_{\rm sp}$ as measured by
    \textsc{Shellfish}. The three halos are chosen to be representative of
    the three qualitative classes of halo profiles we identified in section
    \ref{sec:stacked_comparison}. Because angular median profiles are
    designed
    to remove interfering substructure, they have deeper and more well-defined
    points of steepest slope. The level of agreement between the radius of
    steepest slope of the angular median profiles shown here and the
    $R_{\rm sp}$ values derived by \textsc{Shellfish} is typical.}
    \label{fig:prof_comp}
\end{figure*}

There are many possible ways of mitigating the contribution of subhalos to 
the density profiles of their host halos. We choose one of the simplest
methods for doing this, one which does not rely on the availability of robust
subhalo catalogs, and which could, in principle, be adapted for use on
observed galaxy clusters. The idea is to construct density profiles
using the median estimate of density in each radial shell instead of the
mean density. A similar approach has been used in the analysis of the gas
distribution in clusters \citep{zhuravleva_et_al_2013}. 

Namely, we split each radial shell of the density profile into $N$ solid angle
segments, e.g., using a two-hemisphere variation on the algorithm described by
\citet{gringorten_yepez_1992}, or the HEALPix pixelation algorithm
\citep{gorski_et_al_2005}. We then estimate density, $\rho_i(r),$ for each
segment $i$ and construct the halo density profile by taking the median of
these densities in each radial shell,
$\rho_{\rm med}(r) = {\rm med}\left[\rho_i(r)\right]$. This approach is based
on the basic intuition that subhalos are generally much smaller in extent than
the  host and thus contribute to a fraction of the solid angle in a given radial
shell, while most of the solid angle will be dominated by the diffuse matter
of the host halo.  The median density then will estimate the density of that
diffuse component and will be largely insensitive to the outlier solid angle
segments associated with massive subhalos. 

Figure \ref{fig:prof_comp} shows comparisons between usual spherically averaged
mean density profiles, $\rho(r),$ and angular median density profiles
$\rho_{\rm med}(r)$ for three representative halos of the different classes
described in section \ref{sec:stacked_comparison}. The comparison of the
profiles in the two panels of the figure shows that the angular median profiles of the
halos are much more similar to each other than the mean profile. Unlike the
mean density profiles, which have very different shapes, the angular median density
profiles all behave similarly: there is a narrow, sharp steepening region in the
logarithmic profile centered on the radius that \textsc{Shellfish} reports as
$R_{\rm sp}.$ Thus, the diversity of profile types noted in
\ref{sec:stacked_comparison} is largely absent for profiles of this type. We
also note that the point of steepest slope in angular median profiles is significantly
sharper than it is in mean profiles. Thus the signature of the splashback shell
is easier to detect when halos are analyzed in this way.

To compare $R_{\rm sp, shell}$ measured by \textsc{Shellfish} to
$R_{\rm sp,med}$ derived from the individual angular median profiles, we follow the
procedure described above for every halo in the sample described in section
\ref{sec:sample_selection}. We use 50 solid angle segments per halo with 30
logarithmically-distributed radial bins per decade. This relatively coarse
spacing is needed to make up for the fifty-fold loss in number statistics and
has a non-trivial impact on the maximum fidelity of our angular median profiles: the
width of every bin is 8\% of the radius at which is occurs. Once the median
profile is computed from these segments, we apply a Savitzky-Golay smoothing
filter with a window size comparable to the characteristic radial width of the
regions where profile slope steepens quickly. We set the window size to a
$0.33$ dex with the caveat that other reasonable choices, such as a sixth of
half of dex, can induce systematic changes to the mean  $R_{\rm sp,med}$ of a
halo population of $\approx 5\%$. Thus, the population statistics on
$R_{\rm sp,med}$ cannot be trusted to accuracies smaller than 5\% regardless of
any additional statistical error bars, and that individual $R_{\rm sp,med}$
values measured this way cannot be measured more accurately than 13\%,
regardless of additional profile noise. We leave more nuanced accuracy analysis
on this method to a future work, but note that this level of accuracy is
sufficient for our purposes, which is merely to test whether reducing effect
of subhalos on the radial profiles results in $R_{\rm sp}$ estimates which are
qualitatively consistent with the results of \textsc{Shellfish}.

\begin{figure*}
    \centering
    \subfigure[]{
    \label{fig:comp_mass_bin}
    \includegraphics[width=\columnwidth]{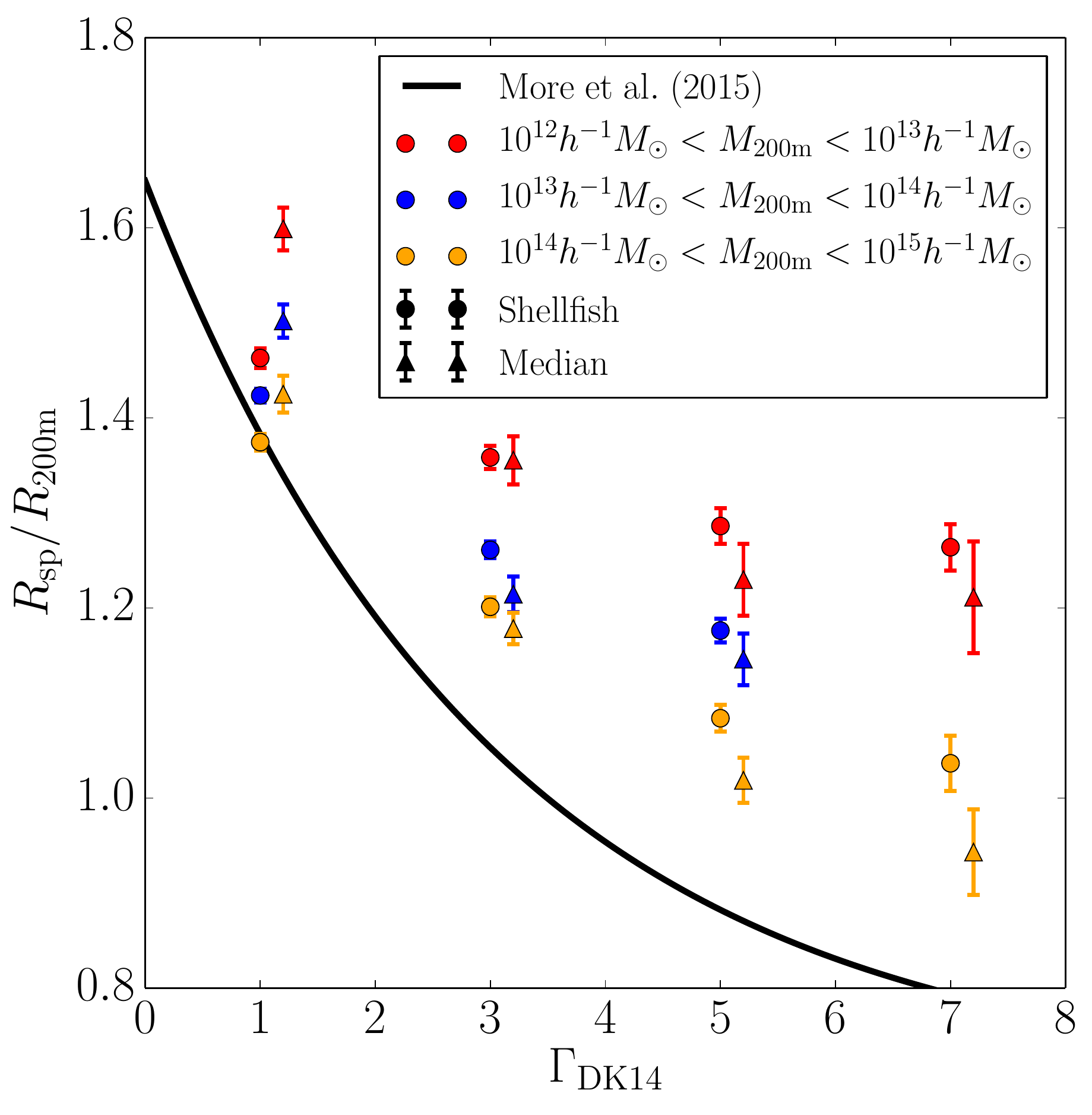}
    }
    \hspace{0.3cm}
    \subfigure[]{
    \label{fig:comp_all_z}
    \includegraphics[width=\columnwidth]{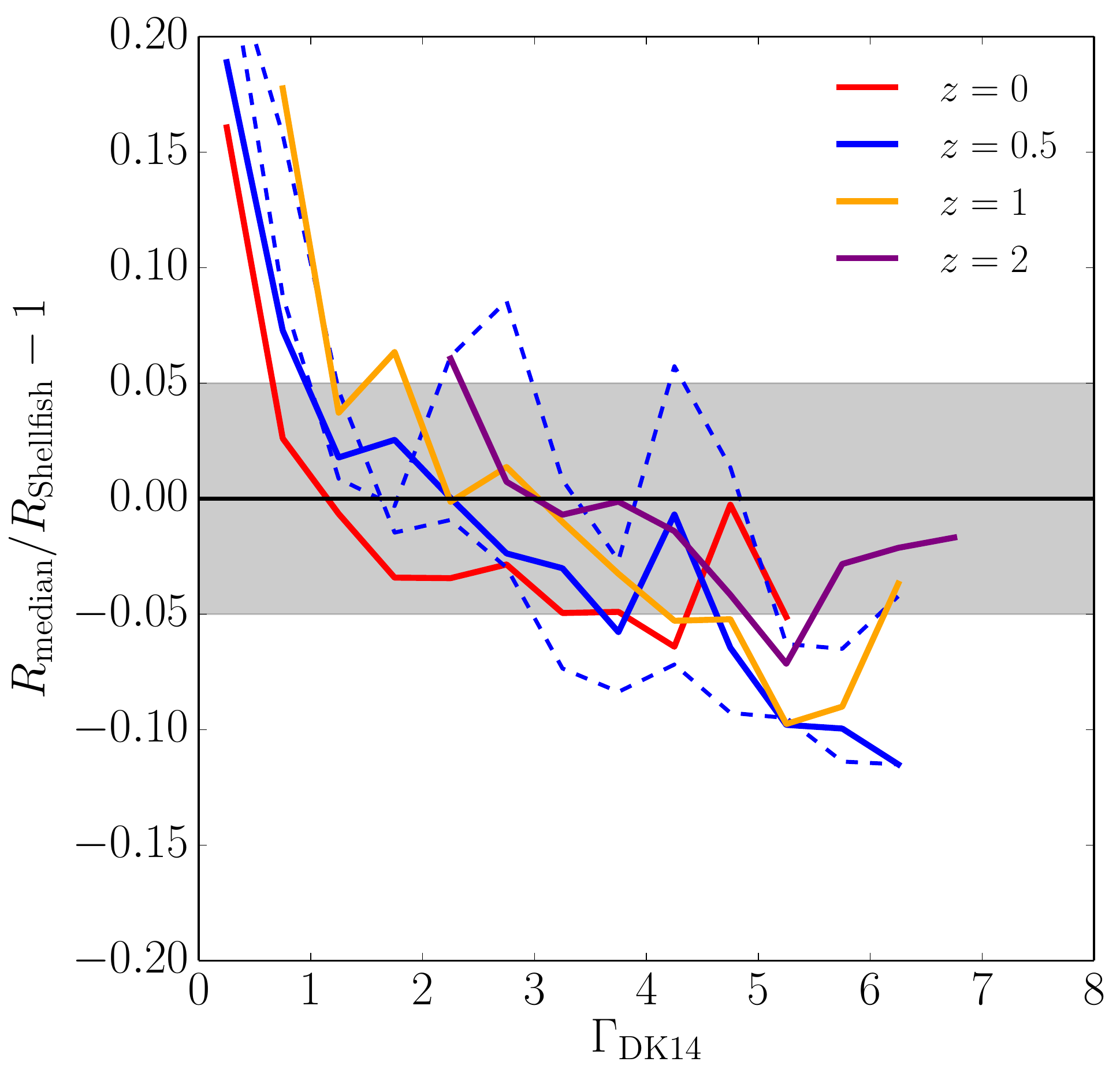}
    }
   \caption{Comparison between the mean $R_{\rm sp}/R_{\rm 200m}$ values
   measured
   by \textsc{Shellfish} and by the angular median profile method described in section
   \ref{sec:median_prof}. The left panel shows measurements made by the two
   methods for different $\Gamma_{\rm DK14}$ and $M_{\rm 200m}$ bins at $z=0.5$. Shellfish
   measurements are shown as circles on the left side of their respective
   $\Gamma_{\rm DK14}$ bins, and angular median profile measurements are shown as triangles on
   the right side of their respective $\Gamma_{\rm DK14}$ bins. Error bars represent only
   the bootstrapped error on the mean and do not account for known systematic
   uncertainty in the angular median profile method (see section 
   \ref{sec:median_prof}). The right panel shows the
   median value of
   $R_{\rm Shellfish}/R_{\rm median} - 1$, for every halo in our sample at
   $z$ = 0, 0.5, 1, and 2. The dashed blue lines show the shape of this curve
   when the angular median profile's Savitzky-Golay window width is varied to the
   edges of its physically reasonable value range to give a sense of the
   systematic variability in this method (see section \ref{sec:median_prof}).
   These two figures illustrate that when large subhalos are removed from
   the density profiles of halos, the location of the point of steepest slope
   becomes consistent with the value of $R_{\rm sp}$ measured by
   \textsc{Shellfish}. They also illustrate that there is a non-trivial
   disagreement between the two methods for very small $\Gamma_{\rm DK14}$.}
   \label{fig:med_shell_comp}
\end{figure*}

We compare the $M_{\rm 200m}$ and $\Gamma_{\rm DK14}$ trends between $R_{\rm sp,shell}$
and $R_{\rm sp,med}$ for our $z=0.5$ halo sample in
Figure \ref{fig:med_shell_comp} and see fairly good agreement. The high
$\Gamma_{\rm DK14}$ disagreement has dropped from $\gtrsim 30\%$ to $\approx 5\%.$ This
is consistent with the known systematic uncertainties in both methods and
confirms that the high $\Gamma_{\rm DK14}$ disagreement with the estimates
of the splashback radius from the stacked mean density profiles is due
to the bias introduced into these profiles by massive subhalos. 

At the same time, at $\Gamma_{\rm DK14}\lesssim 0.5$ there is $\approx 15\%$ 
disagreement between $R_{\rm sp}$ derived from the stacked angular median profiles and
the median measurements of \textsc{Shellfish}. In principle,
this difference could be caused by either the angular median profile method
or \textsc{Shellfish}, but comparison against another
splashback-measuring code, \textsc{Sparta}, which explicitly tracks particle
orbits to find their apocenters,
shows tight agreement with \textsc{Shellfish} at $\Gamma_{\rm DK14} > 0.5$ and a
level of discrepancy comparable to that seen for angular median profiles at
$\Gamma_{\rm DK14} < 0.5$. An extended discussion on how these two methods
compare against one another can be found in \citet{diemer_et_al_2017}.

It is not surprising that the splashback shell is difficult to measure at these
accretion rates. At $z=0$, pseudo-evolution causes static NFW halos with
$c_{\rm vir} \gtrsim 7$ to report $\Gamma_{\rm DK14} > 0.5$ purely due to the cosmological
evolution of $\rho_{\rm m}$ \citep{diemer_et_al_2013}. This means that the
majority of halos with accretion rates this low must be actively losing
particles in order to offset their illusory accretion rates caused by
pseudo-evolution. This particle loss is typically caused by dense environments,
either because the halo is embedded in a massive filament feeding a cluster or
because it is about to merge with a larger halo.

For this reason we believe that our algorithm should not be used to measure
halos with $\Gamma_{\rm DK14} < 0.5$ unless $\gtrsim 15\%$-level systematic errors are
acceptable. We exclude such halos from all subsequent analysis. This is
an aggressive cut for Milky Way-sized halos at low redshifts, where $20\%$
of halos have $\Gamma_{\rm DK14} < 0.5$. The cut is less severe for halos in all other
mass bins and at all other redshifts, affecting less than $5\%$ of halos in
all such parameter slices. Clusters and high redshift halos in particular are
almost completely unaffected by this cutoff.

\subsection{The Relationship Between Mass, Accretion Rate, and Splashback Radius}
\label{sec:g_r_sp}

\begin{figure*}
   \centering
   \includegraphics[width=\columnwidth]{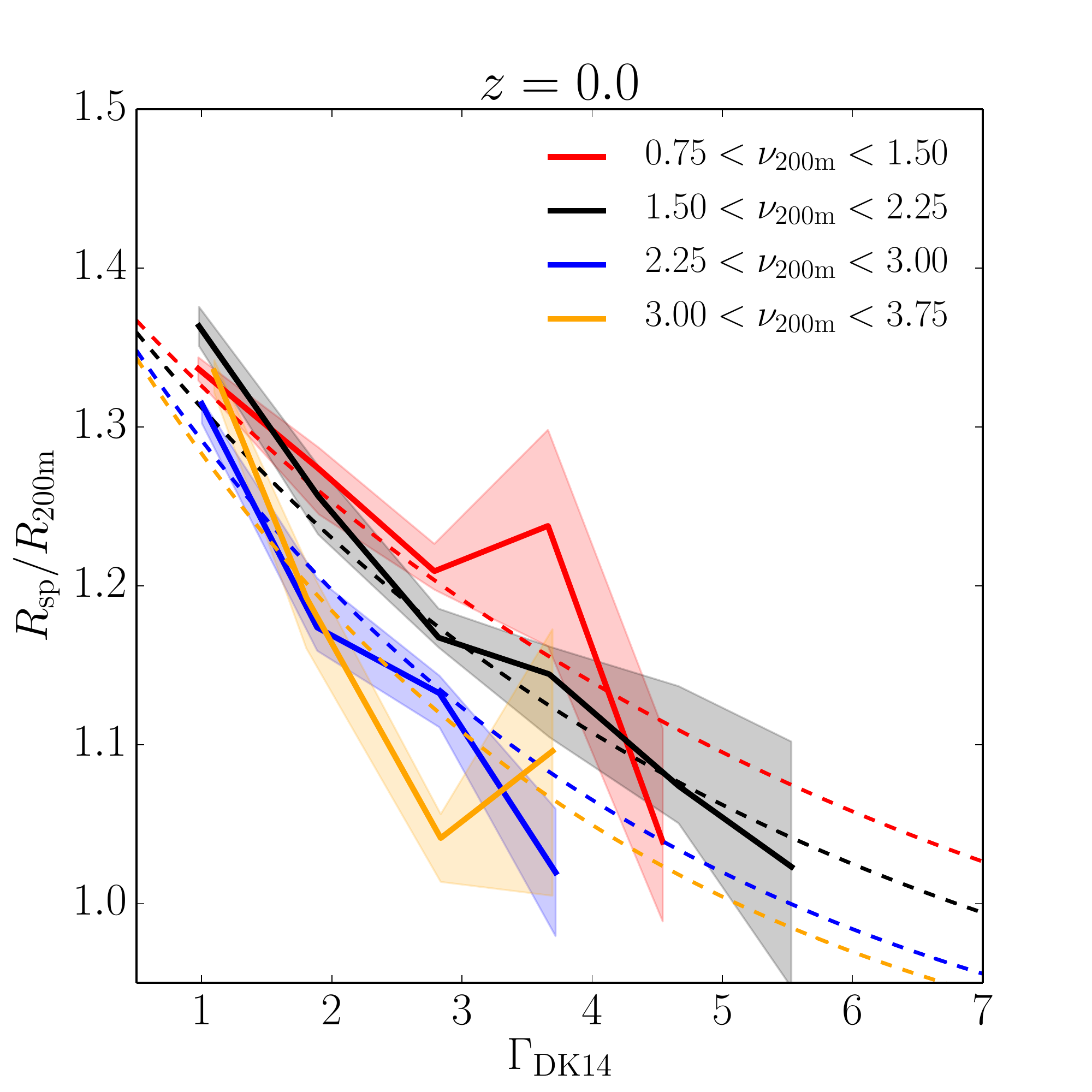}
   \includegraphics[width=\columnwidth]{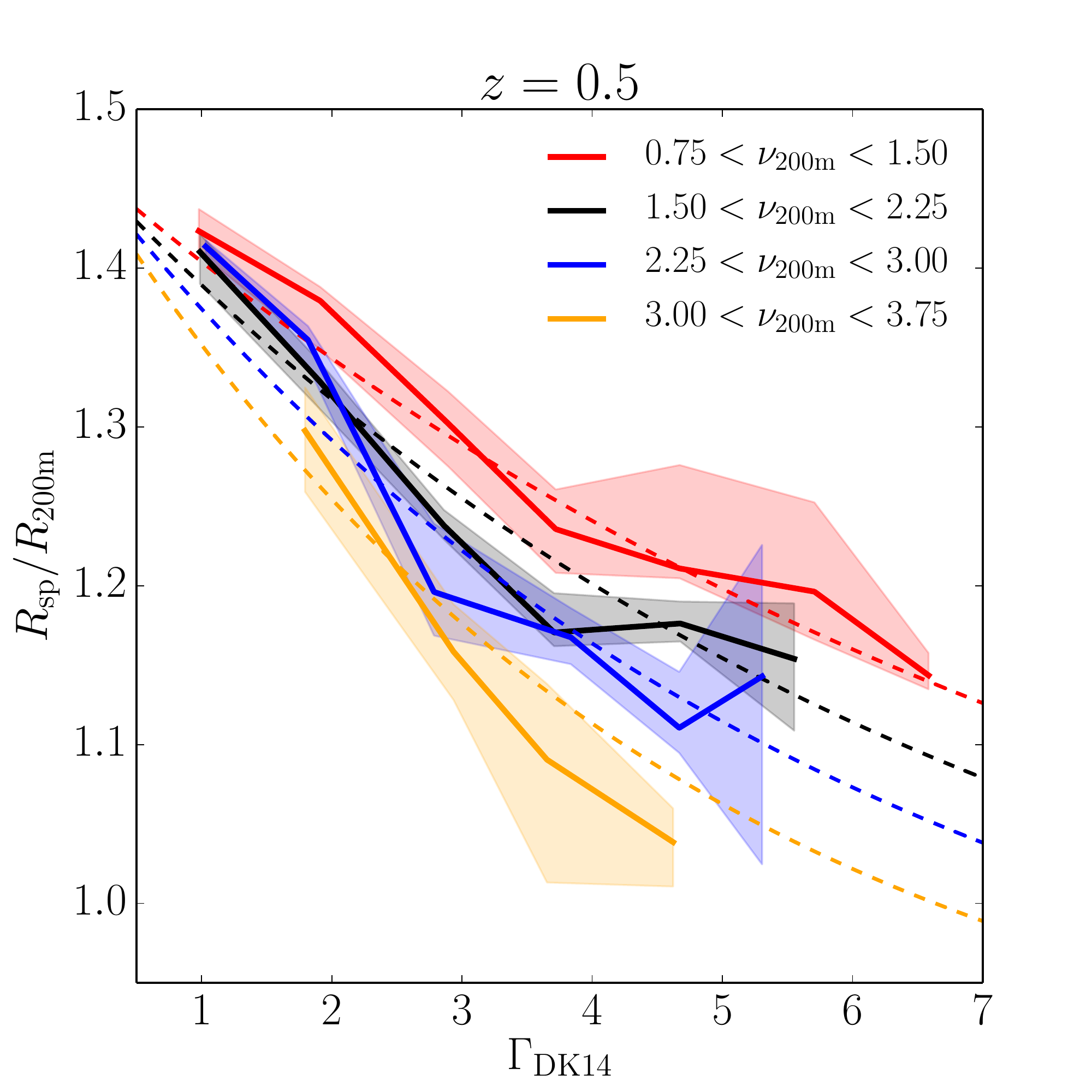}\\
   \includegraphics[width=\columnwidth]{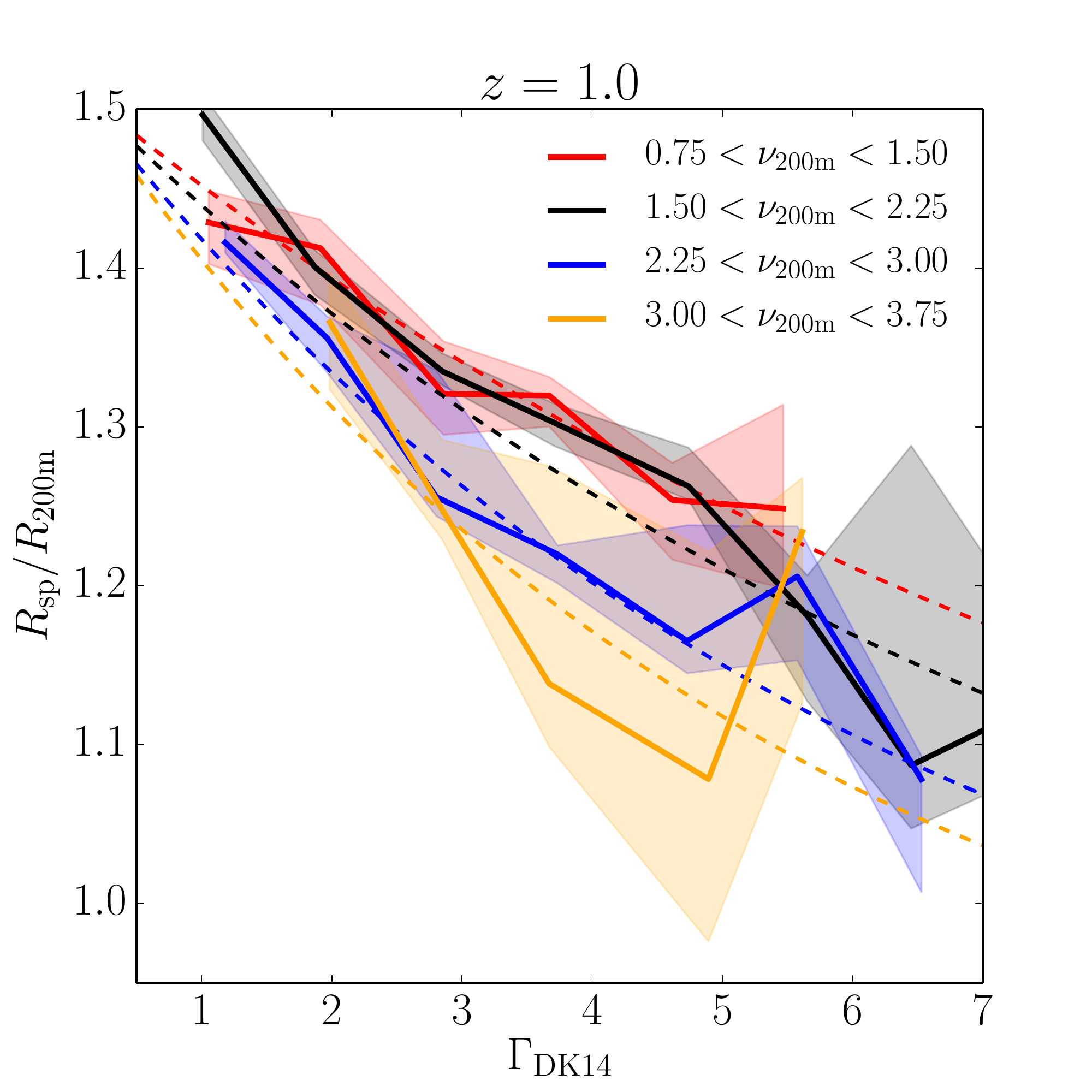}
   \includegraphics[width=\columnwidth]{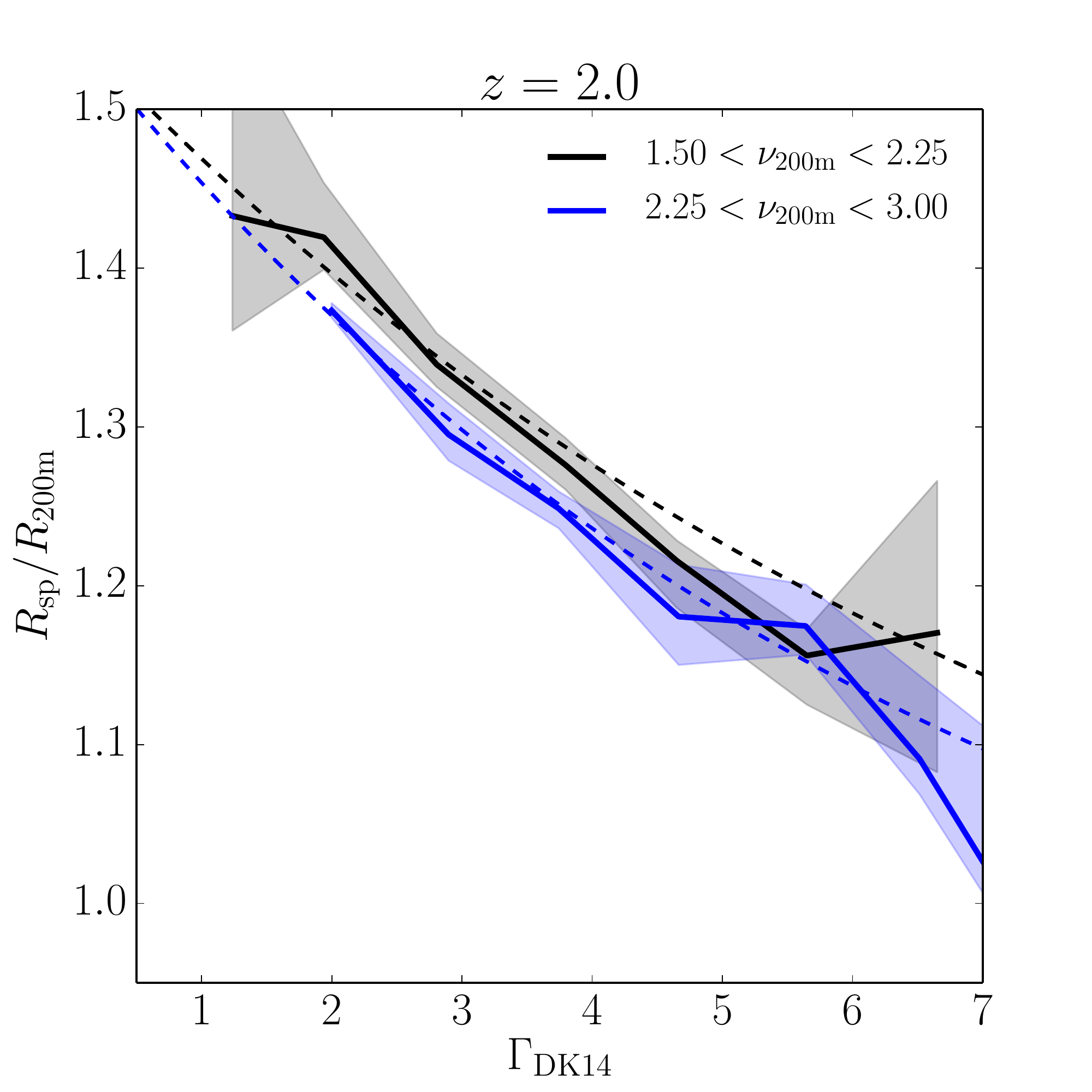}
   \caption{Comparison between our fit and \textsc{Shellfish}'s measurements of $R_{\rm sp}/R_{\rm 200m} \equiv \tilde{R}_{\rm sp}(\Gamma_{\rm DK14}, \nu_{\rm 200m}, z)$. The thick lines represent the median
   value of $R_{\rm sp}/R_{\rm 200m}$ in each $\Gamma_{\rm DK14}$ bin and the
   shaded regions indicate the 68\% errors on those medians, as determined by
   bootstrapping. The thin lines show the median of the distribution 
   given by Equations \ref{eq:r_fit_form} -\ref{eq:r_fit_form_2} evaluated
   at the median $\nu_{\rm 200m}$ value within the corresponding $\nu_{\rm 200m}$ bin.}
   \label{fig:r_centroid}
\end{figure*}

\begin{figure*}
   \centering
   \includegraphics[width=\columnwidth]{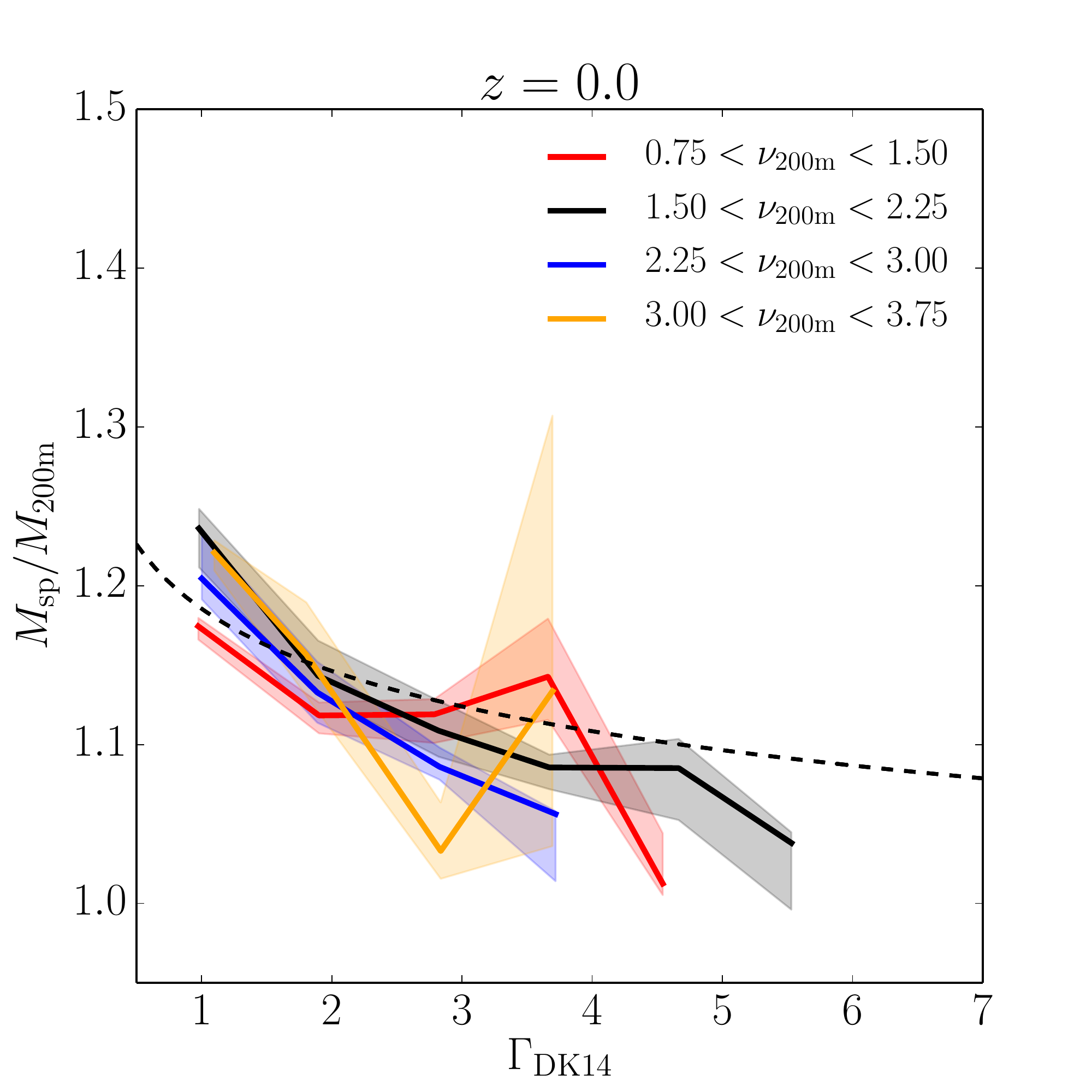}
   \includegraphics[width=\columnwidth]{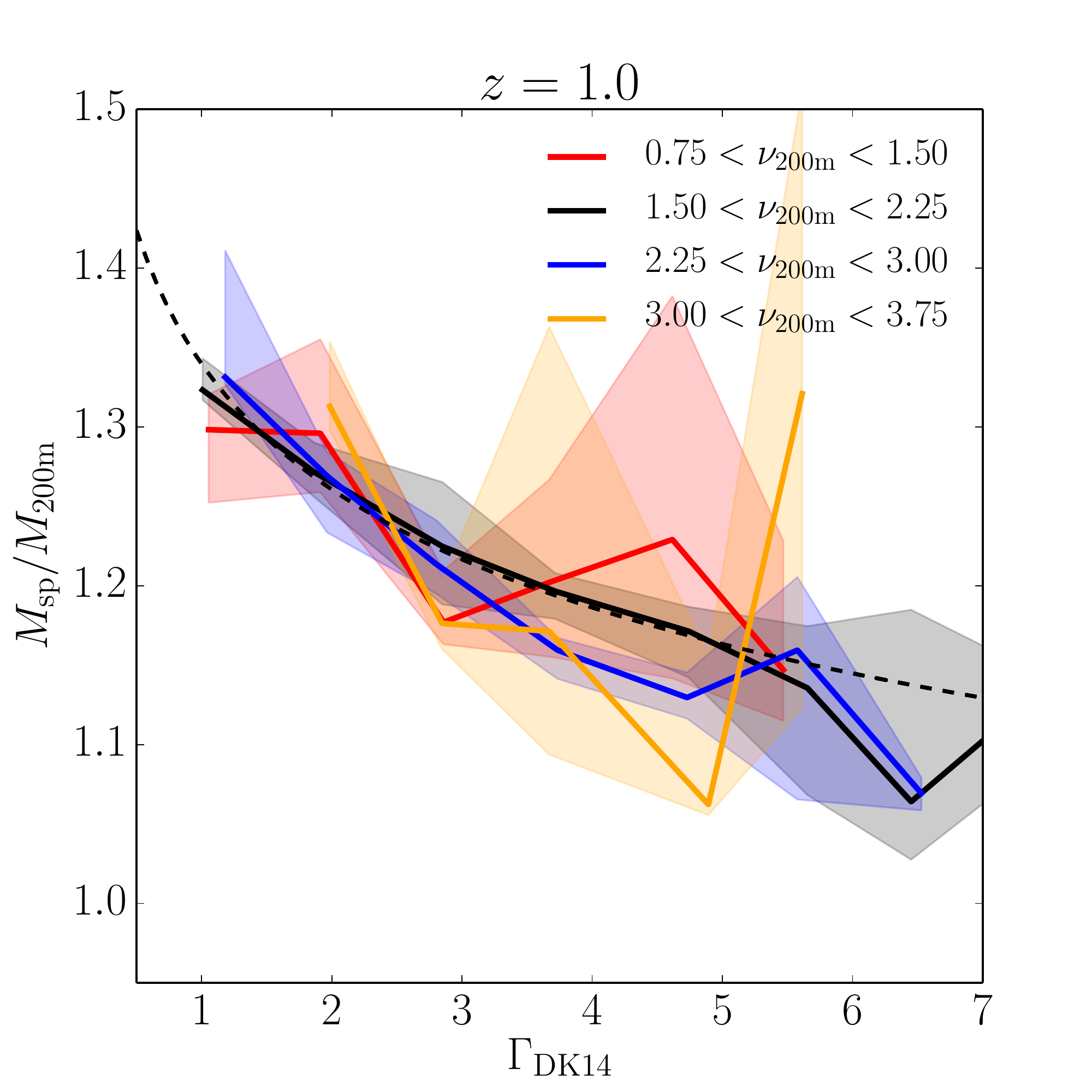}
   \caption{
   Comparison between our fit and \textsc{Shellfish}'s measurements for $M_{\rm sp}/M_{\rm 200m} \equiv \tilde{M}_{\rm sp}(\Gamma_{\rm DK14}, \nu_{\rm 200m}, z)$. 
   The visualization scheme is identical to the one used in Figure
   \ref{fig:m_centroid}, with the thin line corresponding to the median of the
   distribution given by Equations \ref{eq:m_fit_form} and
   \ref{eq:m_fit_form_2}. Note that unlike the fit displayed in Figure
   \ref{fig:r_centroid}, our $\tilde{M}_{\rm sp}$ has no $\nu_{\rm 200m}$
   dependence, so only a single thin line is plotted. There are several
   important caveats to this fit, which we discuss in section \ref{sec:g_m_sp}.}
   \label{fig:m_centroid}
\end{figure*}

\begin{figure}
   \centering
   \includegraphics[width=\columnwidth]{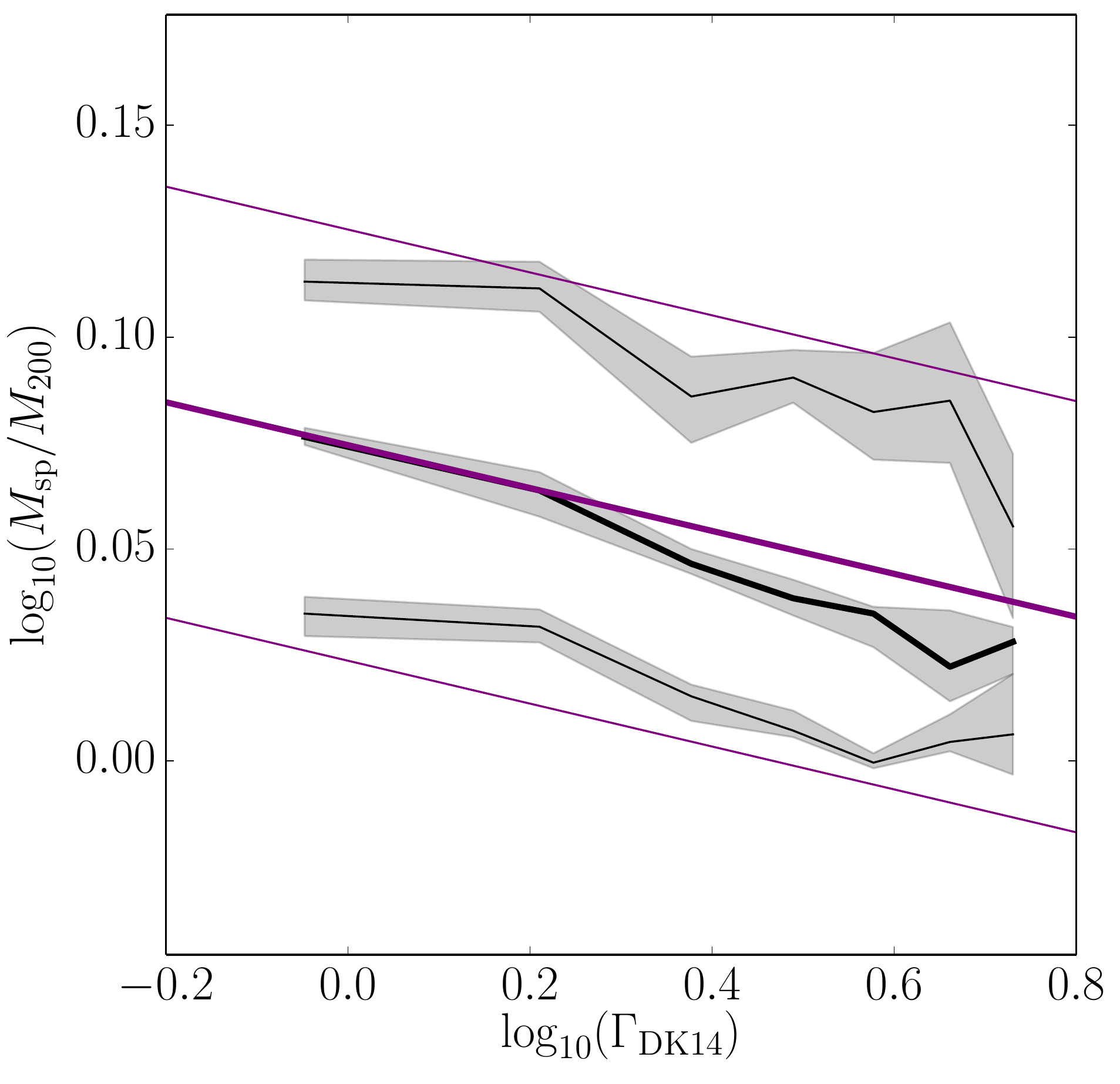}
   \caption{Comparison between the $\tilde{M}_{\rm sp}$ median and $68\%$
   contours for our data and our fit given by Equations \ref{eq:m_fit_form} and
   \ref{eq:m_fit_form_2} at $z=0$. This Figure was made to emphasize the
   weaknesses in our $\tilde{M_{\rm sp}}$ fit and shows an $\approx 2\% - 4\%$
   overestimation of the median at high $\Gamma_{\rm DK14}$ and a similar
   overestimation of the logarithmic scatter, $\sigma_{\rm dex}$. An extended
   discussion of this Figure can be found in section
   \ref{sec:g_m_sp}.}
   \label{fig:msp_distr}
\end{figure}

One of the key results obtained by previous analyses of splashback shells
using stacked radial density profiles \citep{diemer_kravtsov_2014, more_et_al_2015, more_et_al_2016, adhikari_et_al_2016} is the dependence of the
splashback radius in units of the $R_{\rm 200m}$ on the mass accretion rate
$\Gamma_{\rm DK14}$ (see Equation~\ref{eq:Gamma}): halos with larger accretion rates have
smaller values of $R_{\rm sp}/R_{\rm 200m}\equiv\tilde{R}_{\rm sp}$. In this
section we present the result of fits to $\tilde{R}_{\rm sp}$ using the
measurements from \textsc{Shellfish}.

Specifically, we fit the following log-normal distribution to
$\tilde{R}_{\rm sp}$ as a function of $\nu_{\rm 200m},$ $\Gamma_{\rm DK14},$ and
$\Omega_m$:
\begin{align}
        \label{eq:r_fit_form}
    P(\tilde{R}_{\rm sp}) & \propto
            \exp(-\log_{10}^2(\tilde{R}_{\rm sp}/R_{\rm med})/2\sigma_{\rm dex}^2), \\
    R_{\rm med} &= (R_0\Omega_m + R_1) \exp{(\alpha\Gamma_{\rm DK14})} + A, \\
    \alpha &= \eta_0\Omega_m^2 + \eta_1\Omega_m + \eta_2 + \xi\nu_{\rm 200m}.
    \label{eq:r_fit_form_2}
\end{align}
Here $R_0$, $R_1$, $A$, $\xi$, $\eta_0$, $\eta_1$, and $\eta_2$ are fit
parameters.

As discussed above, our sample only includes halos with
$\Gamma_{\rm DK14} > 0.5.$  We fit the functional form given by Equations
\ref{eq:r_fit_form}--\ref{eq:r_fit_form_2} using an implementation of the
affine-invariant Markov Chain Monte Carlo sampling algorithm of
\citet{goodman_weare_2010}. We also adopt a Heaviside prior on the logarithmic 
scatter, $\sigma_{\rm dex},$ to prevent it from becoming non-positive.

We find that the best fit parameters are
\begin{align*}
    R_0 &= 0.2181,
    &\eta_0& = -0.1742, \\
    R_1 &= 0.4996,
    &\eta_1& = 0.3386, \\
    A &= 0.8533,
    &\eta_2& = -0.1929,  \\
    \xi &= -0.04668,
    &\sigma_{\rm dex}& = 0.046.
\end{align*}
The resulting function is plotted against our data in Figure
\ref{fig:r_centroid}.

It is interesting that the radii estimated by \textsc{Shellfish} exhibit
a strong dependence on {\it both} mass accretion rate and peak
height. This trend can also be seen in other methods for measuring individual
splashback shells around halos, such as the median angular
profile method described in section \ref{sec:median_prof} and the
apocenter-based splashback-measuring code \textsc{Sparta}
\citep{diemer_et_al_2017}.
The trend cannot be attributed to convergence trends
because all halos used in the sample have $N_{\rm 200m}$ above the convergence
limit of $5 \times 10^4$ found in section \ref{sec:tests} and because
the mass bounds given in Table \ref{tab:sims} restrict the halos in our sample
to a single decade in particle count.

Previous estimates from stacked density profiles only
found a strong dependence on $\Gamma_{\rm DK14}$, while a $\nu_{\rm 200m}$
dependence was either not apparent or weak \citep[e.g.,][]{more_et_al_2015}.
The $\nu_{\rm 200m}$ dependence is also not predicted in the collapse models of
isolated peaks \citep[e.g.,][]{adhikari_et_al_2014}, even though they
successfully predict a $\Gamma_{\rm DK14}$ dependence. The origin of the
$\nu_{\rm 200m}$ dependence and the seeming discrepancy with the collapse model
is not clear. Additionally, although we have made an empirical argument that
stacked profiles are biased by massive subhalos, we do not yet propose a
physical picture for why this bias should also erase or decrease trends with
$\nu_{\rm 200m}$.

\subsection{Splashback Shell Masses}
\label{sec:g_m_sp}

In contrast to overdensity-based halo definitions, $M_{\rm sp}$ and $R_{\rm sp}$
are independent (albeit correlated) quantities. For this reason we do not fit
the same functional form to both $R_{\rm sp}$ and $M_{\rm sp}.$
We fit the following log-normal distribution to
$\tilde{M}_{\rm sp} \equiv M_{\rm sp}/M_{\rm 200m}$ as a function of
$\Gamma_{\rm DK14}$ and $\Omega_m$:

\begin{align}
        \label{eq:m_fit_form}
    P(\tilde{M}_{\rm sp}) & \propto
            \exp(-\log_{10}^2(\tilde{M}_{\rm sp}/M_{\rm med}) / 
            2\sigma_{\rm dex}^2), \\
    \label{eq:m_fit_form_2}
    M_{\rm med} &= (M_0\Omega_m + M_1) \left(\frac{\Gamma_{\rm DK14}}{\Gamma_{\rm pivot}}\right)^{\alpha_0\Omega_m + \alpha_1}.
\end{align}
Here $\Gamma_{\rm pivot} = 3$ is a characteristic pivot value, and
$M_0,$ $M_1$, $\alpha_0$, and $\alpha_1$ are fit parameters.

Using the same procedure described in section
\ref{sec:g_r_sp} we obtain the parameters
\begin{align*}
    A_0 &= 0.192
    &a_0& = -0.0781 \\
    A_1 &= 1.072
    &a_1& = -0.0284 \\
    \sigma_{\rm dex} &= 0.054
\end{align*}
The median of this fit is shown in Figure \ref{fig:m_centroid}. Note that unlike
our fit to $\tilde{R}_{\rm sp}$, we do not model $\tilde{M}_{\rm sp}$ as having
a $\nu_{\rm 200m}$ dependence because there is not strong evidence for
such a trend in our data. This contrasts with the results of \textsc{Sparta},
which did find a strong $\nu_{\rm 200m}$ trend \citep{diemer_et_al_2017}. It is currently not clear
whether higher quality data would reveal a small mass trend in the
\textsc{Shellfish} data as well.

The left panel of Figure \ref{fig:m_centroid} shows a deviation between our
fit and \textsc{Shellfish}'s measurements at high $\Gamma_{\rm DK14}$ for
$z=0$. We investigate this further in Figure \ref{fig:msp_distr} which shows
the median and 68\% contours of the $\tilde{M}_{\rm sp}$ distribution at $z=0$.
This Figure shows that
although the median of our data is well approximated by a power law, our 
Bayesian fit reports a shallower slope. This results in a $\approx 2\% - 4\%$
overestimation of $\tilde{M}_{\rm sp}$ at high accretion rates for this
redshift.

This overestimation is caused by the fact that at high $\Gamma_{\rm DK14}$
$\tilde{M}_{\rm sp}$ follows an skewed log-normal distribution. Since our model
assumes a log-normal distribution, our fit's median is pulled high relative to
our data's median. The offset between the two medians also leads to an
overestimation of the logarithmic scatter, $\sigma_{\rm dex}$ by a comparable
amount.

Despite this, we deliberately choose not to model the skew for three reasons.
The first reason is simplicity: our experiments with explicitly modeling the
skew show that it has non-linear dependencies on
$\Gamma_{\rm DK14}$ and $z.$ The second reason is that this reduction in
simplicity would result in an increase in accuracy for only a small number of
halos: high accreting halos at $z=0$ are rare. The third reason is that this
effect is comparable to our stated systematic uncertainty in the radii and
masses reported by \textsc{Shellfish}, so any subsequent analysis which would
reach a qualitatively different conclusion from an improvement in fit modeling
is not respecting the known uncertainty in \textsc{Shellfish} shells. Instead,
we choose to use an extremely simple model - a power law with log-normal
residuals and a linear dependence on $\Omega_m$ - and leave more precise
modeling to future work.

The skew seen in the the low redshift, high $\Gamma_{\rm DK14}$ has a simple
explanation. The scatter in $\tilde{M}_{\rm sp}$ has two sources: the first is
the variation in shell sizes which also causes the scatter in
$\tilde{R}_{\rm sp}$,
the second is the presence or non-presence of high mass subhalos. Since halos
with high accretion rates are more likely to have high mass subhalos than halos
with low accretion rates, the second effect is particularly important for them.
If a halo has a massive subhalo outside of $R_{\rm 200m}$ but inside its
splashback shell, $\tilde{M}_{\rm sp}$ is scattered high. If a halo has a
massive subhalo inside $R_{\rm 200m}$, both $M_{\rm 200m}$ and $M_{\rm sp}$
increase, so $\tilde{M}_{\rm sp}$ scatters towards 1. When the median of the 
of the $\tilde{M}_{\rm sp}$ distribution is close to 1, this means that the
presence of massive subhalos has the effect of reducing down scatter and
increasing upscatter relative to what we would expect from variation in shell
sizes alone.

\subsection{Splashback Shell Overdensities}

We model the distribution of
$\Delta_{\rm sp} \equiv 200 \tilde{M}_{\rm sp}/\tilde{R}_{\rm 200m}$ by taking
the ratios of our mass fit (Equations \ref{eq:m_fit_form} and
\ref{eq:m_fit_form_2}) and our radius fit (Equations \ref{eq:r_fit_form} -
\ref{eq:r_fit_form_2}). Because our $\Delta_{\rm sp}$ model is derived from our
$\tilde{M}_{\rm sp}$ fit, it is subject to the same caveats discussed in section
\ref{sec:g_m_sp}. However, because the dynamic range of $\Delta_{\rm sp}$ is larger
than that of $\tilde{M}_{\rm sp}$, the affect of a few-percent disparity in
masses is minimal.

\begin{figure*}
   \centering
   \includegraphics[width=\columnwidth]{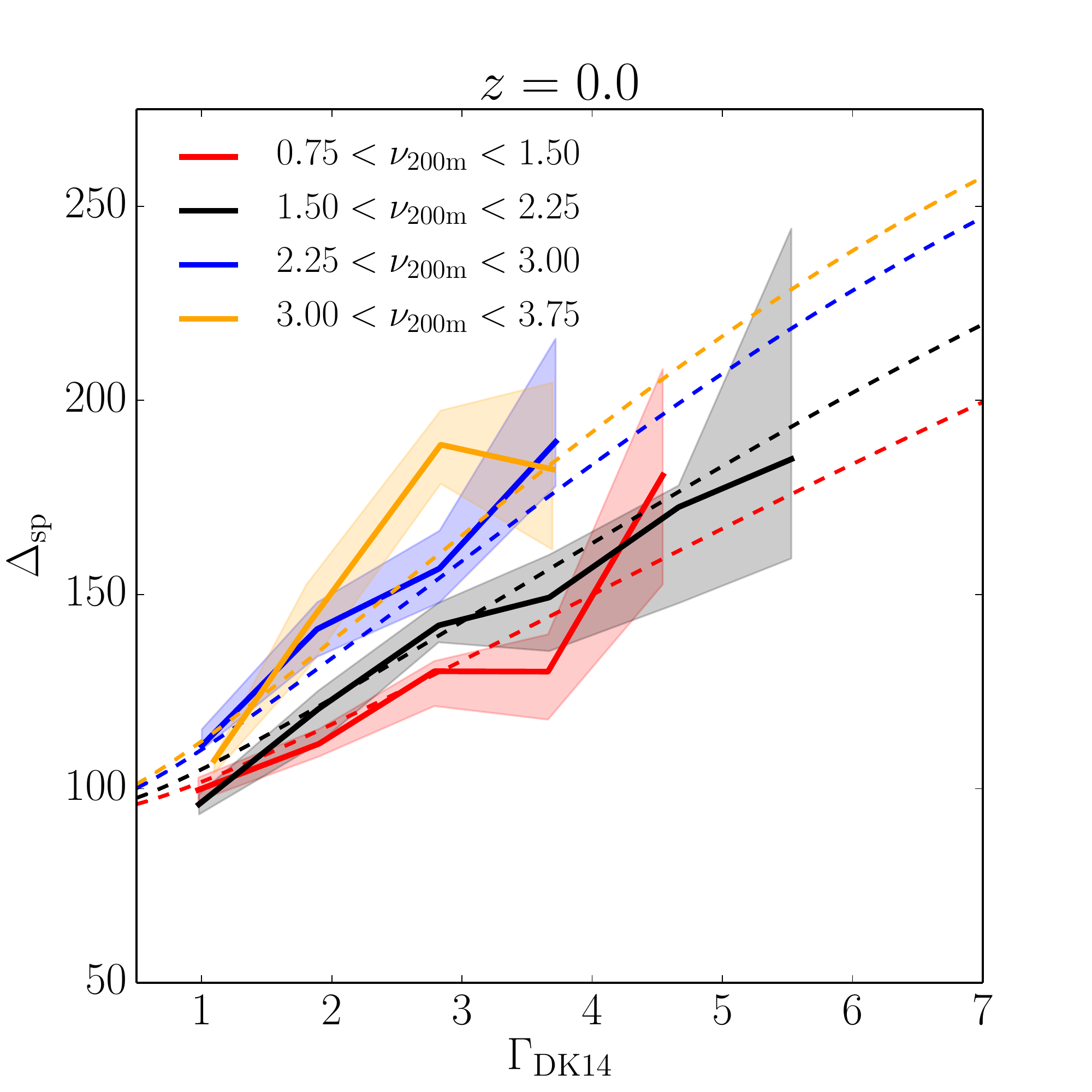}
   \includegraphics[width=\columnwidth]{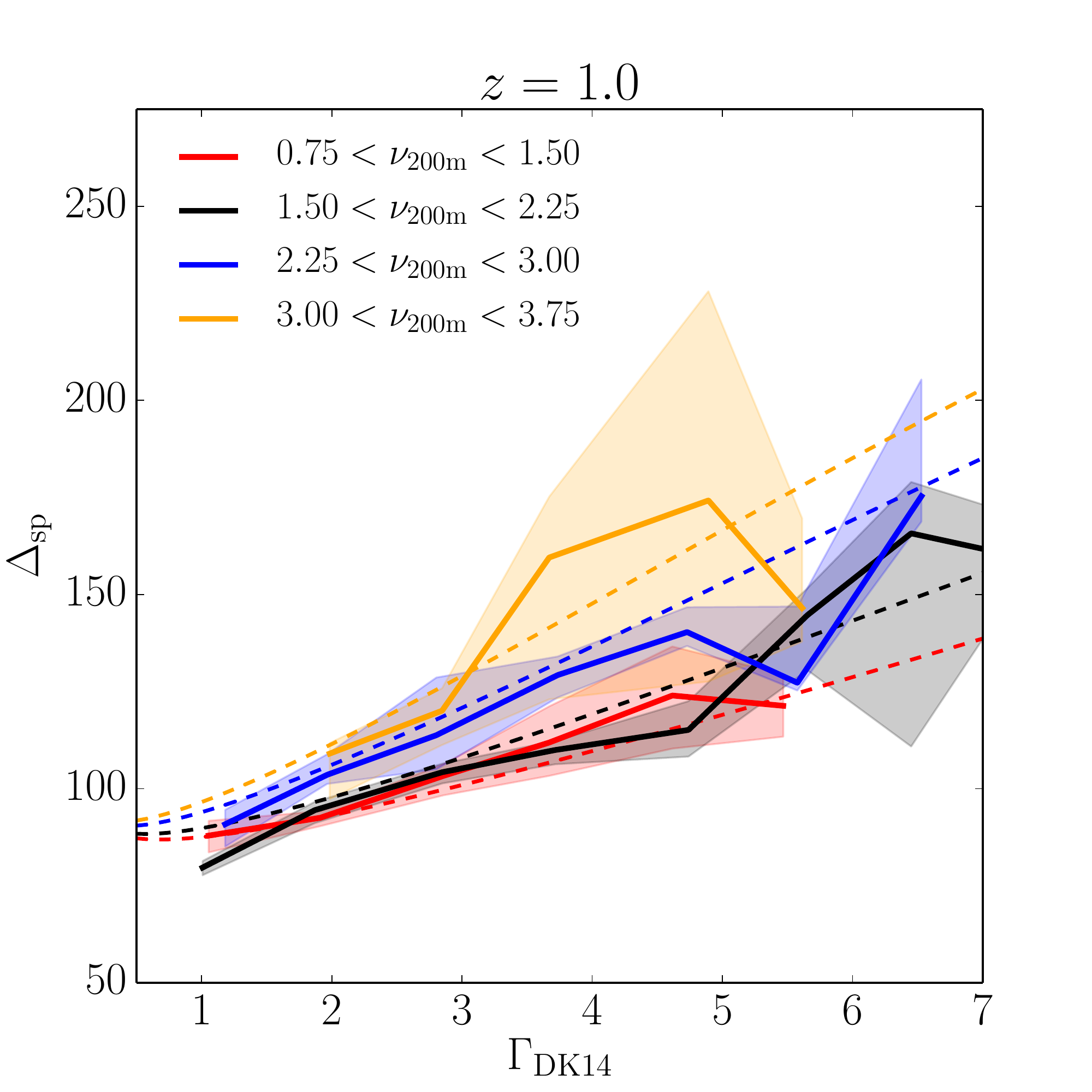}
   \caption{
   Comparison between our fits and \textsc{Shellfish}'s measurements for
   $\Delta_{\rm sp} \equiv 200 \tilde{M}_{\rm sp}/\tilde{R}_{\rm 200m}$ using
   the ratio of our mass and radius fits.
   The visualization scheme is identical to the one used in Figure
   \ref{fig:m_centroid}, with the thin line corresponding to the median of the
   distribution given by Equations \ref{eq:m_fit_form} and
   \ref{eq:m_fit_form_2}. There are several important caveats to this fit, which
   we discuss in section \ref{sec:g_m_sp}.}
   \label{fig:rho_centroid}
\end{figure*}

This ratio is shown in Figure \ref{fig:rho_centroid}. Median overdensities range
between $\approx$ 70 and $\approx$ 200 with strong dependencies on peak height,
accretion rate, and redshift. The most important consequence of these relations
is that \emph{there is not a single classical overdensity boundary which
corresponds to to the splashback shell.}

\subsection{Splashback Shell Shapes}
\label{sec:shapes}

\begin{figure}
   \centering
   \includegraphics[width=\columnwidth]{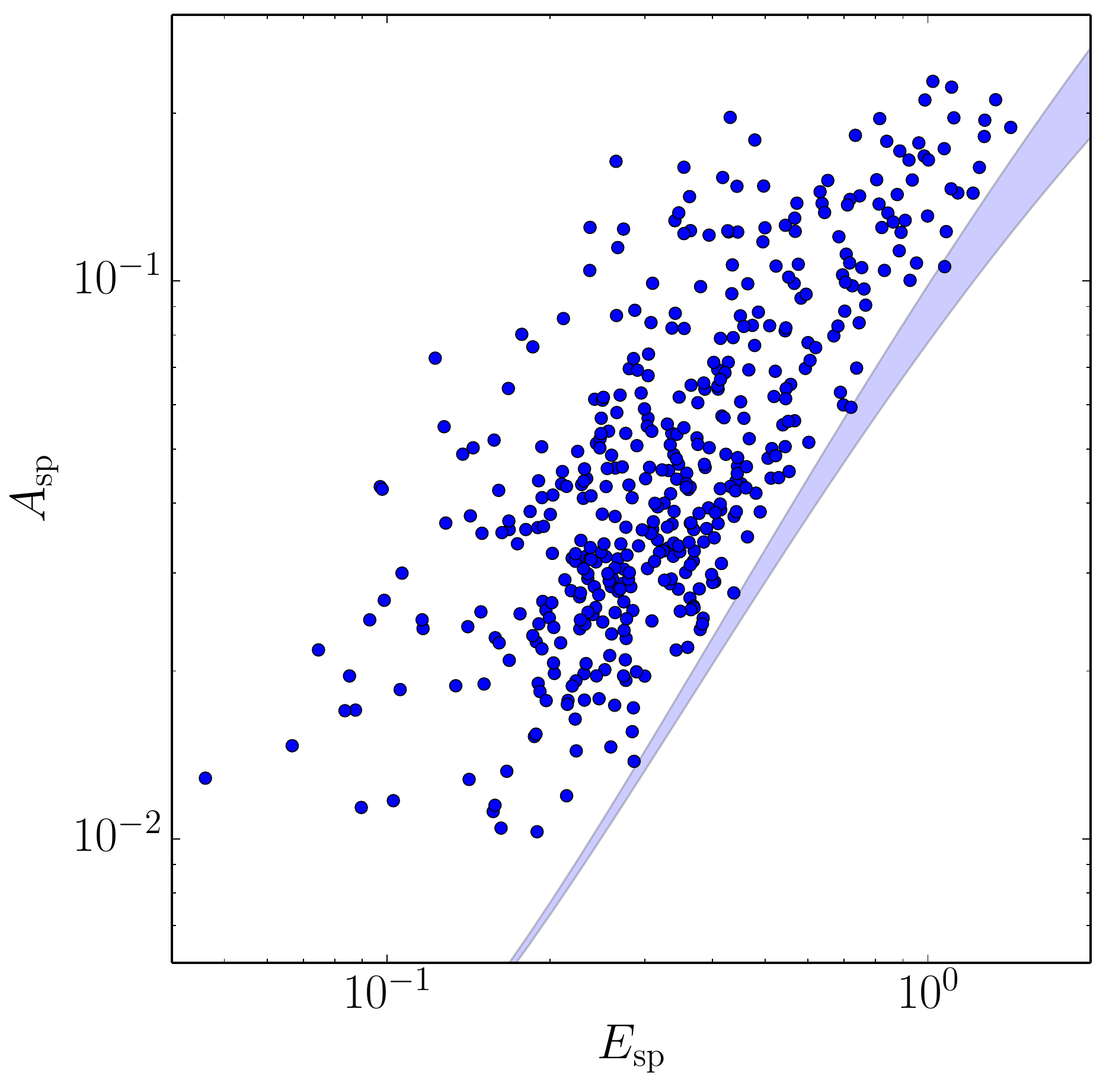}
   \caption{The asphericity parameter, $A_{\rm sp}$, versus 
   the ellipticity parameter, $E_{\rm sp}$ (defined in Equations.~\ref{eq:A_def}
   and Equation~\ref{eq:E_def}, respectively) for our $z=0$ halo sample. The blue
   shaded region shows the range of values of these quantities for ellipsoids
   with different axis ratios. The fact that $A_{\rm sp}$ and  $E_{\rm sp}$
   of the splashback shells lie above the shaded regions means
   that the shells have significantly higher surface areas than ellipsoids of
   similar ellipticity and volume. }
   \label{fig:shape}
\end{figure}

We also investigate the shapes of splashback shells using the asphericity,
$A_{\rm sp}$,  and ellipticity, $E_{\rm sp}$, parameters defined in
Equations~\ref{eq:A_def} and ~\ref{eq:E_def}, respectively. A plot of these two
quantities is shown in
Figure \ref{fig:shape}. The shaded blue region shows the values of these
parameters for ellipsoids with different axis ratios. The fact that $A_{\rm sp}$
and $E_{\rm sp}$ for all splashback shells lie above the shaded regions means
that the shells are significantly more aspherical than ellipsoids.

We perform checks for correlation between $A_{\rm sp}$, $E_{\rm sp}$
and each of $M_{\rm 200m}$, $\Gamma_{\rm DK14}$, $R_{\rm sp}$, and redshift, but find no
evidence of such correlations.

\begin{figure}
   \centering
   \includegraphics[width=\columnwidth]{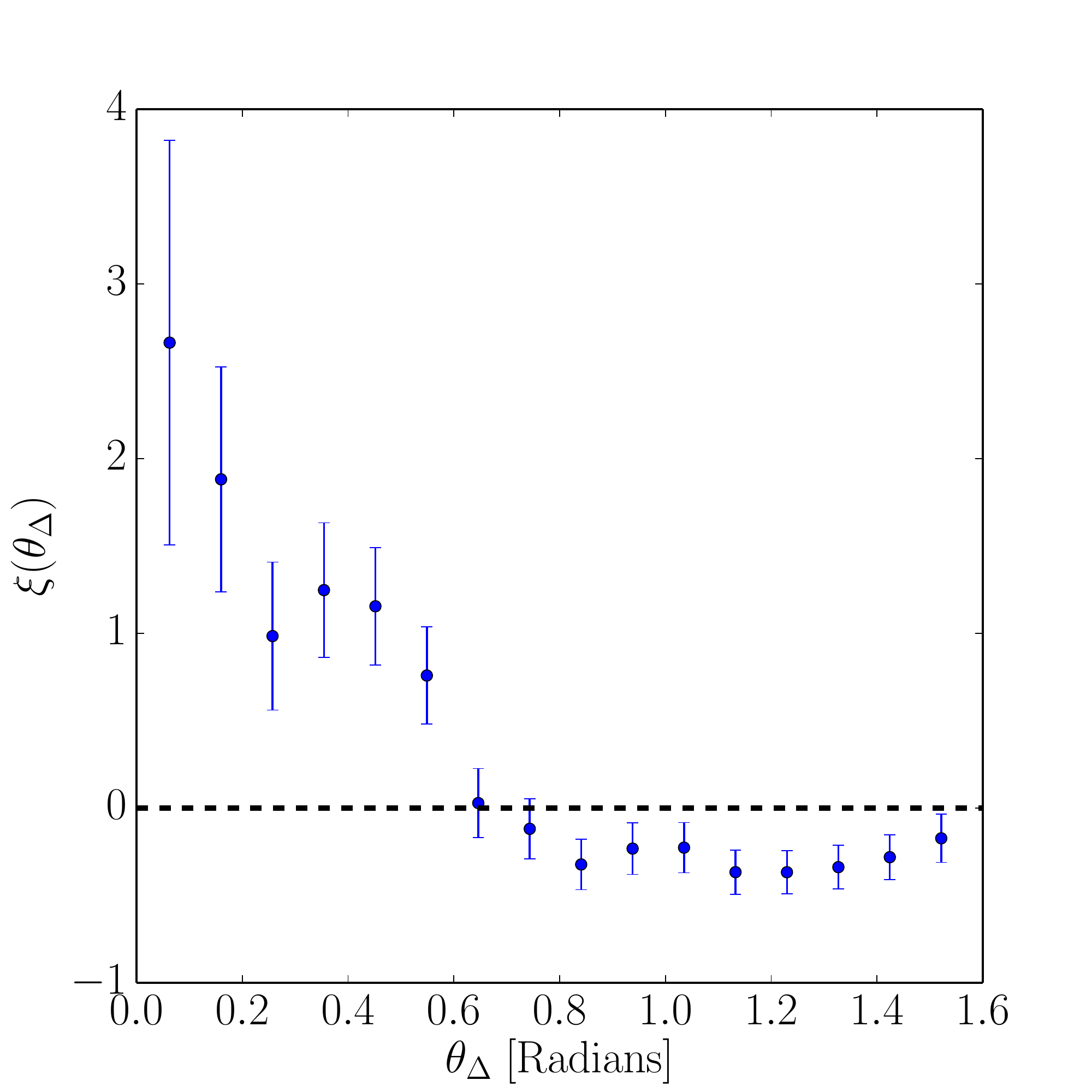}
   \caption{The correlation function, $\xi(\theta_\Delta)$, between
   the major axes of splashback shells and the major axes of total dark
   matter distribution. The dashed black line shows $\xi(\theta_\Delta)=0,$
   and indicates the level of correlation expected for random alignment.}
   \label{fig:align}
\end{figure}

We also calculated the angle $\theta_\Delta$ between the major axis of a
halo's splashback shell and the major axis of the underlying dark matter
distribution, as reported by the Rockstar halo finder
\citep{behroozi_et_al_2013}. In Figure \ref{fig:align} we show the
correlation function for the angle between these two axes, $\theta_\Delta$. We
find anti-correlation at high values of $\theta_\Delta$ and a high degree of
correlation at low values of $\theta_\Delta,$ indicating that splashback
shells are preferentially aligned with major axis of the central dark matter
distribution. This is consistent with earlier studies, which have shown that the
axis ratios of the matter distribution near the centers of halos tend to be
roughly aligned with the axis ratios near the outskirts of halos
\citep{jing_suto_2002}.

\section{Summary and Conclusions}
\label{sec:summary}

In this paper we presented a new algorithm which identifies the splashback
shells around individual halos in simulations. These shells
are caused by the caustics formed by matter at the first apocenter of their
orbits around the halo and correspond to rapid drops in the density field. Our
algorithm relies only on the density distribution within a single simulation
snapshot, and is capable of identifying shells with highly aspherical shapes.

We implemented our algorithm in the publicly
available\footnote{\texttt{github.com/phil-mansfield/shellfish}} code
\textsc{Shellfish} and performed
extensive tests on the correctness of this code. We performed
convergence tests on the splashback shells found by our code and found that
above a convergence limit of $N_{\rm 200m} = 5\times 10^4$, \textsc{Shellfish}
can measure properties of splashback shells with $\lesssim 5\%$ systematic 
error (see Figure \ref{fig:n200_rsp_convergence} and Figure
\ref{fig:comp_all_z}) and percent-level stochastic error
(see Figure \ref{fig:ring_converge}). However,
we identified a sub-population of halos with low mass accretion rates,
$\Gamma_{\rm DK14}\lesssim 0.5$, for which the splashback shell radii estimated by
\textsc{Shellfish} are biased low by $\gtrsim 10\%$. We therefore recommend
that our code not be used for measurements of splashback shells for halos
with $\Gamma_{\rm DK14} < 0.5$. This cutoff removes $20\%$ of Milky Way-sized halos at
$z=0$ and has a negligible effect on all larger mass scales and all earlier
redshift slices.

We presented the first measurements of several basic properties of splashback
shells which are summarized below: 

\begin{enumerate}
    \item[1.] We confirmed that splashback radii generally decrease with
    increasing mass accretion rate, as previously found by analyses of
    stacked halo density profiles. However, we found that the splashback
    radii found
    by \textsc{Shellfish} are larger than these earlier estimates by 20\%-30\%
    for halos with high accretion rates, $\Gamma_{\rm DK14}\gtrsim 3$. We showed that
    the estimate of the splashback radius obtained from the stacked
    density profiles is biased low due to the existence of high-mass
    subhalos in many of these profiles.
    \item[2.] We used a simple method, completely
    independent from \textsc{Shellfish}, for mitigating the
    effect of substructure on density profiles: the so-called 
    ``angular median profile''
    method. In this method, radial shells
    are split into solid angle segments with an estimate of density in each 
    segment. The halo density at a given radius is then taken to be the median
    of all the segments in the corresponding shell. We showed that the effect of
    subhalos on
    these profiles is greatly reduced. Moreover, the angular median profiles are
    more self-similar in their outskirts and exhibit a sharper region of
    profile steepening (i.e., a much more distinct splashback feature). We
    showed that the splashback radii estimated from the stacked angular median
    profiles are in good agreement with the results of \textsc{Shellfish} for
    halos with $\Gamma_{\rm DK14}\gtrsim 0.5$.
    \item[3.] We investigated the correlation between splashback radius and
    mass accretion rate, the scatter around it, and its evolution with redshift.
    We presented the first evidence that the splashback radius depends not only
    on accretion rate, but also has a strong dependence on the
    peak height, $\nu_{\rm 200m}$, with larger $\nu_{\rm 200m}$ halos having systematically smaller
    $R_{\rm sp}/R_{\rm 200m}$ at a fixed $\Gamma_{\rm DK14}$ and $z$. We found that the scatter of
    $R_{\rm sp}/R_{\rm 200m}$ around the median at a given accretion rate is significant, exceeding $10\%$. We provided an accurate fit for
    $R_{\rm sp}/R_{200m}$ and its scatter as a function of $\Gamma_{\rm DK14}$, $\nu_{\rm 200m}$, and
    $\Omega_{\rm m}$ (see Equations \ref{eq:r_fit_form}-\ref{eq:r_fit_form_2}
    and Figure \ref{fig:r_centroid}). We provided a similar fit for
    $M_{\rm sp}/M_{\rm 200m}$ (see Equations \ref{eq:m_fit_form} and \ref{eq:m_fit_form_2}
    and Figure \ref{fig:m_centroid}). Unlike our fit to $R_{\rm sp}$, there are
    several minor caveats to our $M_{\rm sp}$ fit, which we discuss in section
    \ref{sec:g_m_sp}.
    \item[4.] We argued that a single classical overdensity density cannot be
    used as a model of the location of $R_{\rm sp}$ because the overdensity of
    splashback shells have a large dynamic range and have strong dependencies on
    mass, accretion rate, and redshift.
    \item[5.] We studied the shapes of the splashback shells using an
    ellipticity parameter, $E_{\rm sp}$, and an asphericity parameter,
    $A_{\rm sp}$ (defined in Equations~\ref{eq:E_def} and \ref{eq:A_def},
    respectively). We showed that splashback shells are generally highly
    aspherical, with non-ellipsoidal oval shapes being particularly common.
    \item[6.] We investigated potential correlations between splashback shell
    properties and other halo properties, but found no significant correlations
    between $E_{\rm sp}$ and $A_{\rm sp}$ 
    with either mass accretion rate, mass, splashback radius or redshift.
    However, we did find
    that the major axes of splashback shells were correlated with the
    major axis of mass distribution within the inner regions of halos.
\end{enumerate}

This paper is a pilot study of splashback shells of individual halos. Further
applications of the algorithm presented here include investigation of
alternative classifications of isolated halos and subhalos using the
splashback shell instead of the virial radius, investigation of the
systematic differences in halo masses and halo mass accretion histories when
$M_{\rm sp}$ is compared to $M_{\rm \Delta}$, and a comparisons with other
methods for measuring individual splashback shells.

\acknowledgments
We would like to thank Surhud More, Susmita Adhikari, and Neal Dalal for useful discussions
related to the results presented in this paper, as well as Chihway Chang for useful comments on the draft.
This work was supported by the Kavli Institute for Cosmological Physics at
the University of Chicago through grant PHY-1125897 and an endowment from
the Kavli Foundation and its founder Fred Kavli. We have made extensive use of the
NASA Astrophysics Data System and {\tt arXiv.org} preprint server.
The simulations used in this study have been carried out using
the midway computing cluster supported by the University of
Chicago Research Computing Center.

\bibliographystyle{apj}
\bibliography{splashback}

\begin{thebibliography}{}
\expandafter\ifx\csname natexlab\endcsname\relax\def\natexlab#1{#1}\fi
\providecommand{\url}[1]{\href{#1}{#1}}

\bibitem[{{Abel} {et~al.}(2012){Abel}, {Hahn}, \& {Kaehler}}]{abel_et_al_12}
{Abel}, T., {Hahn}, O., \& {Kaehler}, R. 2012, \mnras, 427, 61

\bibitem[{{Adhikari} {et~al.}(2014){Adhikari}, {Dalal}, \&
  {Chamberlain}}]{adhikari_et_al_2014}
{Adhikari}, S., {Dalal}, N., \& {Chamberlain}, R.~T. 2014, JCAP, 11, 019

\bibitem[{{Adhikari} {et~al.}(2016){Adhikari}, {Dalal}, \&
  {Clampitt}}]{adhikari_et_al_2016}
{Adhikari}, S., {Dalal}, N., \& {Clampitt}, J. 2016, ArXiv e-prints,
  arXiv:1605.06688

\bibitem[{{Behroozi} {et~al.}(2013){Behroozi}, {Wechsler}, \&
  {Wu}}]{behroozi_et_al_2013a}
{Behroozi}, P.~S., {Wechsler}, R.~H., \& {Wu}, H.-Y. 2013, \apj, 762, 109

\bibitem[{Behroozi {et~al.}(2013)Behroozi, Wechsler, \&
  Wu}]{behroozi_et_al_2013}
Behroozi, P.~S., Wechsler, R.~H., \& Wu, H.-Y. 2013, The Astrophysical Journal,
  762, 109.
\newblock \url{http://stacks.iop.org/0004-637X/762/i=2/a=109}

\bibitem[{{Behroozi} {et~al.}(2013){Behroozi}, {Wechsler}, {Wu}, {Busha},
  {Klypin}, \& {Primack}}]{behroozi_et_al_2013b}
{Behroozi}, P.~S., {Wechsler}, R.~H., {Wu}, H.-Y., {et~al.} 2013, \apj, 763, 18

\bibitem[{{Bertschinger}(1985)}]{bertschinger_1985}
{Bertschinger}, E. 1985, \apjs, 58, 39

\bibitem[{{Dalal} {et~al.}(2010){Dalal}, {Lithwick}, \&
  {Kuhlen}}]{dalal_et_al_2010}
{Dalal}, N., {Lithwick}, Y., \& {Kuhlen}, M. 2010, ArXiv e-prints,
  arXiv:1010.2539

\bibitem[{{Delaunay}(1934)}]{delaunay_1934}
{Delaunay}, B. 1934, Bulletin de l'Académie des Sciences de l'URSS, Classe des
  sciences mathématiques et naturelles, 6, 793

\bibitem[{{Diemer}(2017)}]{diemer_2017}
{Diemer}, B. 2017, ArXiv e-prints, arXiv:1703.09712

\bibitem[{{Diemer} \& {Kravtsov}(2014)}]{diemer_kravtsov_2014}
{Diemer}, B., \& {Kravtsov}, A.~V. 2014, \apj, 789, 1

\bibitem[{{Diemer} {et~al.}(2013{\natexlab{a}}){Diemer}, {Kravtsov}, \&
  {More}}]{diemer_et_al_2013b}
{Diemer}, B., {Kravtsov}, A.~V., \& {More}, S. 2013{\natexlab{a}}, \apj, 779,
  159

\bibitem[{{Diemer} {et~al.}(2017){Diemer}, {Mansfield}, {Kravtsov}, \&
  {More}}]{diemer_et_al_2017}
{Diemer}, B., {Mansfield}, P., {Kravtsov}, A.~V., \& {More}, S. 2017, ArXiv
  e-prints, arXiv:1703.09716

\bibitem[{{Diemer} {et~al.}(2013{\natexlab{b}}){Diemer}, {More}, \&
  {Kravtsov}}]{diemer_et_al_2013}
{Diemer}, B., {More}, S., \& {Kravtsov}, A.~V. 2013{\natexlab{b}}, \apj, 766,
  25

\bibitem[{{Fillmore} \& {Goldreich}(1984)}]{fillmore_goldreich_1984}
{Fillmore}, J.~A., \& {Goldreich}, P. 1984, \apj, 281, 1

\bibitem[{{Goodman} \& {Weare}(2010)}]{goodman_weare_2010}
{Goodman}, J., \& {Weare}, J. 2010, Communications in Applied Mathematics and
  Computational Science, 5, 65

\bibitem[{{G{\'o}rski} {et~al.}(2005){G{\'o}rski}, {Hivon}, {Banday},
  {Wandelt}, {Hansen}, {Reinecke}, \& {Bartelmann}}]{gorski_et_al_2005}
{G{\'o}rski}, K.~M., {Hivon}, E., {Banday}, A.~J., {et~al.} 2005, \apj, 622,
  759

\bibitem[{{Gringorten} \& {Yepez}(1992)}]{gringorten_yepez_1992}
{Gringorten}, I.~I., \& {Yepez}, P.~J. 1992, Instrumentation Papers, 343, 1

\bibitem[{{Gunn} \& {Gott}(1972)}]{gunn_gott_72}
{Gunn}, J.~E., \& {Gott}, III, J.~R. 1972, \apj, 176, 1

\bibitem[{{Hahn} \& {Angulo}(2016)}]{hahn_angulo_2016}
{Hahn}, O., \& {Angulo}, R.~E. 2016, \mnras, 455, 1115

\bibitem[{{Heath}(1977)}]{heath_1977}
{Heath}, D.~J. 1977, \mnras, 179, 351

\bibitem[{{Jing} \& {Suto}(2002)}]{jing_suto_2002}
{Jing}, Y.~P., \& {Suto}, Y. 2002, \apj, 574, 538

\bibitem[{{Kazantzidis} {et~al.}(2006){Kazantzidis}, {Zentner}, \&
  {Kravtsov}}]{kazantzidis_et_al_2006}
{Kazantzidis}, S., {Zentner}, A.~R., \& {Kravtsov}, A.~V. 2006, \apj, 641, 647

\bibitem[{{Klypin} {et~al.}(2011){Klypin}, {Trujillo-Gomez}, \&
  {Primack}}]{klypin_et_al_2011}
{Klypin}, A.~A., {Trujillo-Gomez}, S., \& {Primack}, J. 2011, \apj, 740, 102

\bibitem[{{Komatsu} {et~al.}(2011){Komatsu}, {Smith}, {Dunkley}, {Bennett},
  {Gold}, {Hinshaw}, {Jarosik}, {Larson}, {Nolta}, {Page}, {Spergel},
  {Halpern}, {Hill}, {Kogut}, {Limon}, {Meyer}, {Odegard}, {Tucker}, {Weiland},
  {Wollack}, \& {Wright}}]{komatsu_et_al_2011}
{Komatsu}, E., {Smith}, K.~M., {Dunkley}, J., {et~al.} 2011, \apjs, 192, 18

\bibitem[{{Kravtsov} \& {Borgani}(2012)}]{kravtsov_borgani_2012}
{Kravtsov}, A.~V., \& {Borgani}, S. 2012, \araa, 50, 353

\bibitem[{{Lahav} {et~al.}(1991){Lahav}, {Lilje}, {Primack}, \&
  {Rees}}]{lahav_et_al_1991}
{Lahav}, O., {Lilje}, P.~B., {Primack}, J.~R., \& {Rees}, M.~J. 1991, \mnras,
  251, 128

\bibitem[{Mansfield {et~al.}(2017)Mansfield, Kravtsov, \&
  Diemer}]{SHELLFISH_2016}
Mansfield, P., Kravtsov, A., \& Diemer, B. 2017, phil-mansfield/shellfish: AAS
  Release, , , doi:10.5281/zenodo.546241.
\newblock \url{https://doi.org/10.5281/zenodo.546241}

\bibitem[{{More} {et~al.}(2015){More}, {Diemer}, \&
  {Kravtsov}}]{more_et_al_2015}
{More}, S., {Diemer}, B., \& {Kravtsov}, A.~V. 2015, \apj, 810, 36

\bibitem[{{More} {et~al.}(2016){More}, {Miyatake}, {Takada}, {Diemer},
  {Kravtsov}, {Dalal}, {More}, {Murata}, {Mandelbaum}, {Rozo}, {Rykoff},
  {Oguri}, \& {Spergel}}]{more_et_al_2016}
{More}, S., {Miyatake}, H., {Takada}, M., {et~al.} 2016, ArXiv e-prints,
  arXiv:1601.06063

\bibitem[{{Patej} \& {Loeb}(2016)}]{patej_loeb_16}
{Patej}, A., \& {Loeb}, A. 2016, \apj, 824, 69

\bibitem[{Penna \& Dines(2007)}]{penna_dines_2007}
Penna, M.~A., \& Dines, K.~A. 2007, IEEE Transactions on Pattern Analysis and
  Machine Intelligence, 29, 1673

\bibitem[{{Powell} \& {Abel}(2014)}]{powell_abel_14}
{Powell}, D., \& {Abel}, T. 2014, ArXiv e-prints, arXiv:1412.4941

\bibitem[{{Rines} {et~al.}(2013){Rines}, {Geller}, {Diaferio}, \&
  {Kurtz}}]{rines_13}
{Rines}, K., {Geller}, M.~J., {Diaferio}, A., \& {Kurtz}, M.~J. 2013, \apj,
  767, 15

\bibitem[{Savitzky \& Golay(1964)}]{savitzky_golay_1964}
Savitzky, A., \& Golay, M. J.~E. 1964, Analytical Chemistry, 36, 1627

\bibitem[{{Shi}(2016)}]{shi_2016}
{Shi}, X. 2016, \mnras, 459, 3711

\bibitem[{{Springel}(2005)}]{springel_2005}
{Springel}, V. 2005, \mnras, 364, 1105

\bibitem[{{Sunayama} {et~al.}(2016){Sunayama}, {Hearin}, {Padmanabhan}, \&
  {Leauthaud}}]{sunayama_et_al_2016}
{Sunayama}, T., {Hearin}, A.~P., {Padmanabhan}, N., \& {Leauthaud}, A. 2016,
  \mnras, 458, 1510

\bibitem[{{Tolman}(1934)}]{tolman_1934}
{Tolman}, R.~C. 1934, Proceedings of the National Academy of Science, 20, 169

\bibitem[{{Tully}(2015)}]{tully_15}
{Tully}, R.~B. 2015, \aj, 149, 54

\bibitem[{{Umetsu} \& {Diemer}(2017)}]{umetsu_diemer_2017}
{Umetsu}, K., \& {Diemer}, B. 2017, \apj, 836, 231

\bibitem[{{Wang} {et~al.}(2009){Wang}, {Mo}, \& {Jing}}]{wang_et_al_2009}
{Wang}, H., {Mo}, H.~J., \& {Jing}, Y.~P. 2009, \mnras, 396, 2249

\bibitem[{{Wetzel} {et~al.}(2014){Wetzel}, {Tinker}, {Conroy}, \& {van den
  Bosch}}]{wetzel_et_al_2014}
{Wetzel}, A.~R., {Tinker}, J.~L., {Conroy}, C., \& {van den Bosch}, F.~C. 2014,
  \mnras, 439, 2687

\bibitem[{{Zemp} {et~al.}(2011){Zemp}, {Gnedin}, {Gnedin}, \&
  {Kravtsov}}]{zemp_et_al_2011}
{Zemp}, M., {Gnedin}, O.~Y., {Gnedin}, N.~Y., \& {Kravtsov}, A.~V. 2011, \apjs,
  197, 30

\bibitem[{{Zentner} {et~al.}(2016){Zentner}, {Hearin}, {van den Bosch},
  {Lange}, \& {Villarreal}}]{zentner_et_al_2016}
{Zentner}, A.~R., {Hearin}, A., {van den Bosch}, F.~C., {Lange}, J.~U., \&
  {Villarreal}, A. 2016, ArXiv e-prints, arXiv:1606.07817

\bibitem[{{Zhuravleva} {et~al.}(2013){Zhuravleva}, {Churazov}, {Kravtsov},
  {Lau}, {Nagai}, \& {Sunyaev}}]{zhuravleva_et_al_2013}
{Zhuravleva}, I., {Churazov}, E., {Kravtsov}, A., {et~al.} 2013, \mnras, 428,
  3274

\end{thebibliography}

\appendix

\section{An algorithm for fast line of sight density estimates}
\label{sec:intersection}

\begin{figure}
\centering
\includegraphics[width=\columnwidth]{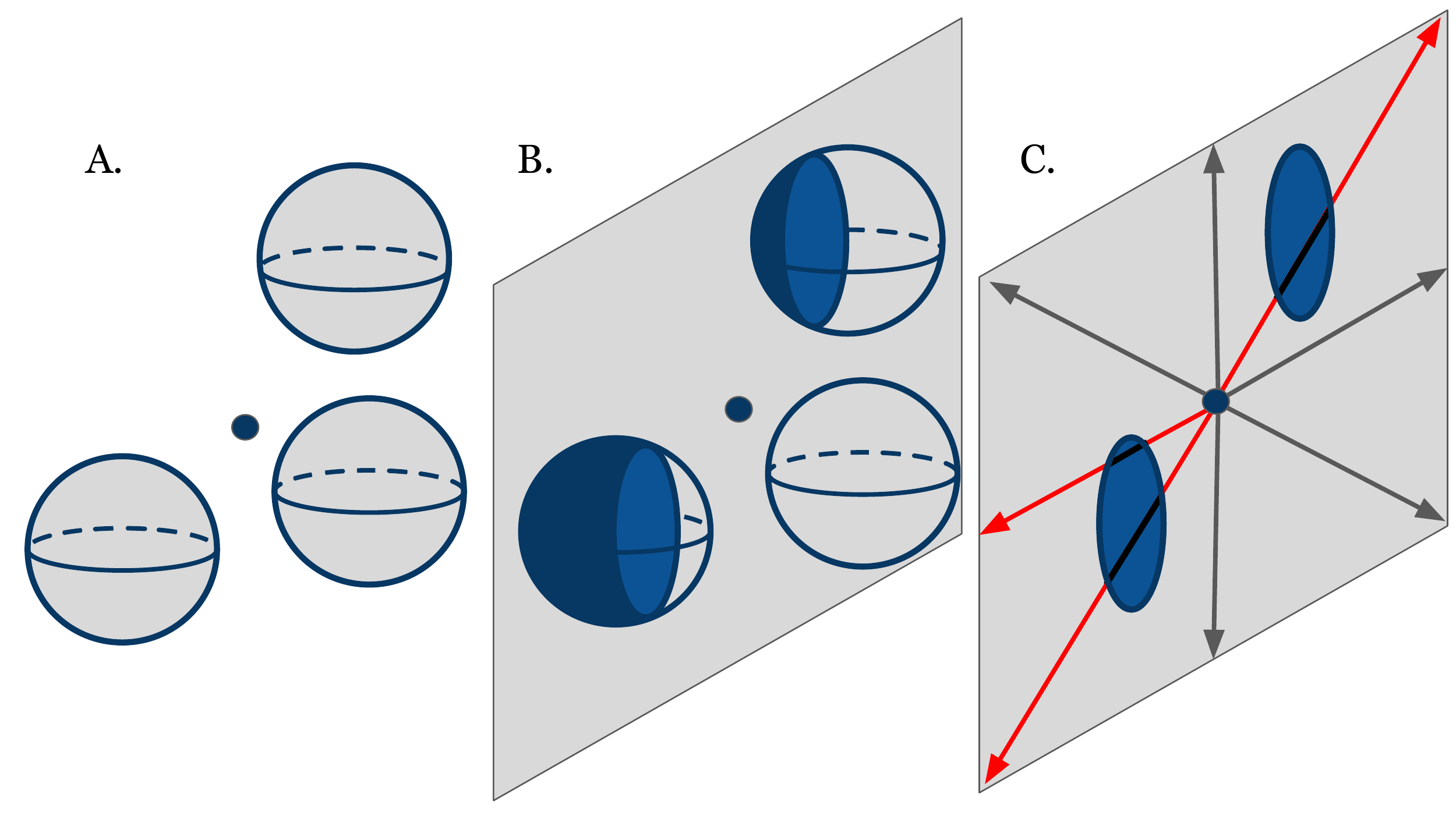}
\caption{
    An illustration of the \textsc{UpdateProfiles'} algorithm described in
    Appendix \ref{sec:intersection}. Panel A shows three, $S_i$, associated
    with three dark matter particles around the center of a halo, shown as a
    solid circle. Panel B shows one of the random planes, $P\in P_H$, passing
    through the halo center and intersections of $P$ with each $S_i$. The top
    and left spheres intersect $P$ while the remaining sphere does not. Panel
    C shows intersection checks being performed between the 2D intersection,
    $C_i$, of $S_i$ and $P$ along a set of lines of sight, $L_{P, i}$, in the
    plane. Inspection of the angular locations of the edges of the $C_i$ shows
    that only the red lines of sight could intersect them, and thus these
    spheres contribute to density profiles only along the red lines. This last
    panel corresponds to the code inside the innermost conditional of the
    algorithm.}  
\label{fig:intr_algo} 
\end{figure}

In this Appendix we describe the method our algorithm uses to construct density
profiles along a set of lines of sight via the evaluation of
Equation~\ref{eq:los_rho}. Generally, this can be broken up into two steps: first,
the set of all spheres which intersect with a particular halo, $H$, is found,
and second, for every sphere, $S,$ which intersects with $H$ a procedure
\textsc{UpdateProfiles}($S$, $H$) is run, which evaluates a term in
Equation~\ref{eq:los_rho} corresponding to $S$ for every line of sight in $H$.

In general, the first step is straightforward to perform efficiently. Even
rudimentary spatial partitioning (such as breaking the simulation's particles
into $\approx 10^2 - 10^3$ spatially coherent segments) results in this step
being highly subdominant to the second, making \textsc{UpdateProfiles} the only
performance bottleneck of this algorithm. A naive implementation of
\textsc{UpdateProfiles} would look like the following:

\begin{algorithmic}
    \Procedure{UpdateProfiles}{$S$, $H$}
    \For{\textbf{each} $L$ in $L_H$}
         \If{$S$ intersects $L$}
             \State $R_{\rm enter},$ $R_{\rm exit}$ $\gets$ \Call{IntersectionRadii}{$L$, $S$}
             \State \Call{InsertToProfile}{$L,$ $\rho_{\rm S},$ $R_{\rm enter},$ $R_{\rm exit}$}
         \EndIf 
    \EndFor
    \EndProcedure
\end{algorithmic}

Here, $L_H$ is the set of all line of sight profiles belonging to the halo
$H$ and $\rho_S$ is the density of the sphere $S.$ The existence of two simple
functions has been assumed: \textsc{IntersectionRadii}($L$, $S$) calculates the
radii at which the line of sight $L$ enters and exits the sphere $S,$
respectively, and \textsc{InsertToProfile}($L$, $\rho$, $R$, $R'$) inserts a
rectangular function with amplitude $\rho$ between $R$ and $R'$
to the profile corresponding to the line of sight $L$.

Because \textsc{UpdateProfiles} performs an intersection check for every line
of sight in $L_H$, the asymptotic cost of this approach is $O(|L_H|)$. Because
$|L_H|$ is on the order of 10$^4$ for the parameter set used in this paper,
this leads to a large number of expensive intersection checks being
performed for every particle, with the vast majority of these checks failing.

We take an alternative approach that allows us to avoid performing explicit
calculations on any line of sight which does not intersect $S.$ We require
that lines of sight exist within a set of planes, $P_H$, that
$|P_H| \ll |L_H|,$ and that lines of sight are oriented in uniformly-spaced
``rings'' within their respective planes. This strong geometric restriction
allows for two optimizations: first, intersection checks are performed on
entire planes before any calculations are done on individual lines of sight,
and second, we calculate the angle subtended by $S$ in intersected planes,
which allows us to find the exact set of lines of sight intersected by $S$ in
the plane. Concretely, our approach is:

\begin{algorithmic}
    \Procedure{UpdateProfiles$'$}{$S$, $H$}
    \For{\textbf{each} $P$ in $P_H$}
        \If{$S$ intersects $P$}
            \State $C$ $\gets$ \Call{SliceSphere}{$S$, $P$}
            \State $\theta_{\rm low}$, $\theta_{\rm high}$ $\gets$ \Call{AngularRange}{$P$, $C$, $H$}
            \State $i_{\rm low}$, $i_{\rm low}'$ $\gets$ \Call{ProfileIndices}{$\theta_{\rm low}$}
            \State $i_{\rm high}$, $i_{\rm high}'$ $\gets$ \Call{ProfileIndices}{$\theta_{\rm high}$}
            \For{\textbf{each} $i$ in [$i_{\rm low}$, $i_{\rm high}'$]}
                \State $R_{\rm enter}$, $R_{\rm exit}$ $\gets$ \Call{IntersectionRadii}{$L_{P, i}$, $S$}
                \State \Call{InsertToProfile}{$L_{P, i},$ $\rho_{\rm S},$ $R_{\rm enter},$ $R_{\rm exit}$}
            \EndFor
        \EndIf
    \EndFor
    \EndProcedure
\end{algorithmic}

Here, $P_H$ is the set of all planes of profiles belonging to the halo $H$
and $L_{P, i}$ is the $i^{\rm th}$ profile within the profile ring
corresponding to the
plane $P$. Here, the existence of several simple functions has been assumed:
\textsc{SliceSphere}($S$, $P$) returns the circle created by slicing the
sphere $S$ by the plane $P$; \textsc{AngularRange}($P$, $C$, $H$) returns two
angles specifying the angular wedge within the plane $P$ which the circle
$C$ subtends relative to the center of the halo $H$; and
\textsc{ProfileIndices}($\theta$) returns the indices of the two nearest
profiles to the angle $\theta,$ with the profile corresponding to the lower
angle being returned first. For ease of reading, the pseudocode which would 
handle the periodicity of angles at $\theta = 0 \equiv 2\pi$ has been omitted.

This method is illustrated in Figure \ref{fig:intr_algo}. Panel A shows a
collection of spheres collected around a halo center, panel B shows the
results of calling \textsc{SliceSphere} on each of these spheres for a
particular plane, and panel C shows the profiles (in red) which would recieve
intersection checks within the innermost loop of \textsc{UpdateProfiles$'$}.

The asymptotic cost of \textsc{UpdateProfiles$'$} is $O(|P_H| + I_{H,S})$,
where $I_{H,S}$ is the number of profiles in the halo $H$ which intersect the
sphere $S$. Since both $|P_H|$ and $I_{H,S}$ are multiple orders of magnitude
smaller than $|L_H|$, this results in a significant increase in performance. In
practice we find that the plane intersection checks are subdominant to the
cost of the innermost loop.

The method described above is further optimized in several ways:
\begin{itemize}
        \item If \textsc{InsertToProfile} is implemented naively - by
            representing
            profiles as arrays containing $\rho(r)$ and updating
            every element of the profile which is within the inserted
            rectangular function - it is the dominant cost of
            \textsc{UpdateProfiles$'$}. To prevent this, we represent our
            profiles as arrays containing $d\rho(r)/dr$. Since the derivative
            of a rectangular function is two delta functions, updating the
            derivative profile only requires updating array elements close to
            the edges of the rectangular function (note that in the discrete
            case this requires four element updates: two for each edge). Once
            \textsc{UpdateProfiles$'$} has been called on every target
            sphere, each derivative profile is integrated to obtain $\rho(r)$.
        \item Instead of explicitly performing the 3D
            \textsc{IntersectionRadii}($L_{P,i}$, $S$), a faster 2D
            analog is used to find the intersection radii of the
            projection of $L_{P,i}$ onto $P$ with the circle $C$.
        \item A successful intersection check between $P$ and $S$ is performed
            in a way which immediately results in the value that would be
            returned by \textsc{SliceSphere}($S$, $P$), as these two
            calculations share many geometric operations.
\end{itemize}

This algorithm is straightforward to generalize to non-constant density spheres
and to density estimates constructed from other geometric solids (most notably
tetrahedra), although the publicly released version of \textsc{Shellfish} does
not allow access to either feature.

\section{Splashback candidate filtering algorithm}
\label{sec:filter_algo}

The Appendix will outline the filtering algorithm which we qualitatively
introduced in section \ref{sec:filter}

The first step of constructing the filtering loop is dividing the point
distribution into $2^{N_{\rm rec}}$ uniformly spaced angular wedges, for some
user-defined $N_{\rm rec}$. We calculate an \emph{anchor point}
for each wedge, which is an estimate of the average location of the splashback
shell within that wedge.

The location of the anchor point within the $i^{\rm th}$ wedge is given by
\begin{equation}
    R_{{\rm anchor},i},~\theta_{{\rm anchor},i}= \textsc{AnchorRadius}(0,~i),~2\pi\frac{i+0.5}{2^{N_{\rm rec}}}.
\end{equation}
Here, $i$ is zero-indexed and \textsc{AnchorRadius} is the following recursive
algorithm:

\begin{algorithmic}
    \Function{AnchorRadius}{$k$, $i$}
        \State $\theta_{\rm low} \gets 2\pi \lfloor i/2^k\rfloor 2^{k - N_{\rm rec}}$
        \State $\theta_{\rm high} \gets 2\pi (\lfloor i/2^k\rfloor + 1) 2^{k - N_{\rm rec}}$
        \State $f \gets$ \Call{WedgeKDE}{$\theta_{\rm low}$, $\theta_{\rm high}$}
    \If{$k$ = $N_{\rm rec}$}
        \State \Return{\Call{GlobalMaximum}{$f$}}
    \Else
        \State $R_{\rm anchor} \gets$ \Call{AnchorRadius}{$k+1$, $i$}
        \State maxes $\gets$ \Call{LocalMaxima}{$f$}
        \State $R_{\rm anchor}'\gets$ min$_R\{|R_{\rm anchor} - R|~\forall~R \in {\rm maxes} \}$
        \If{$|R_{\rm anchor} - R_{\rm anchor}'|~<~ R_{\rm refine}$}
            \State \Return{$R_{\rm anchor}'$}
        \Else
            \State \Return{$R_{\rm anchor}$}
        \EndIf
    \EndIf

    \EndFunction
\end{algorithmic}

We assume the existence of three simple functions:
\textsc{GlobalMaximum}($f$), which returns the global maximum of the function $f$;
\textsc{LocalMaxima}($f$), which returns all the local maxima of the function $f$; and
\textsc{WedgeKDE}($\theta_{\rm low}$, $\theta_{\rm high}$), which returns a
kernel density estimate (KDE)
corresponding to the points contained within the wedge with boundaries
$\theta_{\rm low}$ and $\theta_{\rm high}$.
A KDE is a method
for converting a set of discrete points into a continuous density estimate by
applying a smoothing kernel to every point. It performs much the same role as
a histogram, except that an explicit choice of bin edges is replaced by an
explicit choice of the smoothing kernel.
For our purposes, the most useful property of a KDE is that it provides a
simple way to estimate the point of maximum density.
We define our KDE as the function
\begin{equation}
    {\rm KDE}(r) = \sum_j\exp{\left(-\frac{(r - r_{j})^2}{2R^2_{\rm KDE}}\right)}
\end{equation}
where $R_{\rm KDE}$ is a user-defined smoothing scale, and $r_j$ is a set of
points.

The intuition behind this
approach is that most candidate points in the plane correspond to lines of
sight crossing
the splashback shell, so the maximum of the $k=N_{\rm rec}$ KDE is a good
0$^{\rm th}$ order estimate of of its location. Smaller wedges give more refined
estimates. But if their estimate deviates too far from the coarser estimates,
it's likely that the region corresponds to a filament or a subhalo.

Once anchor points have been found for each wedge, we fit a cubic interpolating
spline to them in the $\theta$ - $R$ plane. This spline is the aforementioned
filtering loop. To remove boundary effects, the range of the anchor points
is extended to $[-2\pi,~4\pi)$ prior to fitting, but the spline
is only ever evaluated in the $[0~2\pi)$ range. We then remove all points which
are further than some distance, $R_{\rm filter}$ from this spline.

This procedure introduces three new free parameters, $R_{\rm KDE}$,
$R_{\rm refine}$, and $R_{\rm filter}$. Tests indicate that the final shells
are robust to changes in $R_{\rm KDE}$ and $R_{\rm filter}$, as long
as they are of the same order of magnitude as $R_{\rm refine}$. For this
reason we simplify parameters by requiring
\begin{equation}
    R_{\rm KDE} = R_{\rm refine} = R_{\rm filter} = R_{\rm max} / \eta.
\end{equation}
Here $\eta$ is a tunable parameter which dictates how strict the filtering
process is. Higher values of $\eta$ are stricter than lower values of $\eta$.

\section{Parameter-specific convergence tests}
\label{sec:setting_params_app}

Most of the fiducial values of parameters of our algorithm listed in
Table \ref{tab:param} were set using one of the following
three approaches, which start with constructing a representative sample
of halos and identifying their splashback shells and estimating their properties 
for a range of values $p_i$ for the selected parameter $p$.
\begin{enumerate}
    \item[A:] Many parameters are known to be optimized when taken to either the
        low value or high value limit, but also decrease the performance
        of the algorithm as the
        parameter approaches this limit. In addition to the shells for the
        $p_{i}$ values, we also fit a shell with $p$ set to some very large
        value, $p_{\rm limit}$. For each halo we calculate a curve
        representing the fractional
        difference between the shells calculated with $p_i$ and with 
        $p_{\rm limit}$ for each of the properties
        defined in Equations~\ref{eq:r_def} - \ref{eq:A_def}. We then set
        the parameter to the lowest $p_i$ which leads to an average
        fractional error of $\lesssim$ 1\%.
    \item[B:] Some parameters are not optimized in either the low or high
        value limit. For each halo we construct curves for each of the
        properties
        defined in Equations~\ref{eq:r_def} - \ref{eq:A_def}. We manually inspect
        these curves: if they generally show an unchanging ``plateau'' for these
        properties over a wide range of $p_i$ values, we set the parameter to an
        arbitrary $p_i$ in the center of the plateau. The existence of a
        plateau over a wide range of $p_i$ indicates that the shell shapes
        depend only weakly on this parameter.
    \item[C:] For parameters where method B was attempted but a no wide 
        plateau was found, we incorporate qualitative assessment of the shells
        into the selection procedure. For a pair of parameter values, $p_i$ and
        $p_j$, we visually inspect every halo in the test set, compare the
        shells produced by both values to the underlying density field, and
        select one of the two as a qualitatively better fit. Once this has
        been completed for every halo, we label the parameter value with more
        successful fits as the better value. This allows us to construct an
        fitness ordering on all the values of $p_i$. We then select the
        maximally fit parameter. In principle, this methodology could lead
        to researcher-dependent results, but for the three parameters where we
        used this method, the optimal value was not ambiguous.
\end{enumerate}
The specific methods we used to set each algorithm parameter are listed in
Table \ref{tab:param}. 
In all cases the halo sample is divided into $M_{\rm 200m}$-selected and
$\Gamma_{\rm DK14}$-selected subsets to test for parameter dependence on halo properties.
In all cases, we found no such dependence.
Parameters which involved additional testing methodology are described below:

\subsection{Setting $R_{\rm kernel}$}
\label{sec:kernel_test}

In order for our algorithm to identify the splashback shell reliably, we need to sample the density distribution around
the shell well.  However, 
typical densities in this region are 
$(0.1 - 10)\times  \rho_{\rm m}$ (see Figure \ref{fig:los_prof}) and there are often relatively few particles. To compensate
for this, we need to make the radius of the spheres associated with particles, $R_{\rm kernel}$ sufficiently large. 

To find the optimal value of $R_{\rm kernel}$, we use an approach similar to the approach A above. We generate
a representative sample of halos and fit Penna-Dines shells
to each halo in the sample for different values of $R_{\rm kernel}$. We then
find the smallest converged value of $R_{\rm kernel}$ for each halo. An example
of this comparison is shown in Figure \ref{fig:kernel_test}. This figure also
illustrates the second test described in section \ref{sec:tests}: in >99\% of
cases, $R_{\rm sp}$ falls within the visual fall-off region of the halo.

We find that for halos with $N_{\rm 200m}~\geq~10^6$, properties of the splashback shells converge for
$R_{\rm kernel}\gtrsim 0.1R_{\rm 200m}$ and for halos with
$N_{\rm 200m}\approx 5\times 10^4$ for 
$R_{\rm kernel} \gtrsim 0.2R_{\rm 200m}$. For simplicity, we set
$R_{\rm kernel}$ to $0.2R_{\rm 200m}$ for all halos.

\subsection{Setting $N_{\rm planes}$}
\label{sec:ring_test}

\begin{figure}
   \centering
   \subfigure[]{
        \label{fig:kernel_test}
        \includegraphics[height=0.4\textwidth]{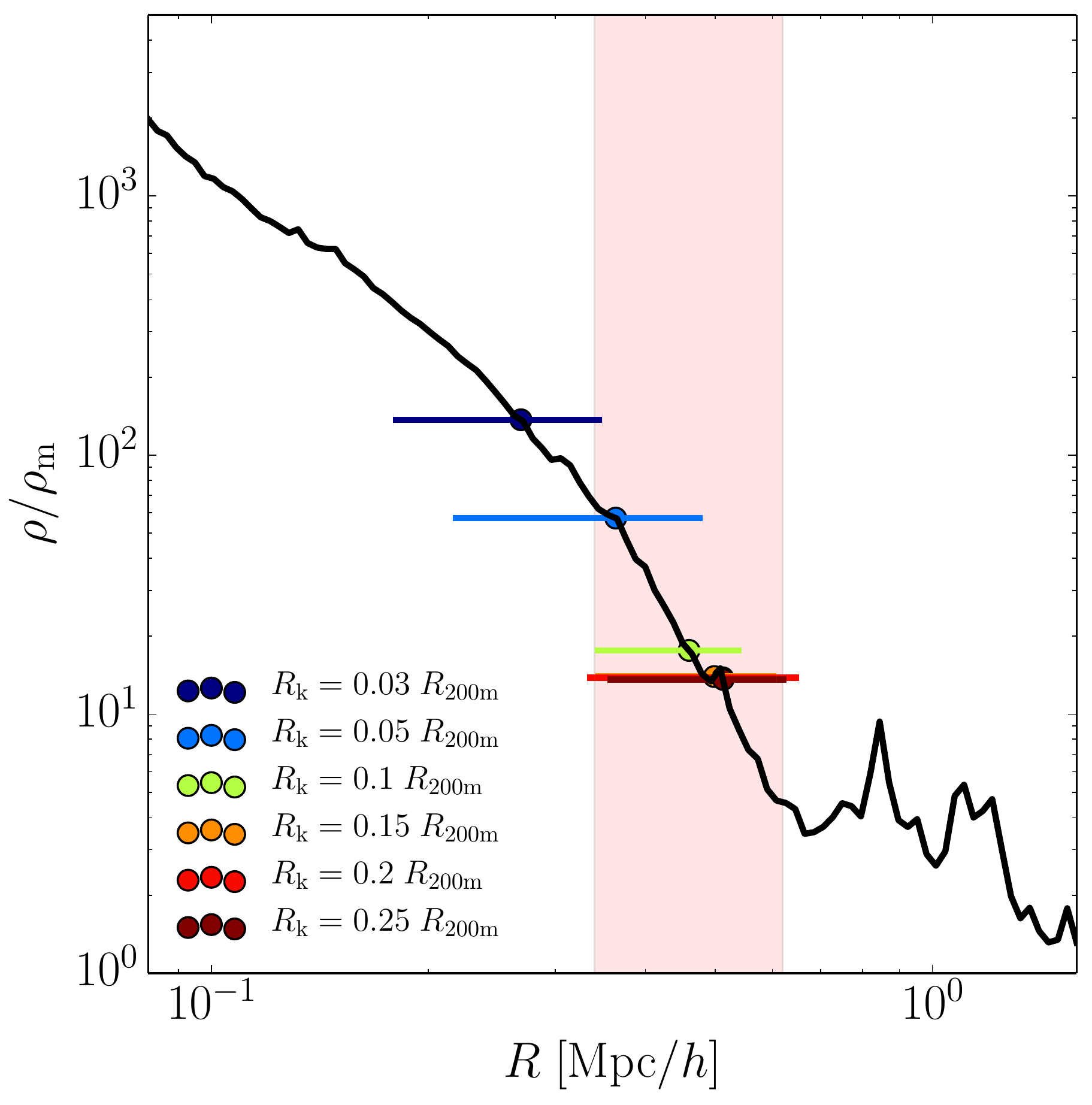}
   }
   \hspace{0.5cm}
   \subfigure[]{
        \label{fig:ring_converge}
        \includegraphics[height=0.4\textwidth]{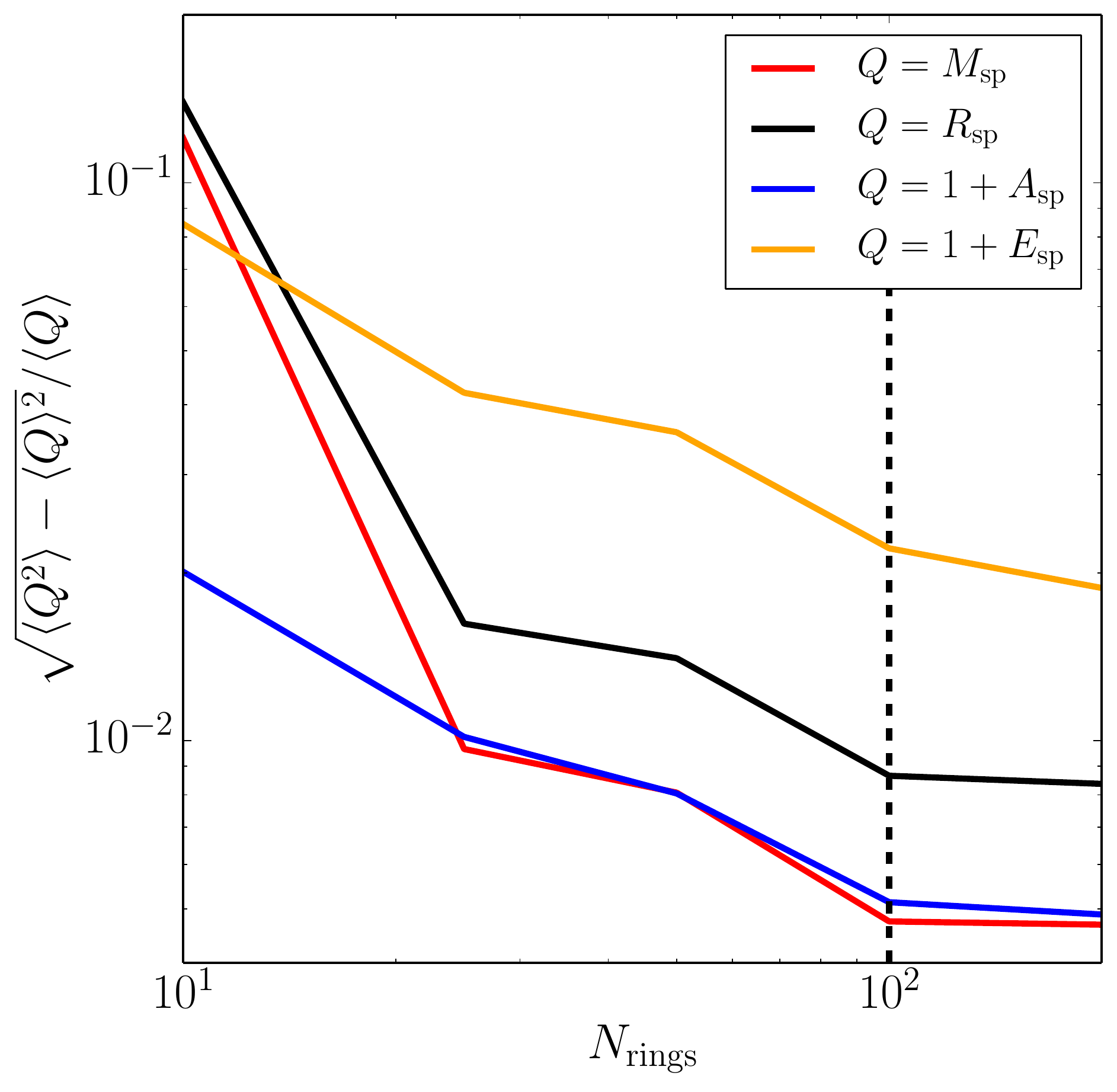}
   }
   \caption{\emph{Left}: Convergence test of $R_{\rm sp}$ as a function of kernel radius for
      a representative halo. The black curve is the density profile of the halo
      obtained through conventional particle binning,
      the points are the $R_{\rm sp}$ values measured from density fields
      generated with different kernel radii,
      the horizontal lines show the range spanned by the minimum and maximum
      radii of these shells, and the shaded red region corresponds
      to the radial range which was visually identified as corresponding to the
      splashback range. This region was found by eye without knowledge of the
      measurements made by \textsc{Shellfish} in accordance with the procedure
      outlined
      in section \ref{sec:tests}. For this halo, $R_{\rm sp}$ is converged
      for kernel radii above $0.15R_{\rm k}$.
            \emph{Right}:The mean fractional stochastic error in shell parameters (defined
       in Equations~\ref{eq:r_def} - \ref{eq:A_def}) as a function of
       $N_{\rm planes}$. The vertical dashed line corresponds to
       $N_{\rm planes} = 100$, the value given in Table \ref{tab:param}.}
\end{figure}

The parameter which has the largest effect on the stochastic error, as opposed
to systematic error, in estimating shell shape is $N_{\rm planes}$. To determine a value
for this parameter, we follow a procedure similar to method A. We identify 
the splashback shells for a
representative sample of halos  for five values of $N_{\rm planes}$, using randomly
oriented normal vectors for each plane so that no lines of sight are
shared between two different realizations. We then calculate
the fractional standard deviation between shell properties determined for
 different random realizations of a given number of planes $N_{\rm planes}$,
$\sqrt{\langle Q^2\rangle -\langle Q\rangle^2} / \langle Q\rangle$ for each
quantity $Q$ defined in Equations~\ref{eq:r_def} - \ref{eq:A_def}. This standard 
deviation is plotted as a function of $N_{\rm planes}$ in Figure \ref{fig:ring_converge}. For
$N_{\rm planes} = 100$, \textsc{Shellfish} achieves sub-percent level per-halo scatter in
$R_{\rm sp}$, $M_{\rm sp}$, and $1 + A_{\rm sp}$ and less than 2\% scatter in
$1 + E_{\rm sp}$. We do not find any evidence that the amplitudes of the curves shown in
Figure \ref{fig:ring_converge} depend on halo mass or accretion rate.

\section{Computing Moment of Inertia-Equivalent Ellipsoidal Shell Axes}
\label{sec:ellipsoid}

It is non-trivial to analytically compute axis ratios from the moments of
inertia for a constant-density ellipsoidal shell. Assuming that the shell
is sampled by some collection of particles with weights $m_k$, the moments can
be obtained by calculating the eigenvalues of the mass-distribution tensor,
\begin{equation}
    \label{eq:mass_distr}
    M_{i,j} = \sum_k m_k (\vec{r}_k)_i(\vec{r}_k)_j.
\end{equation}
The eigenvalues of the mass-distribution tensor are straightforward to
calculate for a homoeoid: the volume enclosed by two ellipsoids with the
same axes ratios and with aligned major axes $a$ and $a^\prime$. In the limit where $a^\prime \rightarrow a$, the
eigenvalues are given by
\begin{equation}
    \label{eq:homoeoid}
    M_i = M_{\rm tot}\frac{a_i^2}{3},
\end{equation}
where $M_i$ and $a_i$ are the moment and ellipsoid axis aligned with the
i$^{\rm th}$ Cartesian axis and $M_{\rm tot} = \sum_k m_k$. Note that this
notation for ellipsoid axes is different that the convention used in
Equation~\ref{eq:axes_def}. Although an infinitely thin homoeoid is often equated with a uniform-density ellipsoid surface
in the literature \citep[see, for example][]{zemp_et_al_2011}, it actually corresponds to an ellipsoid surface with
a non-uniform density. This non-uniformity means that major (minor) axes derived from 
Equation~\ref{eq:homoeoid} are too small (large). This bias increases with increasing ellipticity: for ellipsoidal
shells with axes ratio of $\approx 2:1$, this can bias measured axes ratios by tens of per cent.

A more accurate approximation would be to model a uniform density ellipsoidal
shell by the volume enclosed by two ellipsoids with axes $a$, $b$, $c$ and
$a + \delta$, $b + \delta$, $c + \delta$ and to take $\delta \rightarrow 0$.
This shape gives eigenvalues of
\begin{equation}
    \label{eq:delta_ellipsoid}
    M_i = M_{\rm tot} \frac{a_i^2}{5}\left(\frac{a_ia_j + 3a_ja_k + a_ka_i}
         {a_ia_j + a_ja_k + a_ka_i}\right),
\end{equation}
which can then be numerically solved to obtain ellipsoid axes. Although
for ellipsoids with large axes ratios Equation~\ref{eq:delta_ellipsoid} is a closer approximation than
Equation~\ref{eq:homoeoid},  it still introduces
errors close to our $N_{\rm ring} = 100$ stochastic noise limit. Thus for large
axes-ratio ellipsoids we compute the mapping empirically.

We define the quantities $A_i \equiv \sqrt{M_i/M_{\rm tot}}$ and
$R \equiv (M_iM_jM_k/M_{\rm tot}^3)^{1/6}$. Note that both $A_i$ and $R$ can be
measured directly from the input point distribution. First, we generate a grid
of ellipsoids in $a_0/a_1$ - $a_0/a_2$ space. Next we numerically compute
$A_0/A_1$, $A_0/A_2$, and $a_0/R_{\rm V}$ for each ellipsoid. The resultant
$A_0/A_1$ and $A_0/A_2$ values form a sheared grid, so we Delaunay triangulate
\citep{delaunay_1934}
the $A_0/A_1$ - $A_0/A_2$ plane and perform linear interpolation on the
resulting triangles. We construct three such interpolators which map from
($A_0/A_1$, $A_0/A_2$) pairs to $a_0/a_1$, $a_0/a_2$, and $a_0/R$, respectively.
These interpolators can then be used to find $a_0$, $a_1$, and $a_2$ using only
the eigenvalues of the mass-distribution tensor.

\end{document}